%% file: paper.tex
\documentclass[a4paper,11pt]{article}
\usepackage{jheppub}
\usepackage[english]{babel}
\usepackage{subcaption}
\usepackage{placeins}
\usepackage{xspace}
\usepackage{fancyvrb}
\usepackage{relsize}
\usepackage[utf8]{inputenc}
\usepackage{lineno}
\usepackage{stackrel}
\usepackage{amsmath}
\usepackage{mathtools}
\usepackage{amsmath,amssymb,enumerate,alltt,xspace,multirow}
\usepackage{graphicx}

\input texmacros.tex

\RequirePackage[T1]{fontenc}

\usepackage{newtxtext}
\RequirePackage{flushend}
\RequirePackage[numbers,sort&compress]{natbib}

\begin{document}
\setlength{\abovedisplayskip}{6pt}
\setlength{\belowdisplayskip}{6pt}
\setlength{\abovedisplayshortskip}{6pt}
\setlength{\belowdisplayshortskip}{6pt}

\title{Resonance-aware NLOPS matching for off-shell \boldmath$t\bar t+tW$
production with semileptonic decays}

\preprint{ZU-TH 37/23,\, MS-TP-23-43}

\author[a]{Tom\'a\v{s} Je\v{z}o,} 
\author[b]{Jonas M. Lindert,} 
\author[c]{Stefano Pozzorini}

\emailAdd{tomas.jezo@uni-muenster.de}
\emailAdd{j.lindert@sussex.ac.uk}
\emailAdd{pozzorin@physik.uzh.ch}

\affiliation[a]{Institut f{\"u}r Theoretische Physik, Westf{\"a}lische Wilhelms-Universit{\"a}t M{\"u}nster, \\Wilhelm-Klemm-Stra{\ss}e 9, D-48149 M{\"u}nster, Germany}
\affiliation[b] {Department of Physics and Astronomy, University of Sussex, Brighton BN1 9QH, UK}
\affiliation[c] {Physics Institute, Universit\"at Z\"urich, Z\"urich,
  Switzerland}


\abstract{The increasingly high accuracy of top-quark studies at the LHC
calls for a theoretical description of $t\bar t$ production and decay in
terms of exact matrix elements for the full $2\to 6$ process that includes
the off-shell production and the chain decays of $t\bar t$ and $tW$ intermediate
states, together with their quantum interference.
Corresponding NLO QCD calculations matched to parton showers are
available for the case of dileptonic channels and are implemented in the
\bbfourl{} Monte Carlo generator, which is based on the resonance-aware
\POWHEG method.
In this paper, we present the first NLOPS predictions of this kind
for the case of semileptonic
channels.
In this context, the interplay of off-shell $t\bar t+tW$ production 
with various other QCD and electroweak subprocesses
that yield the same semileptonic final state is discussed in detail.
On the technical side, we improve the resonance-aware \POWHEG procedure by means of
new resonance histories based on matrix elements, which enable a realistic
separation of $t\bar t$ and $tW$ contributions.
Moreover, we introduce a general approach which 
makes it possible to avoid certain spurious
terms that arise from the perturbative expansion of decay widths in any
off-shell higher-order calculation, and which are large enough to jeopardise
physical finite-width effects.
These methods are implemented in a new version of the \bbfourl Monte Carlo generator,
which is applicable to all dileptonic and semileptonic channels, and can
be extended to fully hadronic channels.
The presented results include a NLOPS comparison of off-shell against on-shell
$t\bar t+tW$ production and decay, 
where we highlight various non-trivial aspects related
to NLO and parton-shower radiation in leptonic and hadronic 
top decays.
}

\maketitle

\section{Introduction}
\label{se:intro}

Studies of top-quarks play a key
role in the ongoing physics programme of the Large Hadron Collider (LHC). 
Measurements in the different top-quark production modes, and  especially in
the ubiquitous $\ttbar$ production mode, allow for a detailed
exploration of top-quark interactions, and a precise determination of
fundamental Standard Model (SM) properties such as the top-quark mass.
At the same time, top-quark--pair production represents a
sizeable and challenging background 
for countless measurements and searches at the LHC.
The sensitivity of such analyses critically relies on the precision of
theoretical predictions for the $\ttbar$ production cross section, as well
as for a large variety of kinematic distributions depending on the details
of the nontrivial signatures that result from $t\bar t$ production and
decay.
This calls for the highest possible accuracy 
in the theoretical description of the production 
of top-quark  pairs {\it and} their decays.
In fact, the expected sensitivity of future experimental analyses
requires precise theoretical predictions 
at the level of the 
full $2\to 6$ processes that correspond to $\ttbar$ production 
with dileptonic, semileptonic or fully hadronic decays,
including all relevant off-shell effects, 
irreducible backgrounds and interferences.
Such theoretical predictions provide a unified description of 
off-shell $\ttbar$ and
$\tW$ single-top production, including $\ttbar$--$\tW$ interference
effects~\cite{Cascioli:2013wga,Jezo:2016ujg}.

For the analysis of experimental data, theoretical calculations need to be
matched to parton showers. 
Monte Carlo generators that match NLO QCD calculations 
of on-shell $\ttbar$ production to parton showers
are well established~\cite{Frixione:2003ei,Frixione:2007nw,Alwall:2014hca,Hoeche:2014qda,Cormier:2018tog}.
Such tools describe top-quark decays based on spin-correlated LO matrix
elements~\cite{Frixione:2007zp,Artoisenet:2012st,Garzelli:2014dka,Hoche:2014kca}
and implement a naive modelling of off-shell effects
according to Breit--Wigner distributions.
The emission of QCD radiation 
within top decays is controlled by the 
parton shower~\cite{Sjostrand:2007gs, Sjostrand:2014zea},
which can dispose of built-in matrix-element corrections
that provide a decent approximation of NLO effects.
In the following, tools of this kind are going to be referred to as
on-shell generators.
The first generator that matches 
NNLO QCD calculations\footnote{Fixed-order calculations
are available also at
NLO electroweak (EW)~\cite{
Bernreuther:2008md, 
Kuhn:2006vh, Hollik:2011ps, 
Gutschow:2018tuk, Frederix:2021zsh}
and NNLO~QCD+NLO\,EW~\cite{Czakon:2017wor}. 
}
~\cite{Czakon:2013goa, Czakon:2015owf,Catani:2019iny,
Catani:2019hip}
of on-shell $\ttbar$
production to parton showers 
was presented in~\cite{Mazzitelli:2021mmm,Mazzitelli:2020jio}.

A generator based on NLO QCD calculations where $\ttbar$ production and decay
are both described at NLO in the narrow-width approximation
(NWA)~\cite{Bernreuther:2004jv, Melnikov:2009dn, Campbell:2012uf} was
presented in~\cite{Campbell:2014kua}. Corresponding NNLO QCD implementations
of the NWA are also available, but only at fixed
order~\cite{Gao:2012ja,Brucherseifer:2013iv,Gao:2017goi,Behring:2019iiv,Czakon:2020qbd}.

Concerning $\tW$ single-top production, on-shell NLOPS generators are available in the
five-flavour number scheme
(5FNS)~\cite{Re:2010bp,Frixione:2008yi,Bothmann:2017jfv}. In this scheme,
the NLO QCD corrections\footnote{See~\citere{Beccaria:2006dt,Beccaria:2007tc} for 
$\tW$ production at NLO EW.
} 
involve partonic channels of type $gg\to tW^-\bar b$, which entail resonant
$\ttbar$ topologies that can lead to a double counting of the $\ttbar$ LO
cross section. This issue can be avoided through various methods for the
systematic separation of $\tW$ from $\ttbar$ production~\cite{Zhu:2001hw,
Campbell:2005bb, Frixione:2008yi, White:2009yt, Demartin:2016axk}.
However, such methods are always subject to a degree of arbitrariness
that is either due to ad-hoc prescriptions, violations of gauge invariance,
or to the treatment of interference and off-shell effects.

A fully consistent solution of such issues is provided by calculations
where the production and decays of $\ttbar$ pairs are described in terms of
exact matrix elements for the corresponding $2\to 6$ process, without
relying on the NWA. In this approach, using the complex-mass
scheme~\cite{Denner:2005fg}, all possible $2\to 6$ topologies that involve
off-shell $\ttbar$ and $tW$ intermediate states are handled on the same
footing as contributions to subtopologies with
off-shell $W^+W^-b\bar b$ states.
Corresponding NLO calculations in the 5FNS
are available at NLO QCD for 
dileptonic~\cite{Bevilacqua:2010qb,
Denner:2010jp, Denner:2012yc, Heinrich:2013qaa, Frederix:2013gra,
Cascioli:2013wga} as well as for semileptonic~\cite{Denner:2017kzu}
processes.\footnote{For dileptonic processes
also predictions at NLO EW~\cite{Denner:2016jyo}
and at NLO QCD with one extra jet~\cite{Bevilacqua:2015qha}
are available.}
When performed in the four-flavour number scheme (4FSN),
i.e.~treating $b$-quarks as massive partons, and excluding them
from the inital state,
such off-shell calculations provide a unified NLO
modelling of
$\ttbar$ and $\tW$ production, with a fully consistent treatment of $t\bar
t-tW$ interferences~\cite{Cascioli:2013wga}.

The matching of off-shell NLO calculations to parton showers 
was enabled by nontrivial ``resonance-aware'' extensions of 
the standard matching techniques.
The first resonance-aware matching method
was proposed in~\cite{Jezo:2015aia} as an extension of the
original \POWHEG{} technique~\cite{Nason:2004rx, Frixione:2007vw}.
This approach, which will be referred to as the \POWHEGRES approach, is
based on a probabilistic categorisation of events into different ``resonance
histories'', which correspond to the different combinations of production
and decay subprocess that can contribute to a given off-shell process.
Within each resonance history, \POWHEG radiation is generated in a way that
preserves the virtuality of all resonances. In this way, the
\POWHEGRES approach guarantees that all 
resonances have correct NLO shapes, and also that, in the 
limit of small decay widths, NLOPS predictions
are consistent with the general factorisation properties of the NWA.
An alternative resonance-aware--matching method 
based on the \MCatNLO{}~\cite{Frixione:2002ik} framework,
was proposed in~\citere{Frederix:2016rdc}.

The first NLOPS generator of off-shell $\ttbar+\tW$ production and decay,
based on the 4FNS and \POWHEGRES matching,
was presented in~\citere{Jezo:2016ujg} for the case of
dileptonic final states,
and is available as the \bbfourl generator in the \RES
package~\cite{Jezo:2015aia}.
This generator has been employed and scrutinised
in various experimental studies~\cite{ATLAS:2018ivx,ATLAS:2021pyq,ATLAS:2022jpn}. 
In particular,
in \citere{ATLAS:2018ivx} excellent agreement between \bbfourl and data
has been observed in a phase-space region sensitive to the $\ttbar$--$tW$
interference. Based on this measurement an extraction of the top-quark width
has been proposed in \citere{Herwig:2019obz}. Moreover, \bbfourl{} has been applied 
for the assessment of theoretical uncertainties 
in top-mass measurements~\cite{FerrarioRavasio:2018whr}.

In this paper we present a new \POWHEGRES generator
for off-shell $\ttbar+\tW$ production with semileptonic decays
in the 4FNS. At LO, this corresponds to the process
$pp\to \ell^{\pm} \nu_{\ell} j j b\bar b$, 
which receives contributions from a 
variety of different QCD and EW
subprocesses. As we will see, 
the $\ttbar+\tW$ contribution can be
consistently isolated by imposing---at the level of the theoretical process 
definition---the presence of a $q\bar q'$ pair
with consistent quark flavours as for 
a $W\to q\bar q'$ decay.
The remaining irreducible backgrounds 
of single-top, VBF, and $WZ$ type can be consistently separated
by selecting topologies that are in one-to-one correspondence
with those of the related dileptonic process.
Finally, the process definition needs to be supplemented
by the QCD corrections effects that
arise from the $q\bar q'$ pair in the final state, 
for which we are going to use a $W^+W^-b\bar b$ 
double-pole approximation (DPA).
As we will show, this approach provides a consistent separation 
of off-shell $\ttbar+\tW$ production and decay from all other ingredients of
$pp\to \ell^{\pm} \nu_{\ell} j j b\bar b$, which can be 
described using independent tools.

The new semileptonic generator has been implemented in the same framework as 
the original dileptonic \bbfourl generator, where all
relevant matrix elements are computed with 
\OpenLoops{}~\cite{Cascioli:2011va,Buccioni:2017yxi, Buccioni:2019sur}.
In this framework, we have also introduced the following two methodological novelties,
which are applied both to dileptonic and semileptonic processes.

The first novelty consists of matrix-element--based
resonance-history projectors, which supersede the naive kinematic
projectors used in~\citere{Jezo:2016ujg}.
The new projectors
make it possible to separate histories of $\ttbar$
and $\tW$ types in a reliable way,
and to treat \POWHEG radiation more consistently 
in the case of $\tW$ histories.
The second novelty is related to spurious terms that arise from the
inconsistent perturbative treatment of NLO decay widths,
$\Gamma_\NLO$, in off-shell calculations. In the context 
of the NWA, this issue is rather well know and can be avoided through a
systematic perturbative expansion of terms of the form
$1/\Gamma_{\NLO}$~\cite{Melnikov:2009dn,Hollik:2012rc,Campbell:2012uf}.
In this paper we propose a similar approach for the case of off-shell calculations,
which do not involve explicit $1/\Gamma_\NLO$ terms, but suffer from the same problem.
This method is fully general and should be applied
to any off-shell process, both at fixed-order 
NLO and NLOPS level.
As we will show, in the case of off-shell $\ttbar+\tW$ production and decay
it plays a quite important role, since spurious effects can 
be larger than the entire $\tW$ cross section,
and similarly large as the NLO corrections to top decays.

The paper is organised as follows. 
In \refse{sec:PWGmethod} we review the
\POWRES method in some detail, introducing the notation
that is used in~\refses{se:topwidthcorr}{see:bbfourlsl}. 
In
\refse{se:topwidthcorr} we discuss spurious terms 
and how to avoid them for off-shell processes at 
NLO and NLOPS level.
The new matrix-element--based resonance histories
are presented in \refse{se:bb4lgen}.
The treatment of off-shell $\ttbar +\tW$ production with
semileptonic decays at NLO and NLOPS level 
is discussed in \refse{see:bbfourlsl}.
In \refse{sec:setup} we introduce the setup
used for the numerical studies of \refses{sec:origVsMei}{sec:nlops}, where we 
investigate the impact of the new resonance histories
(\refse{sec:origVsMei}), we study  QCD radiation effects associated with 
hadronic $W$-decays (\refse{sec:lhe}),
and we present a tuned comparison of 
off-shell vs on-shell $t\bar t+\tW$ generators (\refse{sec:nlops}).
We conclude in \refse{se:conclusions},
and in the appendices we present 
kinematic mappings for the new resonance histories 
(\refapp{app:mappings}) 
as well as technical studies that demonstrate the 
consistency of our separation of off-shell $\ttbar+\tW$
production from irreducible backrounds in the
semileptonic channel (\refapp{app:approximation}).

\section{The resonance-aware \POWHEG method} 
\label{sec:PWGmethod}

In this section we review the original \POWHEG method  and its 
resonance-aware \POWHEGRES
extension following Refs.~\cite{Jezo:2015aia,Jezo:2016ujg}.
While this review is entirely based on the 
original literature, in order to 
facilitate the discussion of the new features introduced in 
\refses{se:topwidthcorr}{see:bbfourlsl}, we adopt a new notation
that gives more emphasis to the interplay between resonance histories and
singular regions.

\subsection{The original \POWHEG method}
\label{se:PWGa}

In the \POWHEG approach~\cite{Nason:2004rx, Frixione:2007vw},
the QCD radiation that is emitted in a certain hard process
is generated starting from Born-like events with weights\footnote{As usual, such weights implicitly involve
all relevant NLO squared matrix elements (or interferences) 
as well as convolutions with the PDFs, collinear
factorisation counterterms, and appropriate
normalisation factors for the process at hand.
}
\begin{eqnarray}
\label{eq:PWGa}
  \bar{B} (\Phi_\rB) & = & B (\Phi_\rB) +  V (\Phi_\rB) +  
  \sum_{\labcoll\in \setcoll} \int
R_{\labcoll} (\Phi_{\rR,\labcoll}) 
\,\mathd \Phiradal{\labcoll}\,.
\end{eqnarray}
Here $\Phi_\rB$ describes the Born phase space, and
$B(\Phi_\rB)$ is the usual Born weight, while
$V(\Phi_\rB)$ represents the virtual corrections.
The real corrections are split into a sum of terms that corresponds 
to the various collinear regions for the process at hand. Each 
collinear region is identified by a label $\labcoll\in\setcoll$,
where the set $\setcoll$ corresponds to all possible regions.
Each real-emission contribution 
$R_{\labcoll} (\Phi_{\rR,\labcoll})$
is constructed in such a way that it contains only the collinear singularity 
arising from a specific pair of external partons.\footnote{%
While such regions are often called  
``singular regions'', here and in the following
we denote them ``collinear regions'' or ``collinear sectors''
to underline the fact that, in the 
\POWHEG{} method, such regions/sectors are in one-to-one correspondence 
with the possible collinear splittings for a given process.
}
The weights $R_{\labcoll} (\Phi_{\rR,\labcoll})$
are integrated over the phase space of the unresolved radiation, which 
is parametrised by $\Phiradal{\labcoll}$\,.
For each collinear sector, the full real-emission phase space is 
connected to the Born phase space through a mapping of the form
\begin{eqnarray}
\label{eq:PWGc}
\Phi_{\rR,\labcoll} &\equiv& \Phi_{\rR,\labcoll}(\Phi_{\mathrm{B}}\,,\Phiradal{\labcoll})\,.
\end{eqnarray}
These sector-dependent mappings are defined in such a way that, upon integration over
$\Phiradal{\labcoll}$ and PDF renormalisation, the collinear singularities 
on the rhs of~\refeq{eq:PWGa} undergo a local cancellation in $\Phi_\rB$ space. 
To this end, in each collinear sector,
$\Phi_{\rR,\labcoll}$ and $\Phi_\rB$
should be connected in a way that is consistent with the collinear factorisation identity
\begin{eqnarray}
\label{eq:RESa}
{R_{\labcoll}\left(\Phi_{\rR,\labcoll}\right)}
\;&\xrightarrow[k_{\rT,\labcoll}\ll \sqrt{s}]{}&\;
\frac{{B(\Phi_\rB)}}{k^2_{\rT,\labcoll}}\left[P_{\labcoll}(z)
\,+\,\ord\left(\frac{k_{\rT,\labcoll}}{\sqrt{\hat s}}\right)\right]\,,
\end{eqnarray}
where 
\begin{eqnarray}
\label{eq:kTdef}
k_{\rT,\labcoll}\,\equiv\, k_{\rT,\labcoll}(\Phi_{\rR,\labcoll})
\end{eqnarray}
is the 
transverse momentum of the collinear splitting, 
$P_{\labcoll}(z)$, which is proportional to the 
corresponding splitting function, 
and $z$ is the relevant momentum fraction.
The separation of the real corrections into sectors 
is implemented in such a way that
\begin{eqnarray}
\label{eq:PWGb}
R(\Phi_\rR)
&=&
\sum _{\labcoll\in\setcoll} R_{\labcoll}(\Phi_\rR)\,,
\end{eqnarray}
where $R(\Phi_\rR)$ is the full real-emission weight 
corresponding to the exact squared matrix element for a given real-emission 
partonic subprocess.\footnote{For simplicity, the sum over 
different real-emission partonic subprocesses is kept implicit in our notation. 
}
Note that in~\refeq{eq:PWGb}
all $R_{\labcoll}(\Phi_\rR)$ terms are evaluated at the same 
phase-space point $\Phi_\rR$, and not at the
sector-dependent points $\Phi_{\rR,\labcoll}$ as in~\refeq{eq:PWGa}. 
This guarantees that integrating~\refeq{eq:PWGa}
over the Born phase space yields the exact NLO cross section 
for the process at hand.

In order to fulfill~\refeq{eq:PWGb} one defines
\begin{eqnarray}
\label{eq:PWGf}
R_{\labcoll} (\Phi_\rR) 
&=&
\proj_{\labcoll}(\Phi_\rR)\, 
R (\Phi_\rR)
\,,
\end{eqnarray}
where the ``projectors'' $\proj_{\labcoll}(\Phi_\rR)\in[0,1]$
correspond to the probabilities of hitting the 
various collinear singularities. More precisely, these projectors should fulfill
\begin{eqnarray}
\sum_{\labcoll\in \setcoll} \proj_{\labcoll}(\Phi_\rR)\,=\,1
\qquad\mbox{and}\qquad
\lim_{k_{\rT,\labcoll}\to\, 0}\;
\proj_{\labcoll'}\left(\Phi_{\rR,\labcoll}\right)
\,=\,\delta_{cc'}\,.
\end{eqnarray}
To this end they are defined as
\begin{eqnarray}
\label{eq:PWGg}
\proj_{\labcoll} (\Phi_\rR) 
&=&
\weight_{\labcoll}(\Phi_\rR)\left[
\sum_{\labcoll'\in \setcoll}\weight_{\labcoll'}(\Phi_\rR)\right]^{-1} 
\,,
\end{eqnarray}
where $\rho_{\labcoll}(\Phi_\rR)$ are positive weights proportional to the 
corresponding collinear singularities of $R(\Phi_\rR)$. 
In practice, such weights depend only on the kinematics
of the associated collinear splitting and have the form
\begin{eqnarray}
\label{eq:PWGh}
\weight_{\labcoll}(\Phi_\rR)
&=&
\weightcollal{\labcoll}(\Phiradal{\labcoll})\,,
\end{eqnarray}
where $\Phiradal{\labcoll}$ 
is obtained from  $\Phi_\rR$ 
by inverting the $\labcoll$-dependent\footnote{Note that the denominator of~\refeq{eq:PWGg} 
requires various 
$\weightcollal{\labcoll'}(\Phiradal{\labcoll'})$ depending on 
different $\Phiradal{\labcoll'}$.} mapping~\refeq{eq:PWGc}.
For example, in the case of a collinear splitting that involves two 
final-state partons $i$ and $j$,
\begin{eqnarray}
\label{eq:PWGi}
\weightcollal{\labcoll}(\Phiradal{\labcoll}) &=&
\left[\frac{E_i^2E_j^2}{(E_i+E_j)^2}(1-\cos \theta_{ij})
\right]^{-b}
\,,
\end{eqnarray}
where 
$E_i$, $E_j$ and $\theta_{ij}$ 
are the energies and the angular separation of the two partons
in the partonic CM frame, while
$b$ is a positive constant.

In the \POWHEG approach~\cite{Nason:2004rx, Frixione:2007vw} the matching of NLO calculations
to parton showers is based on the well known formula
\begin{eqnarray}
\label{eq:PWGd}
\mathd \sigma & = & \bar{B} (\Phi_\rB) \,\mathd \Phi_\rB  
\left[
 \Delta (q_{\tmop{cut}}) + 
\sum_{\labcoll\in \setcoll} \Delta (k_{\rT,\labcoll})\, 
 \frac{R_{\labcoll} (\Phi_{\rR,\labcoll})}{B(\Phi_\rB)} 
\,\mathd 
\Phiradal{\labcoll} 
\right]\,,
\end{eqnarray}
where the terms between square brackets 
describe the spectrum of the hardest QCD emission.
The emission probability in each sector is given by the
ratio ${R_{\labcoll} (\Phi_{\rR,\labcoll})}/{B(\Phi_\rB)}$,
and the associated Sudakov form factor
$\Delta (k_{\rT,\labcoll})$ corresponds to the total 
no-emission probability for radiation harder than the
transverse-momentum~\refeq{eq:kTdef}, 
while
$\Delta(q_{\tmop{cut}})$ corresponds to the probability of no emission above the
infrared cutoff $q_{\tmop{cut}}$. The Sudakov form factors are given by
\begin{eqnarray}
\label{eq:PWGsudakova}
  \Delta (q_\rT) & = & 
\exp \left[ {}- \sum_{\labcoll\in \setcoll}\, \int
\theta\left(k_{\rT,\labcoll}- q_\rT\right)\,
  \frac{R_{\labcoll}(\Phi_{\rR,\labcoll})}{B(\Phi_\rB)}\, 
\mathd \Phiradal{\labcoll} \right]\,.
\label{eq:PWGe}
\end{eqnarray}
Events generated according to \refeq{eq:PWGd}--\refeq{eq:PWGsudakova} are known as
Les Houches events (LHEs) and can be directly showered. The only requirement
for the consistent matching to parton showers is the vetoing 
of shower radiation harder than $k_{\rT,\labcoll}$, where the latter variable 
needs to be defined as in \POWHEG.

\subsection{The resonance-aware \POWHEG method}
\label{sec:resmethod}

For processes that involve resonances, the original version of the \POWHEG
method~\cite{Nason:2004rx, Frixione:2007vw} can give rise to serious technical issues and unphysical effects.
This is due to the fact that 
the interplay of
the collinear mappings~\refeq{eq:PWGc} with resonant propagators can lead to
violations of the factorisation identity~\refeq{eq:RESa}. 
A solution to this problem has been introduced in the 
resonance-aware extension of the \POWHEG method~\cite{Jezo:2015aia},
which will be referred to as the \POWHEGRES method in the following.

In order to illustrate the problem and the \POWHEGRES~\cite{Jezo:2015aia}
 solution,
let us consider a mapping $\Phi_{\mathrm{B}}\to \Phi_{\rR,\labcoll}$ 
that corresponds to the splitting of a massless parton of 
momentum $p_{\rB,ij}$ into a pair of massless partons with total momentum 
$p_{\rR,ij}=p_{\rR,i}+ p_{\rR,j}$. In the collinear limit
$p_{\rR,ij}$ and $p_{\rB,ij}$ are identical, while for finite transverse momenta
they differ by a shift
\begin{eqnarray}
\label{eq:RESINTa}
\Delta p_{ij} \,=\, 
p_{\rR,ij} - p_{\rB,ij} \;\propto\; \frac{p_{\rR,ij}^2}{2 E_{ij}}\,,
\qquad\mbox{where}\qquad
p_{\rR,ij}^2 \,=\, \frac{k^2_{\rT,\labcoll}}{z(1-z)}\,,
\end{eqnarray}
and $E_{ij}=E_{\rB,ij}\simeq E_{\rR,ij}$. In general, the shift $\Delta p_{ij}$
results in a corresponding difference in the hard internal momenta,
\begin{eqnarray}
\label{eq:RESINTc}
\Delta q_k  \,=\, 
q_{\rR,k} - q_{\rB,k} \;\propto\; \Delta p_{ij}\,,
\end{eqnarray}
where the internal momentum $q_{\rB,k}$
is a certain combination of external momenta in 
the Born phase space, 
and $q_{\rR,k}$ is the corresponding combination of 
momenta that is obtained by undoing the collinear splitting in the real-emission phase space, i.e.~by replacing 
$p_{\rR,i},p_{\rR,j}\to p_{\rR,ij}$.
As a consequence, the hard kinematic invariants that contribute to 
the squared amplitudes on the lhs and rhs of the 
factorisation identity~\refeq{eq:RESa} can differ by 
\begin{eqnarray}
\label{eq:RESINTd}
\Delta q^2_k  \,=\, 
q^2_{\rR,k} - q^2_{\rB,k} \;\propto\; 2 q_{\rB,k}\cdot \Delta p_{ij}
\propto p_{\rR,ij}^2 \frac{E_k}{E_{ij}}\,.
\end{eqnarray}
In the case of hard scattering processes,
in order to ensure the validity 
of~\refeq{eq:RESa} it is sufficient to require
\begin{eqnarray}
\Delta q^2_k  \,\ll\, \hat s\,,  
\end{eqnarray}
which is always fulfilled in the collinear region 
$k^2_{\rT,c}\ll \hat s$.
Instead, in the presence of intermediate unstable particles,
in order to avoid violations of factorisation in the vicinity of resonance peaks,
the shift in the resonance virtuality should be restricted to
\begin{eqnarray}
\label{eq:RESINTd}
\Delta q^2_r \,\ll \, \Gamma_r M_r\,,
\end{eqnarray}
where $q_r$, $M_r$ and $\Gamma_r$ are the momentum, mass 
and width of the unstable particle.

\newcommand{\graphplot}[1]{
\begin{subfigure}{0.30\linewidth}
\begin{minipage}[t]{.30\linewidth}
\includegraphics[width=45mm,trim= 0 0 0 0, clip]{diagrams/#1} 
\end{minipage}
\caption{\phantom{\hspace{5mm}}}
\label{#1}
\end{subfigure}
}

\begin{figure}
\begin{center}
\graphplot{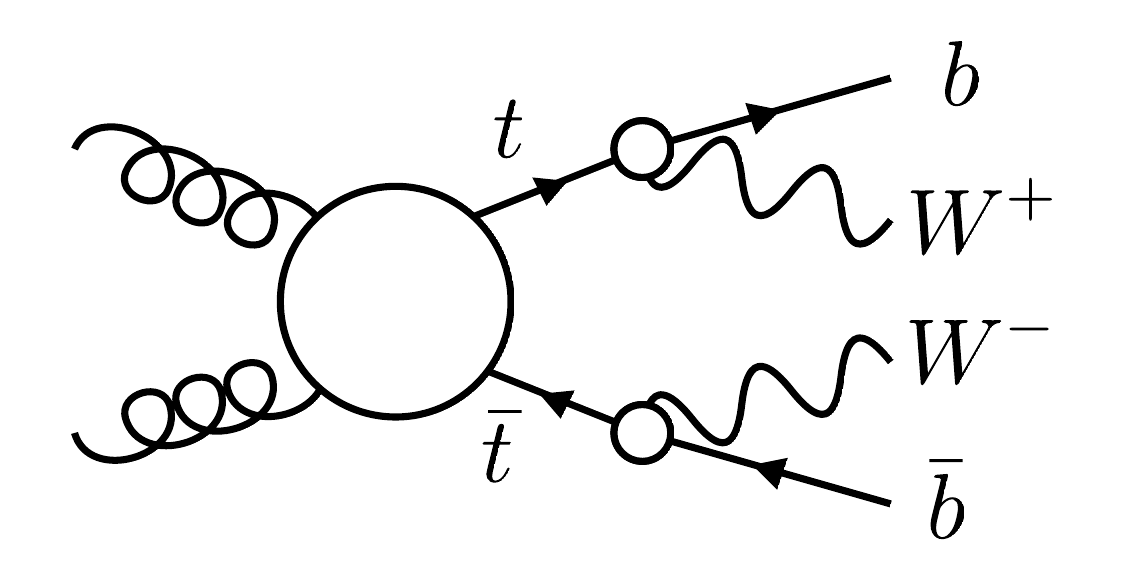}
\graphplot{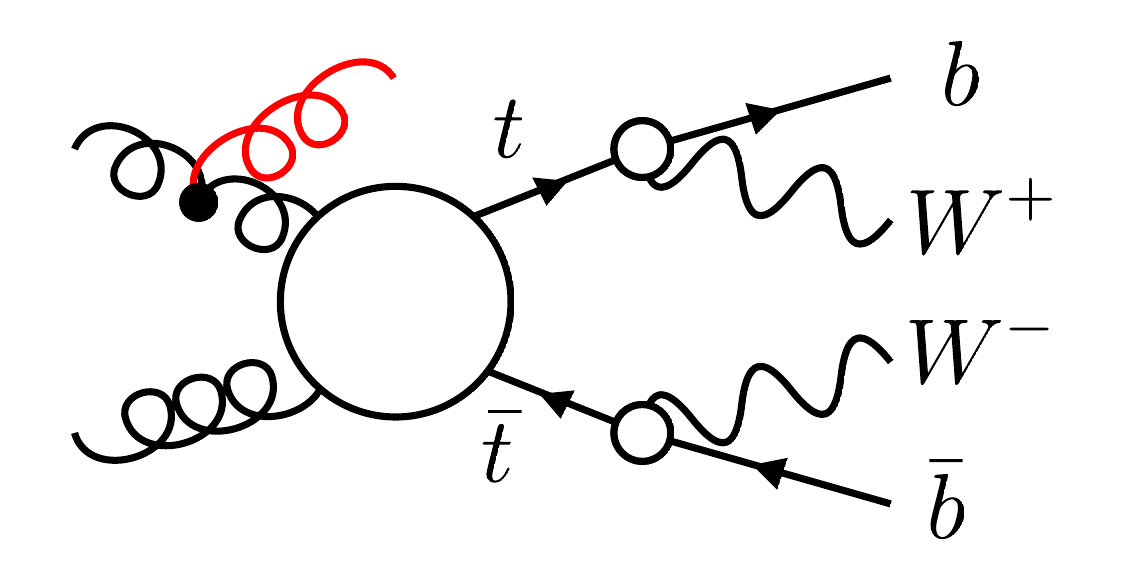}
\graphplot{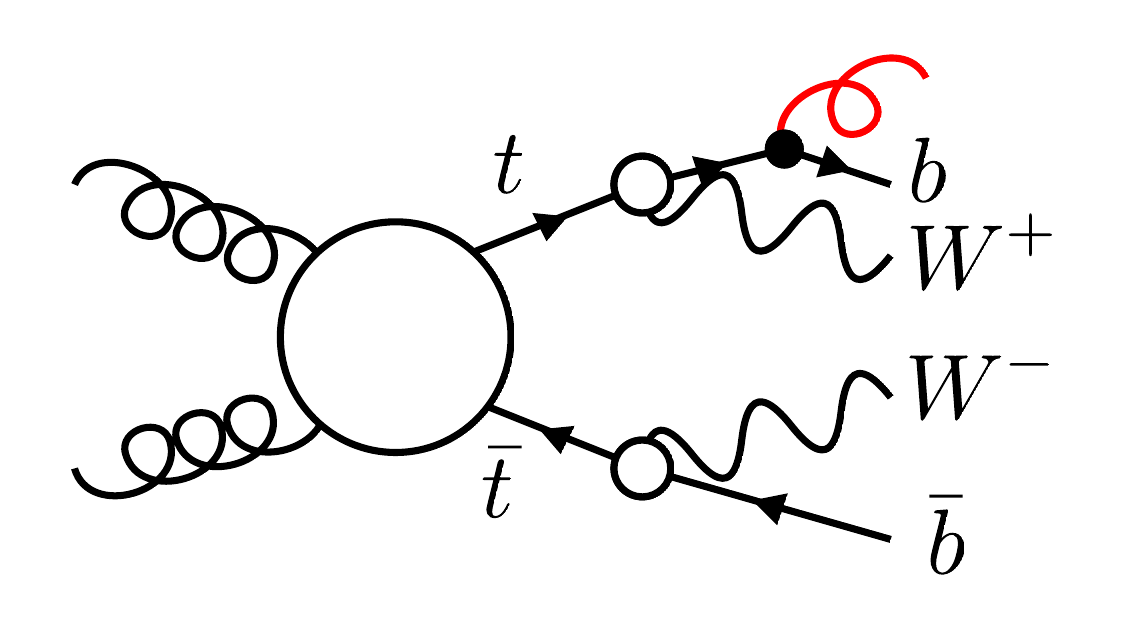}
\\[4mm]
\graphplot{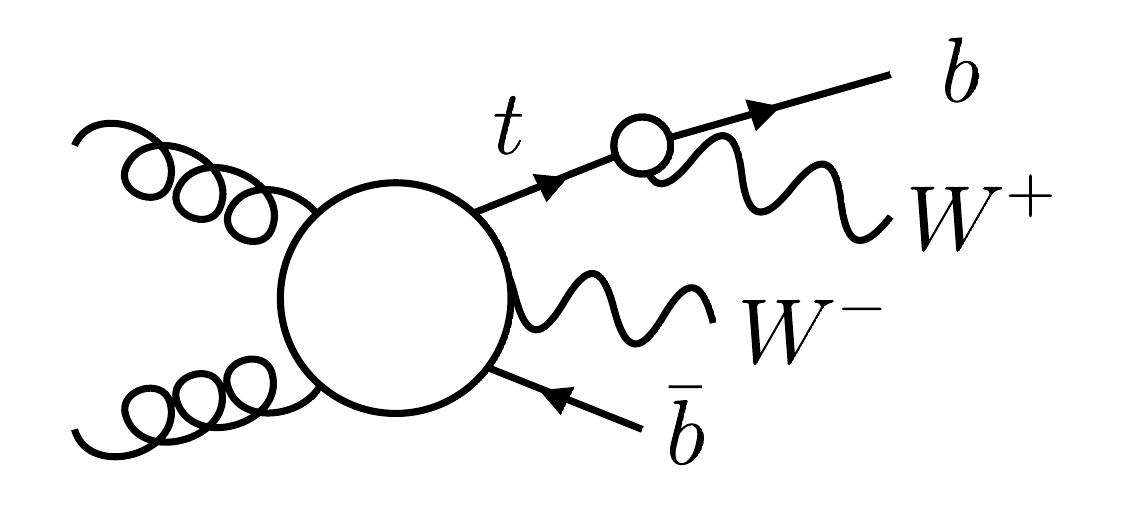}
\graphplot{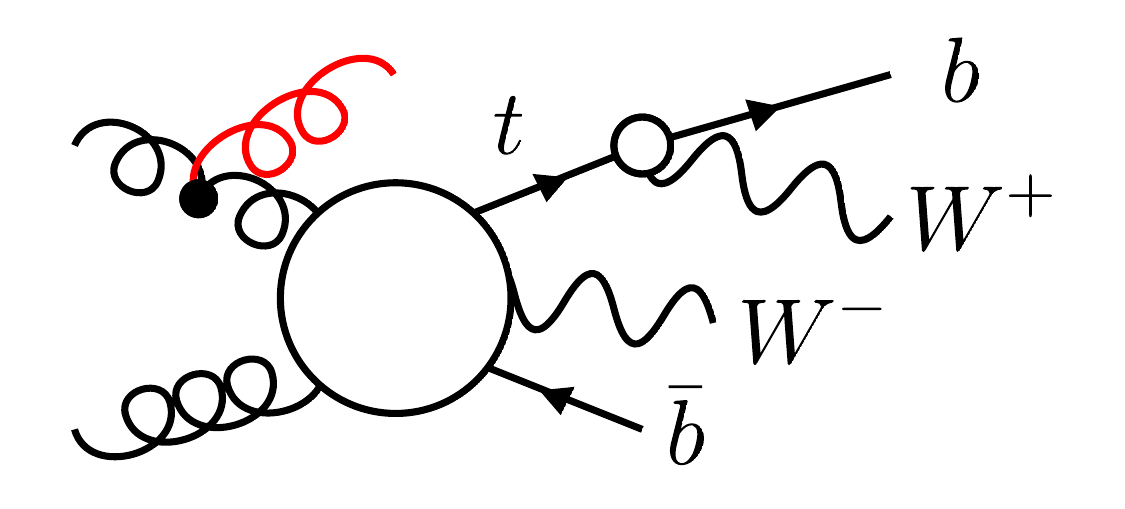}
\graphplot{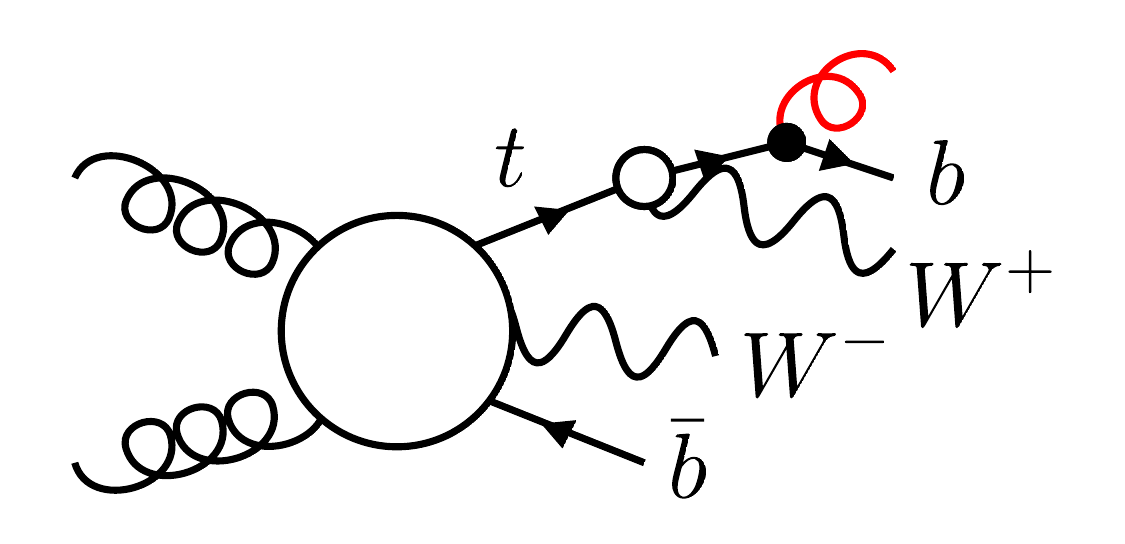}
\\[4mm]
\graphplot{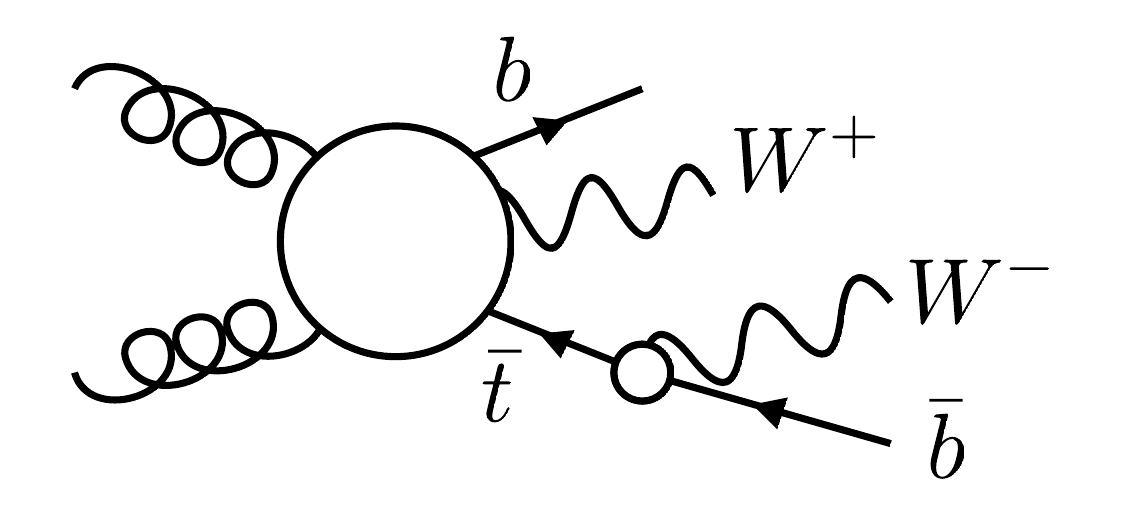}
\graphplot{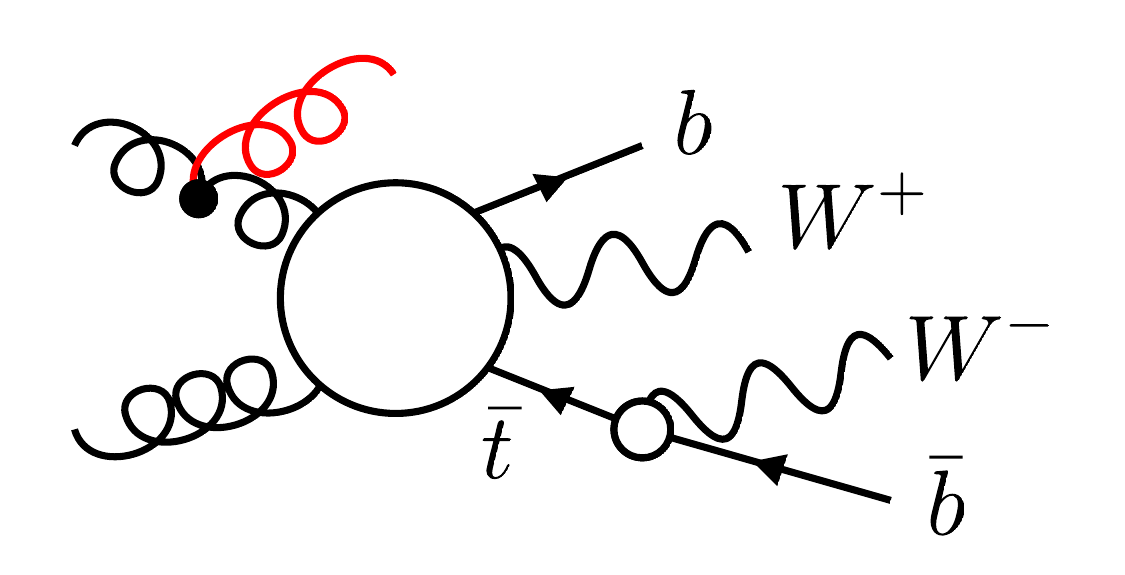}
\graphplot{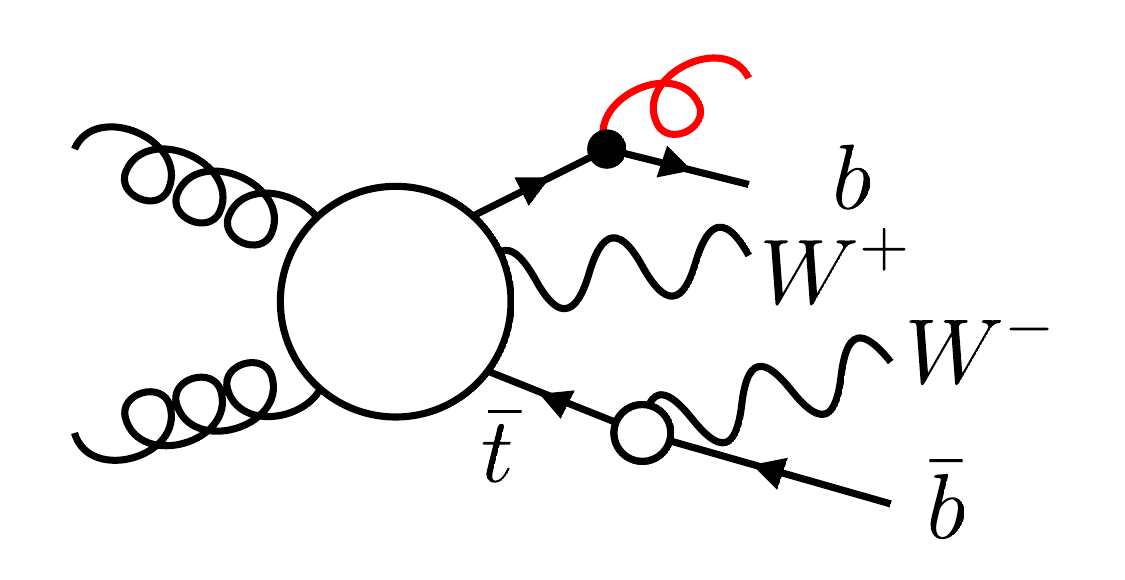}
\\
\end{center}
\caption{Examples of resonance histories 
for the full process $gg\to W^+W^-b\bar b$. The 
three rows correspond to 
Born-level and real-emission histories 
associated with the production subprocesses $gg\to t\bar t$
(diagrams~\subref{hista}--\subref{histc}),
$gg\to t\, W^- \bar b$ 
(diagrams~\subref{histd}--\subref{histf}),
and $gg\to \bar t\, W^+ b$ 
(diagrams~\subref{histg}--\subref{histi}).
In each diagram, the blobs correspond to 
the decomposition of the full process into 
production and decay subprocesses according to a specific resonance history.
Feynman diagrams have been generated with {\tt FeynGame}.~\cite{Harlander:2020cyh}.
}
\label{fig:wwbbhist}
\end{figure}

In the original \POWHEG approach~\cite{Nason:2004rx, Frixione:2007vw} there is no mechanism that guarantees~\refeq{eq:RESINTd}, 
while the key idea behind the POWHEG--RES method~\cite{Jezo:2015aia} is to use alternative 
mappings that satisfy 
\begin{equation}
\label{eq:RESvirt}
\Delta q^2_r=0\,,
\end{equation}
thereby avoiding any unphysical 
effect in the NLOPS modelling of resonant processes.
The strategy of using mappings that preserve the resonance virtualities
can be best understood in the narrow-width approximation (NWA), 
where the production and decay of unstable particles 
are factorised as separate subprocesses, which are connected 
trough intermediate unstable particles that remain exactly on-shell.
For example, let us consider the process
$gg\to t\bar t \to W^+W^-b\bar b$, which factorises into 
the production subprocess $gg\to t\bar t$ 
and the decay subprocesses 
$t\to W^+ b$ and $\bar t\to W^-\bar b$, as illustrated 
in~\reffi{fig:wwbbhist}\subref{hista}--\subref{histc}.
In the presence of QCD radiation, in order to keep the intermediate 
$t$ and $\bar t$ quarks exactly on-shell,
in the $\Phi_{\rR,\labcoll}$ mappings that describe collinear radiation 
emitted by the production subprocess 
(\reffi{fig:wwbbhist}\subref{histb}), the invariant masses of the
$W^+b$ and $W^-\bar b$ systems can be preserved by handling 
the corresponding four-momenta as if they were on-shell 
final-state $t$ and $\bar t$ momenta.
Vice versa, in the case of radiation stemming form a
$t\to W^+b$ decay (\reffi{fig:wwbbhist}\subref{histc})
the virtuality of the $t$ quark can be preserved 
by keeping fixed the full four-momentum of the $W^+b g$
system, as in the case of a pure top-decay process.

In general, in the NWA each process is characterised by a so-called 
{\it resonance history}, which corresponds to a well-defined 
combination of {\it subprocesses} consisting 
of a main production subprocess and a certain number of 
$1\to 2$ decay subprocesses.\footnote{%
Note that in~\cite{Jezo:2015aia}
such ``production subprocesses'' and ``decay subprocesses'' 
are referred to as ``production resonances'' and ``resonance decays'',
while here we prefer 
to use the term ``subprocesses'', since the entities at hand
correspond to the building blocks of a scattering process.
} 
The latter can be independent of each other, or
single steps of a decay chain.
In the NWA, the {\it resonance history} provides the key for the consistent
generation of QCD radiation. In practice, each subprocess radiates as an
independent process, and the momenta of the unstable particles that connect
the various subprocesses are always kept on-shell, 
i.e.~$q_{\rR,r}^2=q_{\rB,r}^2=M_r^2$.
This is achieved by handling the $q_r$ momenta 
on a similar footing as the 
external momenta of a hard scattering process.

Let us now move away from the NWA and consider 
exact NLO calculations of resonant processes, where
finite-width effects are included throughout.
In this case, intermediate unstable particles are no longer exactly on-shell.
Moreover, each process with a well-defined initial and final state can
involve multiple resonance histories that interfere with each other. 
For instance, the process
$gg\to W^+W^-b\bar b$ involves, in addition to the above mentioned
$gg\to t\bar t$ resonance history 
(\reffi{fig:wwbbhist}\subref{hista}--\subref{histc}), also resonance histories 
associated with the single-top production subprocesses
$gg\to t\, W^- \bar b$ 
(diagrams~\subref{histd}--\subref{histf}),
and $gg\to \bar t\, W^+ b$ 
(diagrams~\subref{histd}--\subref{histf}).
In this context, the strategy of the 
\POWHEGRES
method is based on a probabilistic splitting of 
the full process into contributions
that are each dominated by a well defined 
resonance history. 
In this way, within each resonance history, 
the condition~\refeq{eq:RESvirt} can be ensured
by treating QCD radiation in a way that 
corresponds to an off-shell continuation 
of NWA approach.
More explicitly, when unstable particles are off-shell,
i.e.~$q^2_{\rB,r}\neq M_r^2$, within each resonance history 
the emission of 
QCD radiation is organised in the same way as in 
corresponding NWA, but using NWA mappings that correspond to 
unstable particles with ``modified masses'' 
$M_r\to \sqrt{q^2_{\rB,r}}$.
This automatically guarantees that $q^2_{\rR,r}=q^2_{\rB,r}$
in the vicinity of each resonance.

At Born level, the splitting into resonance histories is implemented as
\begin{eqnarray}
\label{eq:RESza}
B (\Phi_\rB)\,=\,
\sum_{\labhist\in\,\sethist}B_{\labhist} (\Phi_\rB)\,,
\qquad\mbox{with}\qquad
B_{\labhist}(\Phi_\rB)\,=\,
\proj^{(\mathrm{hist})}_{\labhist} (\Phi_\rB)\,
B(\Phi_\rB)\,,
\end{eqnarray}
where the labels $r\in \sethist$ correspond to the 
different histories, i.e.~$\sethist$ is the full set of resonance histories
for the process at hand, while 
$\proj^{(\mathrm{hist})}_{\labhist} (\Phi_\rB)\in[0,1]$
denote their phase-space dependent 
probabilities. The latter are determined as
\begin{eqnarray}
\label{eq:RESc}
\proj^{(\mathrm{hist})}_{\labhist} (\Phi_\rB) 
&=&
\weighthistal{\labhist}(\Phi_\rB)\left[
\sum_{\labhist'\in \sethist}\weighthistal{\labhist'}(\Phi_\rB)\right]^{-1}
\,,
\end{eqnarray}
where the weights $\weighthistal{\labhist}(\Phi_\rB)$ should mimic the 
relative probabilities of the different histories
in the limit of small width.
The first implementations of the \POWHEGRES method~\cite{Jezo:2015aia,Jezo:2016ujg}
are based on resonance weights of the form 
\begin{eqnarray} \label{eq:RESb}
\weighthistal{\labhist}(\Phi_\rB) &=& 
\prod_{r\in\calR(h)} \frac{M_r^4}{(q^2_{\rB,r} - M_r^2)^2
+
  \Gamma_r^2 M_r^2}\,,
\end{eqnarray}
where the labels $r\in\calR(h)$ correspond to the various resonances that 
contribute to the history $h$, while
$M_r$, $\Gamma_r$, and $q_{\rB,r}\equiv q_{\rB,r}(\Phi_\rB)$
are, respectively, 
the rest mass, the width, and the four-momentum that flows through the 
propagator of the resonance $r$. 
A more realistic implementation of 
resonance histories based on Born matrix elements
is presented in~\refse{se:newreshist}.
In general, in the limit of vanishing decay widths,
where all unstable particles go on-shell,
any implementation of the resonance weights should obey 
\begin{eqnarray} \label{eq:RESbb}
\lim_{\Gamma_r\,\to\, 0}\; \weighthistal{\labhist}(\Phi_\rB) &\propto& 
\prod_{r\in\calR(h)}  \delta(q_{\rB,r}^2 - M_r^2)\,,
\end{eqnarray}
which guarantees an exact correspondence of the probabilistic 
histories~\refeq{eq:RESc} 
with the uniquely-defined resonance histories of the NWA.

For the resonance-aware extension of the \POWHEG
formula~\refeq{eq:PWGd},
the weights~\refeq{eq:PWGa} are split into 
different resonance-history contributions as
\begin{eqnarray}
\label{eq:RESzb}
\bar{B} (\Phi_\rB)\,=\,
\sum_{\labhist\in\,\sethist}\bar{B}_{\labhist} (\Phi_\rB)
\qquad\mbox{with}\qquad
\bar{B}_{\labhist}(\Phi_\rB)\,=\,
\proj^{(\mathrm{hist})}_{\labhist} (\Phi_\rB)\,
\bar{B}(\Phi_\rB)\,,
\end{eqnarray}
using the probabilities~\refeq{eq:RESc}. As for the 
real-emission probabilities that are used to generate 
\POWHEG{} emissions, the splitting~\refeq{eq:PWGb} into 
collinear sectors is implemented at the level of individual 
resonance histories, i.e.~the full real-emission weight is split 
into 
\begin{eqnarray}
\label{eq:RESca}
R(\Phi_\rR) 
&=&
\sum_{\labhist\in \sethist}
\sum_{\labcoll\in \setcoll(\labhist)}
R_{\labhist,\labcoll} (\Phi_\rR)\,,
\end{eqnarray}
where the labels 
$
\labcoll\in\setcoll(r)
$
correspond to the various collinear sectors that are consistent with the
resonance history $h$. Such collinear sectors are in one-to-one 
correspondence with those for the on-shell production and decay process with 
history $h$ in the NWA.
More explicitly, the sectors $\labcoll\in\setcoll(\labhist)$
correspond to all possible ways of emitting collinear radiation
in the various production 
or decay ``subprocess'' of the history $\labhist$.
As discussed above, the associated 
$\Phi_{\rR,\labcoll}$ mappings are chosen in a way that 
preserves the virtuality of all resonances that belong to
the history $\labhist$.
In this respect, we note that 
collinear sectors associated with the same collinear splitting but 
different resonance histories require, in general, 
different collinear mappings.
Therefore, in order to avoid confusion,
it is convenient to always label collinear sectors that are 
associated with different histories in a different way,
i.e.~all sectors should be labelled in such a way that 
$\setcoll(\labhist)\cap \setcoll(\labhist')\,=\, \varnothing$\,
for $\labhist'\neq \labhist$.

In analogy with~\refeq{eq:PWGf} and~\refeq{eq:PWGg},
the contributions of the individual 
sectors in~\refeq{eq:RESca} are constructed as
\begin{eqnarray}
\label{eq:REScb}
R_{\labhist,\labcoll} (\Phi_\rR)
\,=\,
\proj_{\labhist,\labcoll}(\Phi_\rR)\,
R(\Phi_\rR)\,,
\end{eqnarray}
with projectors
\begin{eqnarray}
\label{eq:REScc}
\proj_{\labhist,\labcoll} (\Phi_\rR) 
&=&
\weight_{\labhist,\labcoll}(\Phi_\rR)\left[
\sum_{\labhist'\in \sethist}
\sum_{\labcoll'\in \setcoll(\labhist')}
\weight_{\labhist',\labcoll'}(\Phi_\rR)\right]^{-1} 
\,,
\end{eqnarray}
and weights of the form
\begin{eqnarray}
\label{eq:RESd}
\weight_{\labhist,\labcoll}(\Phi_\rR)
&=&
\weighthistal{\labhist,\labcoll}(\Phi_{\rR})\,
\weightcollal{\labcoll}(\widetilde{\Phi}_{\rad,\labcoll})\,.
\end{eqnarray}
Here $\weightcollal{\labcoll}$ are the usual collinear weights,
with the only difference that, in the case of 
collinear splitting stemming from the decay of a resonance,
the kinematic variables 
$\widetilde{\Phi}_{\rad,\alpha}\equiv (\tilde E_i, \tilde E_j, \tilde \theta_{ij})$ 
are determined in the rest frame of the relevant resonance.
As for the resonance weights $\weighthistal{\labhist,\labcoll}(\Phi_{\rR})$,
in~\citeres{Jezo:2015aia,Jezo:2016ujg} they have been chosen as
\begin{eqnarray} \label{eq:RESdb}
\weighthistal{\labhist,\labcoll}(\Phi_\rR) &=& 
\prod_{r\in\calR(\labhist,\labcoll)} \frac{M_r^4}{(q^2_{\rR,r} - M_r^2)^2 
+
\Gamma_r^2 M_r^2}\,,
\end{eqnarray}
which is the natural real-emission generalisation of~\refeq{eq:RESb}.
In this case, 
the labels $r\in\calR(\labhist,\labcoll)$ correspond to all
resonances that are present in the 
resonance--collinear history $(\labhist,\labcoll)$, while
$q_{\rR,r}\equiv q_{\rR,r}(\Phi_\rR)$
is the four-momentum associated with 
a certain resonance.
In practice, $q_{\rR,r}$ is the 
sum of all final-state momenta that flow 
through the propagator of the resonance 
$r$, including also real radiation 
that is emitted by the decay products of the
resonance at hand in the collinear sector
$\labcoll$.
A matrix-element--based extension of such resonance weights will be presented
in \refse{se:newreshist}.
In general, the real-emission resonance weights should always obey 
\begin{eqnarray} \label{eq:RESdc}
\lim_{\Gamma\to 0}\; \weighthistal{\labhist,\labcoll}(\Phi_\rR) &\propto& 
\prod_{r\in\calR(\labhist,\labcoll)}  \delta(q^2_{\rR,r} - M_r^2)\,,
\end{eqnarray}
which guarantees an exact correspondence 
of the probabilistic 
histories~\refeq{eq:REScc}
with the uniquely-defined histories of the NWA.
Beyond LO, this correspondence is crucial in order to 
ensure that, in the limit of small width, 
the full process factorises 
into separate production and decay subprocesses
as in the NWA.

In the \POWHEGRES approach~\cite{Jezo:2015aia}, the \POWHEG formula for the generation 
of LHEs assumes the form~\refeq{eq:RESzb}
\begin{eqnarray}
\label{eq:RESea}
\mathd \sigma & = & 
\sum_{\labhist\in\,\sethist}\,
\bar{B}_{\labhist} (\Phi_\rB)
 \,\mathd \Phi_\rB  
\left[
\Delta_{\labhist} (q_{\tmop{cut}}) + 
\sum_{\labcoll\,\in\, \setcoll(\labhist)} 
\Delta_{\labhist} (k_{\rT,\labcoll})\, 
 \frac{R_{\labhist,\labcoll}(\Phi_{\rR,\labcoll})}{B_{\labhist}(\Phi_\rB)} 
\,\mathd 
\Phiradal{\labcoll} 
\right]\,,\qquad
\end{eqnarray}
where $B(\Phi_\rB)$, $\bar B(\Phi_\rB)$ and $R(\Phi_\rR)$ are split into 
resonance histories according to \refeq{eq:RESza}, 
\refeq{eq:RESzb} and \refeq{eq:RESca}, and LHEs are 
generated in the same way as in~\refeq{eq:PWGd}
but on a history-by-history basis.
Similarly as in~\refeq{eq:PWGsudakova},
the relevant Sudakov form factors are given by
\begin{eqnarray}
\label{eq:RESs}
\Delta_{\labhist} (q_\rT) & = & 
\exp \left[ {}- \sum_{\labcoll\in \setcoll(\labhist)}\, \int
\theta\left(k_{\rT,\labcoll}- q_\rT\right)\,
  \frac{R_{\labhist,\labcoll}(\Phi_{\rR,\labcoll})}{B_{\labhist}(\Phi_\rB)}\, 
\mathd \Phiradal{\labcoll} \right]\,.
\label{eq:PWGe}
\end{eqnarray}
For the separation of resonance histories, the \POWHEG formula~\refeq{eq:RESea} 
relies on the projectors~\refeq{eq:RESc} and~\refeq{eq:REScc}, which are constructed in an
approximate way. In this respect, we note 
that such projectors are only used to
split the cross section into separate contributions, while the combination of all
such contributions is consistent with the full $B(\Phi_\rB)$, $\bar
B(\Phi_\rB)$ and $R(\Phi_\rR)$ weights, which are based on the 
exact matrix elements for the full set of resonant and non-resonant Feynman diagrams for
the process at hand. In particular, the
expansion of~\refeq{eq:RESea} to first order in $\as$ beyond the leading order is 
identical to the result of a
full NLO calculation.

The resonance-aware description of the hardest radiation can be
further improved in a way that reflects the factorisation of 
higher-order radiative corrections into production and decay subprocesses 
in the narrow-width limit.
This factorisation property allows one 
to generate one \POWHEG
emission from each production and decay subprocess 
that belongs to the resonance history of a
given event~\cite{Campbell:2014kua,Jezo:2015aia,Jezo:2016ujg}. 
Therefore, in this approach, which is dubbed 
``all radiation'' or {\tt
allrad} option in the \POWHEG{} jargon, a LHE with history $\labhist$ includes
up to
$n(\labhist)$ \POWHEG emissions, where $n(\labhist)$ is 
one plus the number of decay subprocesses
in the history $\labhist$.
For the bookkeeping of these multiple emissions, it is convenient
to identify the associated production and decay subprocesses 
through history-dependent\footnote{For different histories we 
use different labels,
i.e.~$\setsubp(\labhist)\cap \setsubp(\labhist')= \varnothing$
for $\labhist'\neq \labhist$.
}
labels $\labsubp\in\setsubp(\labhist)$, where
$\setsubp(\labhist)$ corresponds to the set of all 
subprocesses within the history $\labhist$. In this way,
the full set of collinear sectors $\setcoll(\labhist)$
can be split into subsectors
$\setcoll(\labsubp)\subset\setcoll(\labhist)$, 
where $\labcoll\in\setcoll(\labsubp)$
corresponds to collinear radiation emitted within 
the subprocess $\labsubp\in\setsubp(\labhist)$,
and
$\setcoll(\labhist)\,=\,{\cup}_{\labsubp\in\setsubp(\labhist)}\,
\setcoll(\labsubp)\,$.
With this notation, the {\tt allrad} extension of the
\POWHEGRES formula reads
\begin{eqnarray}
\label{eq:RESe}
\mathd \sigma & = & 
\sum_{\labhist\in\,\sethist}\,
\bar{B}_{\labhist} (\Phi_\rB)
 \,\mathd \Phi_\rB  
\hspace{-.3em}
\prod_{\labsubp \,\in\,\setsubp(\labhist) }
\left[
\Delta_{\labsubp} (q_{\tmop{cut}}) + 
\sum_{\labcoll\,\in\, \setcoll(\labsubp)} 
\hspace{-.5em}
\Delta_{\labsubp} (k_{\rT,\labcoll})\, 
 \frac{R_{\labhist,\labcoll}(\Phi_{\rR,\labcoll})}{B_{\labhist}(\Phi_\rB)} 
\,\mathd 
\Phiradal{\labcoll} 
\right]\,,\qquad
\end{eqnarray}
where each term between squared brackets
corresponds to a separate \POWHEG{} emission stemming from 
the subprocess $\labsubp\in\setsubp(\labhist)$ of the resonance history $\labhist$.
The Sudakov form factors $\Delta_{\labsubp} (q_\rT)$
are defined in the same way as
in~\refeq{eq:RESs} but with 
$\setcoll(\labhist)$ replaced by $\setcoll(\labsubp)$,
i.e.~including only radiation stemming 
from a specific production or decay subprocess.
In the \POWHEGRES approach 
the $\Phi_{\rad,\labcoll}$ radiation phase space 
is split into soft and hard parts.
Similarly as in the original \POWHEG method, 
this separation is controlled by 
the so-called ${\tt hdamp}$
parameter~\cite{Alioli:2008tz},
and the \POWHEGRES formulas~\refeq{eq:RESea}
and~\refeq{eq:RESe} are used only to generate soft radiation.
Instead, hard radiation is emitted without
applying any Sudakov form factor and always as a single hard emission,
i.e.~disabling the {\allrad} option.
For simplicity, this different treatment of
soft and hard radiation is 
not explicitly shown in the above formulas.

As for the matching of LHEs to parton showers, in the POWHEG--RES method~\cite{Jezo:2015aia} the standard 
matching approach is adapted to the resonance structure of the events.
Specifically, for each LHE,
all final-state partons are attributed to a 
corresponding production or decay subprocess $\labsubp\in\setsubp(\labhist)$, 
based on the resonance history $\labhist$
of the event. The parton shower is then instructed to radiate 
in each production and decay subprocess 
in a way that preserves the virtuality of each
resonance. 
Moreover, in the {\tt allrad} approach the shower radiation that
is emitted in the various subprocess $\labsubp$
is subject to different veto scales, which are given by the transverse momenta of the 
corresponding \POWHEG emissions.

\section{Spurious width effects in off-shell NLO and NLOPS cross sections}
\label{se:topwidthcorr}

The \POWHEGRES method is designed according to the
general factorisation properties of higher-order corrections in the NWA,
which must be fulfilled also by off-shell calculations 
in the limit of small decay widths.
In particular, when decay widths become small, 
the radiative corrections should factorise 
to all orders into contributions associated with the various 
production and decay subprocesses.
Moreover, upon integration over the full phase space and summation 
over all possible decay channels, 
all decay probabilities must be equal to one. 
Thus, in the zero-width limit, the total cross section 
of the full off-shell process 
should be identical to the one of the associated production subprocess.

As pointed out in~\citeres{Melnikov:2009dn,Hollik:2012rc,Campbell:2012uf}, 
this consistency property is not fulfilled 
by naive fixed-order implementations of the NWA. However,
it can be ensured by means of a rigorous
perturbative treatment of all terms that are inversely proportional 
to the decay widths 
of the resonant particles.
In the following, this approach will be dubbed ``inverse-width expansion''.
Its application in the context of the NWA is discussed in
\refse{se:NWAexp}, while in 
\refse{se:offshellexp} we propose an extension of the
inverse-width expansion to 
off-shell NLO
calculations and their matching to parton showers in the
\POWHEGRES framework.

\subsection{Inverse-width expansion in the NWA}
\label{se:NWAexp}

For simplicity, let us start with the case of a process that involves 
the production and decay of a single resonance. 
In the NWA, to all orders in perturbation theory,
its differential cross section factorises as
\begin{equation}
\label{eq:GEXPa}
{\rm d} \sigpxd  \,=\, {\rm d} \sigp \frac{{\rm d} \Gamma}{\Gamma}\,,\quad\mbox{where}\quad
\Gamma \,=\, \int \limits_{\dec}{} {\rm d} \Gamma\,.
\end{equation}
Here $\sigp$ denotes the cross section for the 
production subprocess, while the integration is over the full decay phase space,
and the sum over different decay channels is implicitly understood.
As a result of this factorisation property, to all orders of perturbation theory
the integral of the
production$\times$decay cross section over the 
full decay phase space is equal to the production cross section, i.e.
\begin{equation}
\label{eq:GEXPb}
  \int \limits_{\dec}^{} {\rm d} \sigpxd \,=\, {\rm d} \sigp\,.
\end{equation}
Let us now consider NLO calculations in the NWA, where the ingredients of
\refeq{eq:GEXPa} can be split into LO parts and NLO corrections as
\begin{eqnarray}
\label{eq:GEXPc}
{\rm d} \sigpord{\NLO} \,=\, {\rm d} \sigpord{0} + {\rm d} \sigpord{1}\,,\qquad
{\rm d} \Gamma_\NLO \,=\, {\rm d} \Gamma_0 + {\rm d} \Gamma_1\,, 
\qquad
\Gamma_\NLO \,=\, \Gamma_0 + \Gamma_1\,.
\end{eqnarray}
If the $1/\Gamma$ term in~\refeq{eq:GEXPa} is not expanded,
the NLO cross section is given by
\begin{equation}
\label{eq:GEXPd}
   {\rm d} \sigpxdord{\NLO}  \,=\, 
{\rm d} \sigpord{0}\, \frac{{\rm d} \Gamma_0 }{\Gamma_\NLO}
+ {\rm d} \sigpord{1}\, \frac{{\rm d} \Gamma_0 }{\Gamma_\NLO} 
+ {\rm d} \sigpord{0}\, \frac{{\rm d} \Gamma_1}{\Gamma_\NLO}
\,,
\end{equation}
and integrating over the decay phase space yields
\begin{equation}
\label{eq:GEXPe}
\int\limits_{\dec}{} {\rm d} \sigpxdord{\NLO}
\, =\,
\rd  \sigpord{0}\;\int\limits_{\dec}{} \frac{\rd\Gamma_\NLO}{\Gamma_\NLO}
\,+\, \rd \sigpord{1} \; \int\limits_{\dec}{}\frac{\rd(\Gamma_{\NLO}-\Gamma_1) }{\Gamma_\NLO} 
\, =\,  \rd \sigpord{0}+\rd \sigpord{1} - \rd \sigpord{1}
\frac{\Gamma_1}{\Gamma_\NLO}\,.
\end{equation}
At variance with this naive approach, a consistent perturbative expansion of 
the inverse decay width~\cite{Melnikov:2009dn}, 
\begin{equation}
\label{eq:GEXPf}
\frac{1}{\Gamma_\NLO}\,\to\,
\frac{1}{\Gamma_0}\left(1-\frac{\Gamma_1}{\Gamma_0}\right)\,,
\end{equation}
and its extension to the full production$\times$decay cross section, 
yields
\begin{eqnarray}
\label{eq:GEXPg}
\int\limits_{\dec}{} {\rm d} \sigpxdord{\NLOexp}
&=&\
\rd \sigpord{0}\;\left[\left(1-\frac{\Gamma_1}{\Gamma_0}\right)
\int\limits_{\dec}{} \frac{\rd\Gamma_0}{\Gamma_0}
\,+
\int\limits_{\dec}{} \frac{\rd\Gamma_1}{\Gamma_0}
\right]
\,+\, \rd \sigpord{1} \; \int\limits_{\dec}{}\frac{\rd\Gamma_0 }{\Gamma_0} 
\; =\;  \rd \sigpord{0}+\rd \sigpord{1}\,,\qquad
\end{eqnarray}
which is identical to the  NLO production cross section,
as expected from~\refeq{eq:GEXPb}. 
In contrast, the ``unexpanded''  implementation~(\ref{eq:GEXPd}) of the NWA
deviates form the ``expanded'' one by the contribution
\begin{equation}
\label{eq:GEXPh}
{\rm d} \sigfake
\,=\, - {\rm d} \sigma_1 \frac{\Gamma_1}{\Gamma_\NLO}\,,
\end{equation}
which violates the factorisation property~\refeq{eq:GEXPb}
and is thus going to be denoted as ``spurious''.
Since this difference between the 
expanded and unexpanded versions of the NWA
is of $\ord(\as^2)$, 
in principle one may object that
it does not violate~\refeq{eq:GEXPb} at NLO, and that 
it could be regarded as part of the NLO uncertainty.
This perspective would make sense only if it was uncertain
whether terms of the form~\refeq{eq:GEXPh} can contribute at NNLO or not.
However, this is not the case, since in the perturbative expansion of
~\refeq{eq:GEXPb},
\begin{equation}
\label{eq:GEXPi}
\int \limits_{{\rm dec}}^{} {\rm d} \sigma = {\rm d} \sigma_0 + {\rm d}
\sigma_1 + {\rm d} \sigma_2+...\,,
\end{equation}
it is clear that there is no room for terms that depend on the decay width. 
This can be easily understood also in the standard approach, where
$1/\Gamma_\NLO$ is kept unexpanded. In this case,
the spurious term~\refeq{eq:GEXPh} is exactly 
cancelled by a corresponding NNLO term, 
\begin{equation}
\label{eq:GEXPj}
\int\limits_{\dec}{} 
{\rm d} \sigma^{\NLO\times \NLO}_{\pxd}
 \,=\,{\rm d} \sigma_1 \frac{\Gamma_1}{\Gamma_\NLO}\,,
\end{equation}
which consists of the product 
of the $\ord(\alpha_s)$ corrections to 
$\rd \sigma_{\rm prod}$ and $\rd \Gamma$.
For these reasons, the term~(\ref{eq:GEXPh})
can be regarded as a spurious $\ord(\alpha_s^2)$ contribution, 
while the expanded result~(\ref{eq:GEXPg}) 
guarantees a consistent implementation of the corrections 
to the decay subprocess,
in the sense that  
the general properties~(\ref{eq:GEXPa})--(\ref{eq:GEXPb}) of the NWA
are fulfilled order by order in perturbation theory.

We note that the consistency of the inclusive NWA cross section
could also be guaranteed by keeping $1/\Gamma_\NLO$ unexpanded and 
including the product 
of the NLO corrections to production and decay,
$\rd \sigpord{1}\rd\Gamma_1/\Gamma_{\NLO}$, 
into the NLO calculation.
As compared to the
inverse-width expansion~\refeq{eq:GEXPg}, this alternative approach
would yield the same inclusive cross section, but an improved 
differential description of NLO radiation in decays.
However, its implementation is technically more involved. 

Let us now consider the NWA for the production and decay of multiple resonances.
In this case, to all orders in perturbation theory\footnote{In the following 
we use a notation corresponding to a single decay 
channel per resonance. However, all formulas are also 
applicable to resonances with more than one decay channel.
In general, each $\rd \Gamma_r$ should be understood as the 
partial decay width $\rd \Gamma_{r,c_r}$ for a specific 
decay channel $c_r$ of the resonance $r$,
while the total decay width corresponds to 
$\Gamma_r = \sum_{c_r}\int_{\dec}\Gamma_{c_r}$.
}
\begin{equation}
\label{eq:GEXPk}
{\rm d} \sigpxd  = {\rm d} \sigp 
\prod_{r\in\calR}\frac{{\rm d} \Gamma_{r}}{\Gamma_r}\,,
\end{equation}
where $\calR$ is the set of intermediate unstable particles 
in the process at hand.
The naive NLO implementation of the NWA corresponds to 
\begin{equation}
\label{eq:GEXPl}
\rd\sigpxdord{\NLO}
\, =\,
\left[
 \rd\sigpord{0}
+ \sum_r \rd \sigpord{0}
\frac{\rd \Gamma_{r,1}}{\rd \Gamma_{r,0}} 
\,+\, \rd \sigpord{1}
\right]\left(\prod_{r\in\calR}\frac{{\rm d} \Gamma_{r,0}}{\Gamma_{r,\NLO}}\right)\,.
\end{equation}
Instead, extending the expansion to all $1/\Gamma_{r,\NLO}$ terms yields
\begin{equation}
\label{eq:GEXPm}
\rd\sigpxdord{\NLOexp}
\, =\,
\left[
 \rd\sigpord{0}
+
\sum_r
\left(\rd \sigpord{0}\frac{\rd \Gamma_{r,1}}{\rd \Gamma_{r,0}} 
-
\rd \sigpord{0}\frac{\Gamma_{r,1}}{\Gamma_{r,0}} 
\right)
\,+\, \rd \sigpord{1}
\right]\left(\prod_{r\in\calR}\frac{{\rm d} \Gamma_{r,0}}{\Gamma_{r,0}}\right)\,,
\end{equation}
and it is easy to show that, similarly as in~\refeq{eq:GEXPg},
integrating~\refeq{eq:GEXPm} over the decay phase space and summing 
over all decay channels, yields exactly the NLO cross section for the 
production process.
In the context of top-quark pair production,
this expansion approach was first proposed in~\citere{Melnikov:2009dn}
and was also applied at NLO in~\citere{Campbell:2012uf},
and at NNLO in~\citere{Czakon:2020qbd}.

At NLO, the expanded version of the NWA can be related to the
unexpanded one through
\begin{eqnarray}
\label{eq:GEXPn}
\rd \sigpxdord{\NLOexp}
&=&
\left(\prod_{r\in\calR}\frac{\Gamma_{r,\NLO}}{\Gamma_{r,0}}\right)
\rd \sigpxdord{\NLO}
-
\left(\sum_{r\in\calR}\frac{\Gamma_{r,1}}{\Gamma_{r,0}} 
\right)
\rd \sigpxdord{\LO}\,,
\end{eqnarray}
where 
\begin{equation}
\label{eq:GEXPnb}
\rd \sigpxdord{\LO}\,=\,
\rd\sigpord{0} \left(\prod_{r\in\calR}\frac{{\rm d} \Gamma_{r,0}}{\Gamma_{r,0}}\right)
\end{equation}
is the LO cross section evaluated using the LO decay widths as input parameters.
Using~\refeq{eq:GEXPn} one can express the
spurious contribution for the case of multiple resonances as
\begin{eqnarray}
\label{eq:GEXPo}
\rd \sigfake
&=& \rd \sigpxdord{\NLO} - \rd \sigpxdord{\NLOexp}  
\,=\,
\delta\kappafake\,\rd \sigpxdord{\LO}
\,,
\end{eqnarray}
where
\begin{eqnarray}
\label{eq:GEXPp}
\delta\kappafake
&=& 
\left[1-
\left(\prod_{r\in\calR}\frac{\Gamma_{r,\NLO}}{\Gamma_{r,0}}\right)
\right]\kappa^{\pro\times\dec}_{\NLO}
\;-\;
\sum_{r\in\calR}\frac{\Gamma_{r,1}}{\Gamma_{r,0}} 
\,,
\end{eqnarray}
with
\begin{eqnarray}
\label{eq:GEXPq}
\kappa^{\pro\times\dec}_{\NLO}&=&\frac{\rd \sigpxdord{\NLO}}{\rd \sigpxdord{\LO}}
\,.
\end{eqnarray}
Here we see that the spurious correction factor~\refeq{eq:GEXPp} 
is a simple linear function of the differential NLO $K$-factor~\refeq{eq:GEXPq} 
with constant coefficients. Thus, $\delta\kappafake$
should feature a similarly mild kinematic dependence as the
differential $K$-factor. 

In order to estimate the typical size of
$\delta\kappafake$, it is useful to simplify~\refeq{eq:GEXPp} 
by retaining only  $\ord(\as^2)$ terms of 
type $\rd\sigpord{1} \Gamma_{r,1}$ 
and discarding terms beyond first order in $\Gamma_{r,1}$. 
This is justified by the fact that
the QCD corrections to production processes are typically 
significantly bigger as compared to those for decays.
In this approximation one can show that 
\begin{eqnarray}
\label{eq:GEXPr}
\delta\kappafake
&\simeq& {}- \frac{\rd\sigpord{1}}{\rd\sigpord{0}}
\left(\sum_{r\in\calR}\frac{\Gamma_{r,1}}{\Gamma_{r,0}} 
\right)
\,,
\end{eqnarray}
where the sum between brackets corresponds to the
relative effect of the corrections to all 
decay subprocesses.
For processes that involve 
$t\bar t$ production, possibly in association with other particles,
we have
\begin{eqnarray}
\label{eq:GEXPrb}
\delta\kappafake^{t\bar t+X}
&\simeq& {}- 2\,\frac{\rd\sigpord{1}}{\rd\sigpord{0}}
\frac{\Gamma_{t,1}}{\Gamma_{t,0}} 
\simeq {}+17\%\,\frac{\rd\sigpord{1}}{\rd\sigpord{0}}\,,
\end{eqnarray}
which can be a quite significant effect, 
depending on the size of the 
$K$-factor for the production subprocess.
In general,
if the corrections to the production subprocess
are close to $100\%$, 
as in the case of $t\bar t b\bar b$ production~\cite{Jezo:2018yaf}, 
then $\delta\kappafake$ is as large 
as the total NLO correction to all 
decay subprocess. Thus, for any processes or kinematic region
where $\rd\sigpord{1}/\rd\sigpord{0}$ is large,
the NLO corrections to the involved decays 
are strongly distorted in the unexpanded NWA.

As we will see in \refse{sec:nlops}, in the 
case of $t\bar t+tW$ production $\delta\kappa_{\rm spurious}$ is around $7\%$,  
which exceeds the size of the full 
$\tW$ contribution. Thus, in order to avoid this misleading 
effect, in the next section we propose a generalisation
of the expansion~\refeq{eq:GEXPm} to off-shell calculations.

\subsection{Inverse-width expansion in off-shell calculations and \POWHEGRES matching}
\label{se:offshellexp}

Let us now consider an off-shell process 
that involves (or is dominated by) a single resonance history, i.e.~a 
single production subprocess and a single combination of chain decays.%
\footnote{The inverse-width expansion for fixed-order NLO calculations 
with multiple resonance histories can be implemented in a similar way 
as discussed below in the \POWHEGRES framework, namely through a
separation of the various resonance histories.}
In the limit of small widths, the off-shell LO and NLO 
cross sections tend to the respective NWA cross sections, i.e.
\begin{equation}
\label{eq:GEXPt}
\rd \sigofford{\mathrm{(N)LO}}
\,\underset{\Gamma_r\to 0}{\longrightarrow}\,
\rd \sigpxdord{\mathrm{(N)LO}}\,.
\end{equation}
Thus the inverse-width expansion~\refeq{eq:GEXPn} 
can be easily extended to the off-shell
case by replacing the LO and NLO cross sections in the NWA
by their off-shell counterparts.
In addition, in order to ensure that the
shapes of the various resonances are perfectly consistent with the
corresponding NLO decay widths,
$\rd \sigofford{\mathrm{LO}}$ should be replaced by 
the Born contribution to the NLO cross section,
which is computed using NLO instead of LO 
widths as input parameters. 
For this quantity we use the 
symbol $\rd \sigofford{(0)}$,
and the off-shell extension of the expansion~\refeq{eq:GEXPn} 
is given by 
\begin{equation}
\label{eq:GEXPs}
\rd \sigofford{\NLOexp}
\, =\,
\left(\prod_{r\in\calR}\frac{\Gamma_{r,\NLO}}{\Gamma_{r,0}}\right)
\left[\rd \sigofford{\NLO}
-
\left(\sum_{r\in\calR}\frac{\Gamma_{r,1}}{\Gamma_{r,0}} 
\right)
\rd \sigofford{(0)}\right]\,,
\end{equation}
where $\calR(h)$ is the set of resonances
that occur in the resonance history $\labhist$.
Note that here, at variance with~\refeq{eq:GEXPn}, 
the product of $\Gamma_{r,\NLO}/\Gamma_{r,0}$ factors
is applied both to 
$\rd \sigofford{\mathrm{NLO}}$ and 
$\rd \sigofford{(0)}$ 
since these two ingredients are
both computed using 
NLO decay widths.
By construction, the zero-width limit of~\refeq{eq:GEXPs}
is equivalent to~\refeq{eq:GEXPn}, and is thus consistent with the
correct NLO production cross section. 

Now we turn our discussion to an extension of the inverse-width expansion to the matching 
of off-shell NLO calculations in the \POWHEGRES framework.
The aim of the prescription~\refeq{eq:GEXPs} is to restore the correct
normalisation of the cross section, while
\POWHEG's emission probabilities, 
i.e.~the terms between square brackets
in~\refeq{eq:RESea} and~\refeq{eq:RESe}, 
have no net effect on the normalisation.
Thus, in~\refeq{eq:RESea} and~\refeq{eq:RESe} 
the inverse-width expansion can be restricted to
the $\bar B_{\labhist}(\Phi_\rB)$ terms and implemented,
in analogy with~\refeq{eq:GEXPs}, as
\begin{eqnarray}
\label{eq:GEXPu}
\bar{B}_{\labhist}(\Phi_\rB)\Big|_{\rm exp}
&=& 
\left(\prod_{r\in\calR(\labhist)}\frac{\Gamma_{r,\NLO}}{\Gamma_{r,0}}\right) 
\left[
\bar{B}_{\labhist}(\Phi_\rB)
- \left(\sum_{r\in\calR(\labhist)} \frac{\Gamma_{r,1}}{\Gamma_{r,0}}\right)
B_{\labhist}(\Phi_\rB)
\right]
\,.
\end{eqnarray}
Since $\bar B_{\labhist}(\Phi_\rB)$ is the 
contribution of a specific resonance history
$\labhist\in\sethist$, its inverse-width expansion
involves the widths of the corresponding resonances 
$r\in \calR(\labhist)$.
Here we assume that the relevant building blocks,
$B(\Phi_\rB)$, $V(\Phi_\rB)$ and $R(\Phi_\rR)$, 
have been constructed as usual, i.e.~without inverse-width expansion,
and using NLO widths as input parameters.
Thus, all terms in~\refeq{eq:GEXPu}, including the Born term
$B_{\labhist}(\Phi_\rB)$,
are multiplied by the products of $\Gamma_{r,\NLO}/\Gamma_{r,0}$
factors between squared parentheses.
This factor acts also to the virtual and real contributions to 
$\bar B(\Phi_\rB)$. Moreover it should be applied also 
to hard radiation as
\begin{eqnarray}
\label{eq:GEXPv}
R^{(\rm hard)}_{\labhist,\labcoll}
(\Phi_\rR)\Big|_{\rm exp}
&=& 
\left(\prod_{r\in\calR}\frac{\Gamma_{r,\NLO}}{\Gamma_{r,0}}\right) 
R^{(\rm hard)}_{\labhist,\labcoll}(\Phi_\rR)\,,
\end{eqnarray}
where $R^{(\rm hard)}_{\labhist,\labcoll}(\Phi_\rR)$ corresponds to 
the so-called hard remnant, i.e.~real radiation 
``harder than ${\tt hdamp}$'', 
which is handled as fixed-order NLO radiation.

As demonstrated in~\refse{sec:nlops}, this inverse-width expansion has a 
quite significant impact on off-shell $\ttbar+\tW$ production and decay cross sections. In
particular, when comparing off-shell against on-shell
$\ttbar+\tW$ generators,
the inverse-width expansion is absolutely crucial in order
to avoid large spurious differences and
to identify the remaining small differences 
that are due to physical off-shell effects.

\section{Off-shell \boldmath$t\bar t+tW$ production with dileptonic decays}
\label{se:bb4lgen}

In this section we briefly review the treatment of resonance histories in the
original version of the \bbfourl{} generator~\cite{Jezo:2016ujg}, and we then introduce a new
version of \bbfourl{}, which implements an extended set of resonance
histories as well as improved resonance projectors based on matrix-element
information, 
and the inverse-width expansion introduced in
\refse{se:topwidthcorr}.

\begin{table}
  \centering
\renewcommand{\arraystretch}{1.3}
  \begin{tabular}{c|c|c|c}
resonance history &  production subprocess & decay subprocesses & 
examples
\\\hline
$t\bar t$      &  $pp\to t\bar t$        &  $t\to W^+b$          & Fig.~\ref{fig:wwbbhist}\subref{hista}--\subref{histc}\\%
               &                         &  $\bar t \to W^-\bar b$ &\\\hline
$t W^-$        &  $pp\to t\, W^-\bar b$  &  $t\to W^+b$  & Fig.~\ref{fig:wwbbhist}\subref{histd}--\subref{histf}\\\hline
$\bar t \,W^+$ &  $pp\to \bar t\, W^+b$  &  $\bar t \to W^-\bar b$ & Fig.~\ref{fig:wwbbhist}\subref{histg}--\subref{histi}\\\hline
$Z$            &  $pp\to Z/H+b\bar b$ &  $Z/H \to \ell^+ \nu_\ell \,\ell^{\prime\,-}\bar{\nu}_{\ell'}$& \\   
  \end{tabular}
  \caption{List of the Born resonance histories for the off-shell 
dileptonic process~\refeq{eq:dileptprocdef} with corresponding 
labels and decomposition into production and decay subprocesses. 
The leptonic $W$-boson decays
$W^+\to \ell^+ \nu_\ell$  and
$W^-\to \ell^{\prime\,-} \bar{\nu}_{\ell'}$ are kept implicit.
}
\label{tab:bb4lreshist}
\end{table}

The \bbfourl{} generator describes the family of dileptonic processes
\begin{eqnarray}
\label{eq:dileptprocdef}
pp \to \ell^+ \nu_\ell \,\ell^{\prime\,-} 
\bar{\nu}_{\ell'} b \bar{b}\,.
\end{eqnarray}
All process-dependent ingredients for the generation 
of events, \ie the terms $B$, $V$ and $R$ in the various 
\POWHEG formulas, are based on exact matrix-element input of 
$\ord(\alpha_s^2\alpha^4)$ at LO and 
$\ord(\alpha_s^3\alpha^4)$ at NLO.
At Born  level, the full process~\refeq{eq:dileptprocdef} 
involves four different resonance histories, 
which are listed in Tab.~\ref{tab:bb4lreshist} and 
correspond to the production subprocesses
$pp\to t\bar t$,\,
$pp\to t W^-\bar b$,\,
$pp\to\bar t\, W^+b$,\, and $pp\to Z/H+ b\bar b$, 
together with the respective decay subprocesses.
For simplicity, we will collectively 
denote the 
$t W^-\bar b$ and $\bar t\, W^+b$ histories as 
$\tW$ (or sometimes $tWb$) histories.
In practice the $pp\to Z/H+ b\bar b$ subprocess plays a 
negligible role, and the full process is completely dominated by
the resonance histories of 
$t\bar t$ and $\tW$ type. The latter are illustrated in
\reffi{fig:wwbbhist}, where leptonic $W$-boson decays 
are omitted for simplicity.

In the dileptonic \bbfourl generator the
inverse-width expansion~\refeq{eq:GEXPu}--\refeq{eq:GEXPv}
is restricted to top resonances and is applied,
depending on the
resonance history, to one or two top quarks.
Thus~\refeq{eq:GEXPu} yields correction factors 
$(\Gamma_{t,\NLO}/\Gamma_{t,0})^{n(h)}$
and $n(h)\,(\Gamma_{t,1}/\Gamma_{t,0})$ where 
$n(h)$ is the history-dependent number of top resonances.
As for  $W$ resonances,  in the 
dileptonic process~\refeq{eq:dileptprocdef} they can not
give rise to any spurious effects of type~\refeq{eq:GEXPp}.
This is due to the fact that leptonic $W$ decays do not receive any 
$\ord(\as)$ correction. 
Thus, all matrix elements are 
evaluated using $\Gamma_{W,\NLO}$ throughout as
input parameter, and without expanding $1/\Gamma_{W,\NLO}$.
For a discussion of the inverse-width expansion in the case 
of semileptonic decays see \refse{se:PWGRESsl}.

\subsection[The original \bbfourl{} generator]{The original \bbfourlInTitle{} generator}
\label{se:bb4loriginal}

The implementation of the \POWHEGRES approach 
in the original version of the \bbfourl{} generator~\cite{Jezo:2016ujg}
is based on a simplified treatment of the resonance structure of the
process~\refeq{eq:dileptprocdef}. In particular, instead of the full set of
Born resonance histories listed in Tab.~\ref{tab:bb4lreshist}, 
only the $t\bar t$ and the $Z$ resonance histories 
have been considered. The role of the former was to guarantee a correct
treatment of all top resonances, 
while the $Z$ history was meant to account for non-resonant contributions.

These two resonance histories have been implemented using naive projectors
of type~\refeq{eq:RESb} and~\refeq{eq:RESdb}, including the relevant top, anti-top, $W$- and $Z$-boson
resonances. 
In~\citere{Jezo:2016ujg} it was 
found that the $Z$ resonance history
is suppressed at the sub-percent level.  Thus
almost all LHEs are generated
according to the $\ttbar$ history in the original \bbfourl{} generator.
This implies that, for almost all events, \POWHEG and shower radiation is
emitted in a way that preserves the virtuality of the $W^+b$ and $W^-\bar b$
pairs. Moreover, at the LHE level, the {\tt allrad} approach~\refeq{eq:RESe}
leads to three
``factorised'' QCD emissions: one from the $\ttbar$ production subprocess plus 
two extra emissions from the $t\to W^+b$ and $\bar t\to W^-\bar b$ decays.
As discussed in more detail below, this treatment is well justified only for
events of $t\bar t$ type, while in the original version of \bbfourl{} it is
applied also to events of $\tW$ type.

\subsection[The new \bbfourldl{} generator with improved resonance histories]{New \bbfourldlInTitle{} generator with improved resonance histories}
\label{se:newreshist}

In this section we present a new version of the \bbfourl{} generator that
implements new resonance histories of $\tW$ kind together with a more
accurate determination of the resonance history projectors~\refeq{eq:RESc} and~\refeq{eq:REScc},
based on matrix-element information.  This new version of \bbfourl{} will be referred
to as \bbfourldl{}, where ``{\tt dl}'' stands for dileptonic, while its
single-lepton extension introduced in Sect.~\ref{see:bbfourlsl} will be called \bbfourlsl{}.

At variance with the original \bbfourl{}, in the new \bbfourldl{} generator
the negligible $Z$ resonance history is discarded, while the dominant $\ttbar$
history is supplemented by the two subdominant resonance histories of $\tW$ type, which
were absent in the original version of \bbfourl{}.  
Thus \bbfourldl{} is based on the full set of
resonance histories 
that dominate the process~\refeq{eq:dileptprocdef}
in most regions of the phase space.

As discussed in Sect.~\ref{sec:resmethod}, in each resonance history
the kinematic mappings~\refeq{eq:PWGc} 
need to be defined in such a way that collinear QCD radiation does not 
modify the virtuality of the relevant resonances. 
To this end, all mappings associated with the $t\bar t$ history
have been kept as in the original \bbfourl{} generator, 
while in the case of the
$tW^-\bar b$\;($\bar t W^+ b$) histories novel mappings are used, where
the non-resonant $W^-\bar b$\;($W^+ b$) pairs are handled as part of the hard
production subprocess, and only the virtuality of the resonant $t$\,($\bar
t$) quarks is preserved.

The most significant difference between the original and the new version of
{\tt bb4l} lies in the treatment of QCD radiation stemming from events of
$\tW$ kind.  On the one hand, the new $\tW$ histories imply that 
events of $\tW$ kind generate only two \POWHEG emission in the {\tt allrad} approach: one
from the $pp\to tWb$ production subprocess and a second one from the decay
of the top resonance. The generation of these two independent 
\POWHEG emissions is justified by the factorisation properties 
of QCD radiation in the presence of a top resonance.
On the other hand, in the original approach without $\tW$ 
histories, events of $\tW$ kind 
are treated in the same way as $\ttbar$ events.
In the {\tt allrad} approach, this leads to 
a third \POWHEG emission that is generated by 
an off-shell $Wb$ pair in a way that 
is not supported by theoretical arguments.
In fact, from the viewpoint of the correct $\tW$ resonance structure, 
this third emission can violate the consistent ordering of 
QCD radiation in the $pp\to tWb$ production subprocess.
Vice versa, with the new $\tW$ resonance histories,
the $pp\to tWb$ subprocess leads to a single \POWHEG radiation 
followed by consistently ordered shower emissions.

For what concerns the implementation of the new resonance histories, in
order to assign realistic $\ttbar$ and $\tW$ probabilities and to assess the
related ambiguities, we have considered two different types of history projectors: naive projectors 
of type~\refeq{eq:RESb} and improved projectors based on matrix elements.

\subsubsection*{Naive resonance projectors}
As a first option we have considered resonance projectors of the same 
form~\refeq{eq:RESc}--\refeq{eq:RESb} as in the original {\tt bb4l} generator. 
At Born level, the weight of the $\ttbar$ history reads
\begin{eqnarray} 
\label{eq:naiveresA}
\weighthistal{\ttbar}(\Phi_\rB)\Big|_{\naivehist} 
&=&  W_t (p_t) W_t (p_{\bar
t})\,,
\end{eqnarray}
with
\begin{eqnarray}
\label{eq:naiveresB}
W_t(p)&=& \frac{M_t^4}{(p^2 - M_t^2)^2+\Gamma_t^2 M_t^2}\,,
\end{eqnarray}
where the top and anti-top momenta are
given by $p_t=p_b+p_{\ell^+}+p_{\nu_\ell}$ and 
$p_{\bar t} = p_{\bar b}+p_{\ell^{'-}}+p_{\bar\nu_{\ell'}}$.
In the case of $\tW$ histories, one of the top propagators is absent. At
the same time, in the dominant $gg$ channel, and in the 
dominant phase-space region for $\tW$ production, the
$gg\to W^+W^- b\bar b$ matrix elements involve an enhanced $t$-channel 
propagator that is associated with initial-state $g\to b\bar b$ splittings.
In the collinear regions, the virtualities of such enhanced propagators 
correspond to the transverse energies of the $b$-quark spectators that are involved in the 
initial-state splittings, \ie
\begin{eqnarray} 
\label{eq:naiveresC}
E^2_{\rT, b}      &=& m_b^2 + p_{\rT,b}^2\,,
\qquad\mbox{or}\qquad
E^2_{\rT, \bar b} \,=\, m_{b}^2 + p_{\rT,{\bar b}}^2\,.
\end{eqnarray}
Thus the weights for the $\tW$ histories are defined as the 
product of the weights~\refeq{eq:naiveresB}
for a single top or anti-top propagator
combined with the $1/E^2_{\rT}$ enhancements
form the 
associated $g\to b\bar b$ splittings, \ie
\begin{eqnarray} \label{eq:naiveresD}
\weighthistal{\bar t W^+}(\Phi_\rB) 
\Big|_{\naivehist} 
&=&  
\frac{\chitw\,m_t^2}{E^2_{\rT,b}}\,
{W_t (p_{\bar t})}
\,,
\qquad
\weighthistal{t W^-}(\Phi_\rB) 
\Big|_{\naivehist} 
\,=\,  
\frac{\chitw\,m_t^2}{E^2_{\rT,\bar b}}\,
{W_t (p_{t})}\,.
\end{eqnarray}
To be precise, in the resonance weights~\refeq{eq:naiveresA} and~\refeq{eq:naiveresD}
 we have also included extra 
contributions corresponding to the $W^+$ and $W^-$ propagators. However,
such $W$ contributions are identical for all considered histories, and thus they 
cancel out in the resonance projectors~\refeq{eq:PWGg} 
and~\refeq{eq:RESc}.
For what concerns~\refeq{eq:naiveresD}, note that  
we have introduced a factor $\chitw$, which can be adjusted in
a way that the relative weights of the $\ttbar$ and $\tW$ histories are in
reasonably good agreement with the corresponding physical cross sections.
This is not guaranteed for $\chitw=1$, since the different nature of the 
Breit--Wigner and initial-state $g\to b\bar b$ enhancements implies the presence of 
different extra prefactors in~\refeq{eq:naiveresB} and \refeq{eq:naiveresD}.

The real-emission resonance weights 
$\weighthistal{\labhist,\labcoll}(\Phi_{\rR})$
that enter~\refeq{eq:RESd} have been constructed as in 
\refeq{eq:naiveresA}--\refeq{eq:naiveresD}, but 
replacing $p_{b}\to p_{b}+p_g$ or $p_{\bar b}\to p_{\bar b}+p_g$
in the collinear sectors that involve the corresponding 
splittings.

\subsubsection*{Matrix-element based resonance projectors}

For a more accurate separation of the resonance histories of 
$\ttbar$ and $\tW$ type we have 
implemented improved resonance projectors 
based on matrix elements.\footnote{This strategy was first suggested
in~\cite{Jezo:2015aia} but not implemented so far.} 
To this end, we have split the $q\bar q$ and gluon--gluon Born amplitudes
of the full process~\refeq{eq:dileptprocdef}
according to their resonance structure
as
\begin{eqnarray} 
\label{eq:MEresA}
\calA_\full\,=\, 
\calA_{\ttbar}
\,+\,\calA_{\bar t W^+}
\,+\,\calA_{t W^-}
\,+\,\calA_{\mathrm{rem}}
\,.
\end{eqnarray}
Here the summands on the rhs correspond to the subsets of Feynman diagrams 
that contain: two (anti)top resonances ($\calA_{\ttbar}$), 
a single top ($\calA_{\bar t W^+}$)
or anti-top ($\calA_{t W^-}$) resonance, 
and no top resonance at all ($\calA_{\mathrm{rem}}$).
In general, this splitting is not gauge invariant. However, it ensures a 
gauge-invariant separation of double and single-top processes in 
the phase-space regions that are strongly dominated by on-shell $t\bar t$ or $\tW$ production.
For the separation of the various resonance histories, we 
have employed the corresponding squared matrix elements, \ie
\begin{eqnarray} 
\label{eq:MEresB}
\weighthistal{\ttbar}(\Phi_\rB)\big|_{\MEhist} 
&=&  
|\calA_{\ttbar}|^2
\,,\qquad
\nonumber\\[2mm]
\weighthistal{\bar t W^+}(\Phi_\rB) 
\big|_{\MEhist} 
&=&
|\calA_{\bar t W^+}|^2
\,,
\nonumber\\[2mm]
\weighthistal{t W^-}(\Phi_\rB) 
\big|_{\MEhist} 
&=&
|\calA_{t W^-}|^2
\,.
\end{eqnarray}
In this definition of resonance projectors, 
all non-resonant (\ie free from any top resonance) effects,
as well as interferences between $\ttbar$, $\tW$ and non-resonant
contributions are neglected under the assumption that they play only a
subleading role. 
In the following%
, all non-resonant and interference contributions 
will be simply referred to as interference effects,
since they are dominated by the interference between $\ttbar$ and $\tW$ channels.

In the phase-space regions that are dominated either by 
$\ttbar$ or $\tW$ production, interference effects are expected to be
suppressed. However, in the off-shell phase space that separates 
the $\ttbar$ from the $\tW$ dominated regions,
interference effects can play a significant role.
In this case, the resonance weights~\refeq{eq:MEresB} 
may involve a significant ambiguity due to the fact
that interference effects are assigned to the
$\ttbar$ or $\tW$ histories in an uncontrolled way.
In order to quantify this
ambiguity we have considered two alternative 
resonance-history definitions, 
where the interference effects are explicitly assigned 
to one of the histories.
The first alternative is given by the weights
\begin{eqnarray} 
\label{eq:MEresC}
\weighthistal{\ttbar}(\Phi_\rB)\big|_{\MEhist'} 
&=&  
|\calA_\full|^2 - |\calA_{\bar t W^+}|^2 - |\calA_{\bar t W^+}|^2
\,,
\nonumber\\[2mm]
\weighthistal{\bar t W^+}(\Phi_\rB) 
\big|_{\MEhist'} 
&=&
|\calA_{\bar t W^+}|^2
\,,
\nonumber\\[2mm]
\weighthistal{t W^-}(\Phi_\rB) 
\big|_{\MEhist'} 
&=&
|\calA_{t W^-}|^2
\,,
\end{eqnarray}
which effectively assign 
all interference effects 
to the 
$\ttbar$ history, while the second alternative is given by
\begin{eqnarray} 
\label{eq:MEresD}
\weighthistal{\ttbar}(\Phi_\rB)\big|_{\MEhist''} 
&=&  
|\calA_{\ttbar}|^2
\,,\qquad
\nonumber\\[2mm]
\weighthistal{\bar t W^+}(\Phi_\rB) 
\big|_{\MEhist''} 
&=&
|\calA_{\bar t W^+}|^2
\,,
\nonumber\\[2mm]
\weighthistal{t W^-}(\Phi_\rB) 
\big|_{\MEhist''} 
&=&
|\calA_{t W^-}|^2
\,,
\nonumber\\[2mm]
\weighthistal{\mathrm{rem}}(\Phi_\rB) 
\big|_{\MEhist''} 
&=&
|\calA_{\full}|^2-
|\calA_{\ttbar}|^2-
|\calA_{\bar t W^+}|^2-
|\calA_{t W^-}|^2
\,,
\end{eqnarray}
where interference effects are excluded from the $\ttbar$ and
$\tW$ histories, and are assigned to an additional
``remainder'' history (rem).

\begin{table}
  \centering
\renewcommand{\arraystretch}{1.3}
  \begin{tabular}{c|cc|ccc|c}
 & \multicolumn{2}{c|}{naive} &\multicolumn{3}{c|}{matrix-element--based} & extrapolation
\\
         & $\chitw=1$& $\chitw=0.1$&$\MEhist$ &$\MEhist'$&$\MEhist''$ & $\Gamma_t \to 0$          \\\hline 
    $\ttbar$ & 90.6\%    & 95.3\%      & 94.2\%   & 93.7\%   &  95.3\% & 96.0\%  \\
    $tW $    & 9.4\%     & 4.7\%       & 5.8\%    &  6.3\%   &  6.2\%  & \multirow{ 2}{*}{$4.0\%$ } \\
rem      &           &            &         &         & ${}-1.5$\%    \\
  \end{tabular}
  \caption{Relative contributions of different resonance histories
to the total LO cross section for 
the full processes 
$pp \to \ell^+ \nu_\ell \,\ell^{\prime\,-} 
\bar{\nu}_{\ell'} b \bar{b}$ at 13\,TeV. 
The first two columns correspond to the ``naive'' $\ttbar$ and $\tW$
resonance histories defined in 
\refeq{eq:naiveresA}--\refeq{eq:naiveresD}
for two different values of the normalisation parameter $\chitw$.
The last three columns correspond to the 
matrix-element resonance histories defined in 
\refeq{eq:MEresB}--\refeq{eq:MEresD}. The $\MEhist''$ variant
defined in~\refeq{eq:MEresD} involves also a remainder history (``rem''), which
embodies all interference and non-resonant contributions.
\label{tab:resfractions}}
\end{table}

For the separation of real emission into resonance histories,
the Born weights~\refeq{eq:MEresB} and their
variants~\refeq{eq:MEresC}--\refeq{eq:MEresD}
have been extended to the real-emission phase space 
according to
\begin{eqnarray}
\label{eq:RESdd}
\weighthistal{\labhist,\labcoll}(\Phi_{\rR})\big|_{\MEhist} 
&=&
\weighthistal{\labhist}
(\widetilde{\Phi}_{\rB,\labcoll})\big|_{\MEhist} \,,
\end{eqnarray}
where the lhs corresponds to the weight that enters~\refeq{eq:RESd},
and the Born events $\widetilde{\Phi}_{\rB,\labcoll}$
on the rhs are defined through special \rtb
mappings,
\begin{eqnarray}
\label{eq:RESde}
\widetilde{\Phi}_{\rB,\labcoll}\,\equiv\,
\widetilde{\Phi}_{\rB,\labcoll}(\Phi_\rR)\,,
\end{eqnarray}
which are defined in a way that preserves the relative probabilities of
$\ttbar$ and $\tW$ histories.
This can not be achieved by inverting the standard collinear mappings, since
such mappings can lead to severe distortions of 
the virtualities 
of the enhanced $t$-channel propagators 
associated with initial-state $g\to b\bar b$ splittings
within $tW^-\bar b$\,($\bar t W^+ b$) resonance histories.
For this reason we have designed dedicated \rtb mappings~\refeq{eq:RESde}
that simultaneously preserve the 
virtualities of the resonant top quarks and $W$ bosons, and also, as 
far as possible, the
transverse energies of the $b$- or $\bar b$-quark emitters,
according to the collinear sector at hand.
Such mappings can be found in~\refapp{app:mappings}
and are only used for the 
construction of the resonance weights~\refeq{eq:RESdd}.

In order to quantify the ambiguity that is related to the treatment of 
interference effects, we have compared the contributions of the 
various resonance histories defined in~\refeq{eq:naiveresA}--\refeq{eq:naiveresD} 
and~\refeq{eq:MEresB}--\refeq{eq:MEresD} to the LO cross section
for the full process~\refeq{eq:dileptprocdef}. 
As shown in 
Tab.~\ref{tab:resfractions}, the naive resonance 
histories~\refeq{eq:naiveresA}--\refeq{eq:naiveresD} 
yield $\ttbar$ and $\tW$ fractions that are strongly sensitive to the choice of
the free normalisation parameter $\chitw$. At the level of the total cross
section,  setting $\chitw\simeq 0.1$ yields 
a reasonable $\tW$ fraction.  However,
this choice is not guaranteed to provide a consistent 
$\ttbar$\,--\,$\tW$ separation in the presence of arbitrary cuts and for
any differential observable. For this reason, matrix-element 
based resonance histories 
are certainly preferable. In this case, the
three different options defined in~\refeq{eq:MEresB}--\refeq{eq:MEresD}
yield fairly consistent $\tW$ fractions, which vary between 
$5.8\%$ and $6.3\%$. 
Comparing the $\MEhist'$ and $\MEhist''$ histories, as expected
we observe  
almost identical $\tW$ fractions, while 
the different treatments of $\ttbar$ histories and interference effects
give rise to significant deviations in the corresponding fractions.
In the $\MEhist''$ case, the ``rem'' channel embodies all 
interference effects, which turn out to be negative and amount  
to $-1.5\%$, while the ``pure'' $\ttbar$ channel corresponds to $95.3\%$.
Vice versa, in the $\MEhist'$ case 
all interference effects are attributed to the $\ttbar$ channel, which is
thus shifted by about $-1.5\%$ as compared to the $\MEhist''$ case. 
As for the $\MEhist$ case, comparing against $\MEhist''$ we observe that the 
interference effects are shared between the $\ttbar$ and $\tW$ channels 
with contributions that amount, respectively, to $-1.1\%$ and $-0.4\%$ 
of the total LO cross section.

In order to demonstrate that the new matrix-element--based projectors 
provide a reasonably well defined separation between $\ttbar$ and $\tW$
contributions, we have compared the various matrix-element--based fractions reported in 
Tab.~\ref{tab:resfractions} against an alternative separation 
based on the $\Gamma_t\to 0$ extrapolation
(see e.g.~\cite{Cascioli:2013wga})
of the LO cross section for the full process~\refeq{eq:dileptprocdef}. 
The key idea is that, 
in the limit where the total top-decay width is sent to zero,
the $\ttbar$ contribution ($|\calA_{\ttbar}|^2$) 
to the integrated cross section scales like $1/\Gamma_t^2$, 
while $\tW$ contributions and $\ttbar$--$\tW$ interferences scale like
$1/\Gamma_t$. Thus, the $\ttbar$ contribution can be 
defined in a gauge-invariant way as%
\begin{eqnarray} 
\label{eq:GammatexA}
\sigma_{\ttbar}\,=\,
\lim_{\xi_t\to
0}\left(\xi_t^2\,\sigma_{\mathrm{bb4l}}\Big|_{\Gamma_t\to \xi_t\Gamma_t}
\right)\,.
\end{eqnarray}
For the LO cross section at 13\,TeV, performing a 
numerical $\Gamma_t\to 0$ extrapolation we found 
\begin{eqnarray} 
\label{eq:GammatexB}
\frac{\sigma_{\ttbar}}{\sigma_{\mathrm{bb4l}}} \,=\,
  96.0\%\,,\qquad 
\frac{\sigma_{\mathrm{non-}\ttbar}}{\sigma_{\mathrm{bb4l}}} \,=\,
1-\frac{\sigma_{\ttbar}}{\sigma_{\mathrm{bb4l}}} 
\,=\,
  4.0\%\,.
\end{eqnarray}
This separation is expected to be equivalent to the one provided by the 
$\MEhist''$ resonance histories, since in both cases 
the $\ttbar$ channel and its complement correspond to the 
contributions stemming from 
$|\calA_{\ttbar}|^2$ and $|\calA_{\full}-\calA_{\ttbar}|^2$, 
respectively. 
However, contrary to the $\MEhist''$ approach, the 
definition of the $\ttbar$ contribution~\refeq{eq:GammatexA} involves also the 
 narrow-width limit. Still, the $\ttbar$ fractions obtained with the
$\MEhist''$ resonance histories and the $\Gamma_t\to 0$ extrapolation 
turn out to agree at the few permil level. Of course, the same level of 
agreement is also found between 
the non-$\ttbar$ parts in~\refeq{eq:GammatexB} 
and the combination of the $\tW$ and remainder histories of $\MEhist''$ type.

These findings, together with the comparison in
Tab.~\ref{tab:resfractions},
demonstrate that matrix-element--based resonance histories provide a sound
separation of the full process~\refeq{eq:dileptprocdef} into contributions
of $\ttbar$ and $\tW$ kind.  This separation is not exact due to the
unavoidable ambiguities that are related to the assignment of interference
effects.  However, such ambiguities can be controlled in a systematic way
through the definition of resonance histories, and in the case of the
integrated cross section they turn out to be quite small.

We note in passing that the $\ttbar$ and $\tW$ resonance histories of the new
{\tt bb4l} generator may also be exploited for applications that go beyond
the generation of \POWHEG radiation.
For example, the fact that LHEs are assigned, by construction,
to a specific resonance history\footnote{If needed, also the individual $\ttbar$ and $\tW$
probabilities can be made available on an event-by-event basis.}
makes it possible to split {\tt bb4l} event samples into $\ttbar$ and $\tW$ 
subsamples in a way that bears similarities with the separation of different
processes based on the matrix-element method~\cite{Kondo:1988yd}. 

Finally, we note that the introduction of resonance 
histories with independent $\ttbar$ and $\tW$ channels 
has required some technical improvements in the \RES integrator.
In the original setup the integrator adaptive sampling grids have been 
optimised using an
average of grids over all resonance histories weighted by the cross sections
in the individual resonance histories.
This strategy works well if the average grid is well suited for all 
resonance histories that yield a significant contribution to the 
total cross section.
This is not the case with the new resonance histories,
and with the original integration approach the $\tW$ histories feature a poor
convergence already at LO.
For this reason, we modified the \RES{} integrator such that each
resonance history provides an independent integration grid.
In this way, the relative error in the cross
section is each resonance history is roughly the same, and the total relative
error is a much steeper function of the number of calls as compared to the
case with only one grid.

\section{Off-shell \boldmath$t\bar t+tW$ production with semileptonic decays}
\label{see:bbfourlsl}

In this section we present the new {\tt bb4l-sl} version of the 
{\tt bb4l} generator, which describes  
off-shell $\ttbar$ and $\tW$ production with semileptonic decays. This  
reaction is part of the full process
\beq
\label{eq:sl-inclusive_signature}
pp\to \ell^{\pm} \nu_{\ell} j j b\bar b\,,
\eeq
which involves a variety of other  
QCD and electroweak reactions.
As will be discussed in~\refse{sect:bb4lsldef}, the 
contributions associated with 
$\ttbar$ and $\tW$ production
can be separated from the remaining contributions 
in a way that is free from any significant ambiguity 
due to interferences.
Based on this observation,
which is the outcome of a detailed analysis presented in \refapp{app:approximation},
we will select the physics content of the new \bbfourlsl{} generator,
i.e.~the contributing 
perturbative orders, partonic processes and Feynman diagrams,
in a way that corresponds to a direct generalisation
of the original \bbfourldl{} generator.
The implementation of the \bbfourlsl{} generator is described in
\refses{se:PWGRESsl}{se:technicalimp}.

\subsection[Selection of $t\bar t$ and $tW$ contributions to $\ell^{\pm}
\nu_{\ell} j j b\bar b$ production]{Selection of \boldmath$t\bar t$ and
\boldmath$tW$ contributions to \boldmath$\ell^{\pm} \nu_{\ell} j j b\bar b$
production}
\label{sect:bb4lsldef}

\begin{table}
  \centering
\renewcommand{\arraystretch}{1.5}
  \begin{tabular}{c|c|c|c}
$\as^n\alpha^m$  & dominant subprocesses & type  & order
\\    \hline
$\as^4\alpha^2$ & $W^\pm b\bar b$\,+\,2 jets & $V$+HF  & NNLO
\\    \hline
$\as^3\alpha^3$ & \multicolumn{3}{c}{tiny interference} 
\\    \hline
\multirow{5}{*}{$\as^2\alpha^4$}  & $\ttbar+tWb$ &  $\ttbar+tW$\;  & 4FNS\;LO 
\\
                &  $gq \to t q'\bar b$\,+\,1 jet & $t$-chanel single-top\; & 4FNS\;NLO 
\\
                &  $q\bar q'\to t\bar b$\,+\,2 jets & $s$-chanel single-top\; & NNLO 
\\
                &  $W^\pm Z$\,+\,2 jets with $Z\to b\bar b$ & $VV$ & NNLO 
\\
                &  $W^\pm jj$\,+\,2 $b$-jets & VBF & NNLO 
\\    \hline    
$\as^1\alpha^5$ & \multicolumn{3}{c}{tiny interference}
\\    \hline
\multirow{2}{*}{$\alpha^6$}      & $W^\pm Zjj$ with $Z\to b\bar b$  & VBS &  LO 
\\
               & $W^\pm ZV$ with $Z\to b\bar b$, $V\to jj$  & $VVV$  & LO 
\end{tabular}
  \caption{Dominant processes in the 
Born cross section for $pp\to \ell^{\pm} \nu_{\ell} j j b\bar b$
at the various orders 
$\as^{4-n}\alpha^{2+n}$ for $0\le n\le 4$.
The last two columns indicate, respectively,
the corresponding hard process (without light-jet emissions) 
and the order in QCD perturbation theory
at which it starts contributing to the 
$pp\to \ell^{\pm} \nu_{\ell} j j b\bar b$
Born cross section.
For instance, the vector-boson plus heavy flavour ($V$+HF) process
$pp\to W^\pm b\bar b$ starts contributing at NNLO.
See the main text for more details.
\label{tab:pertorders}
}
\end{table}

The Born cross section for the process~\refeq{eq:sl-inclusive_signature}
involves a tower of five different perturbative contributions, which range 
from $\ord(\as^4\alpha^2)$ to $\ord(\alpha^6)$,
and originate from the interplay of 
scattering amplitudes of order $\gs^4 e^2$,  $\gs^2 e^4$ and $e^6$.
As summarised in~\refta{tab:pertorders}, the physics content of the
individual squared Born terms is as follows.

\begin{itemize}

\item[(i)] The terms of $\ord(\as^4\aew^2)$ represent the leading QCD contributions
and originate form squared matrix elements of order $\gs^4 e^2$.
They are dominated by 
$W$-boson plus heavy-flavour production  ($W$+HF) 
in association with two additional light jets, \ie
$pp\to W^\pm b\bar b jj$, where the
$W$ boson decays leptonically.

\item[(ii)] The terms of  $\ord(\as^2\aew^4)$ arise from squared 
matrix elements of order $\gs^2 e^4$ as well as from 
the interference between 
matrix elements of order $\gs^4 e^2$ and $e^6$.
Such interferences are strongly colour-suppressed 
and are seven orders of magnitude smaller wrt 
the full $\ord(\as^2\aew^4)$ cross section.
The latter is dominated by $\ttbar$ and $\tW$ production, \ie 
$pp\to W^+W^- b\bar b$, with one leptonic and one hadronic $W$-boson
decay. Further subleading contributions are listed 
in \refta{tab:pertorders} and are discussed in more detail 
below and in \refapp{app:approximation}.

\item[(iii)] The terms of $\ord(\aew^6)$ arise from squared matrix elements of order $e^6$
and represent the lowest order in \as. They are dominated
by the vector-boson scattering (VBS) process
$pp\to W^\pm Z jj$ and the tri-boson production processes 
$pp\to W^\pm Z V$, with $Z\to b\bar b$ and 
a leptonically decaying $W$ boson, while in the tri-boson process
the vector boson $V=Z,W^\pm$ 
decays into two jets.

\end{itemize} 

The additional contributions of $\ord(\as^5\alpha^3)$ and
$\ord(\as^3\alpha^5)$ correspond to pure interferences between matrix
elements of different order. Such interferences are strongly suppressed due (also) to
colour-interference effects. 
Thus, the contributions (i)--(iii) can be regarded as three separate processes, 
and only (ii) is included in 
\bbfourlsl{}, while (i) and (iii) can be described through independent generators.
The exact physics content of the
\bbfourlsl{} generator is defined as the 
subset of the ingredients of the
full process~\refeq{eq:sl-inclusive_signature},
which results from the following three-step selection:
\begin{enumerate}
\item[(S1)] Only terms of $\ord(\as^2\alpha^4)$ at LO 
and $\ord(\as^3\alpha^4)$ at NLO are included;
 
\item[(S2)] The two light jets in the final state are 
required to contain a $q\bar q'$ pair with quark flavours
consistent with a $W^{\mp}\to q\bar q'$ decay;

\item[(S3)] Only LO and NLO topologies that are in one-to-one correspondence 
with those occurring in the related dileptonic process~\refeq{eq:dileptprocdef} are included, with the addition, as detailed below, of  
NLO QCD corrections associated with the $q\bar q'$ pair.

\end{enumerate}
As discussed below, and in more detail in \refapp{app:approximation},
this selection is free from possible ambiguities due to interferences.
Moreover it provides a good approximation of the 
full process~\refeq{eq:sl-inclusive_signature}
in phase-space regions where the 
invariant mass of the dijet system is not 
too far from $m_W$, i.e.~in the regions that are usually selected 
for experimental measurements of $t\bar t$ and/or $\tW$ production.

\newcommand{\bbfourlgraph}[2]{
\begin{subfigure}[b]{0.32\linewidth}
\begin{minipage}[t]{\linewidth}
\includegraphics[width=\textwidth,clip]{diagrams/#1} 
\end{minipage}
\caption{#2}
\label{#1}
\end{subfigure}
}

\begin{figure}
	\begin{center}
\bbfourlgraph{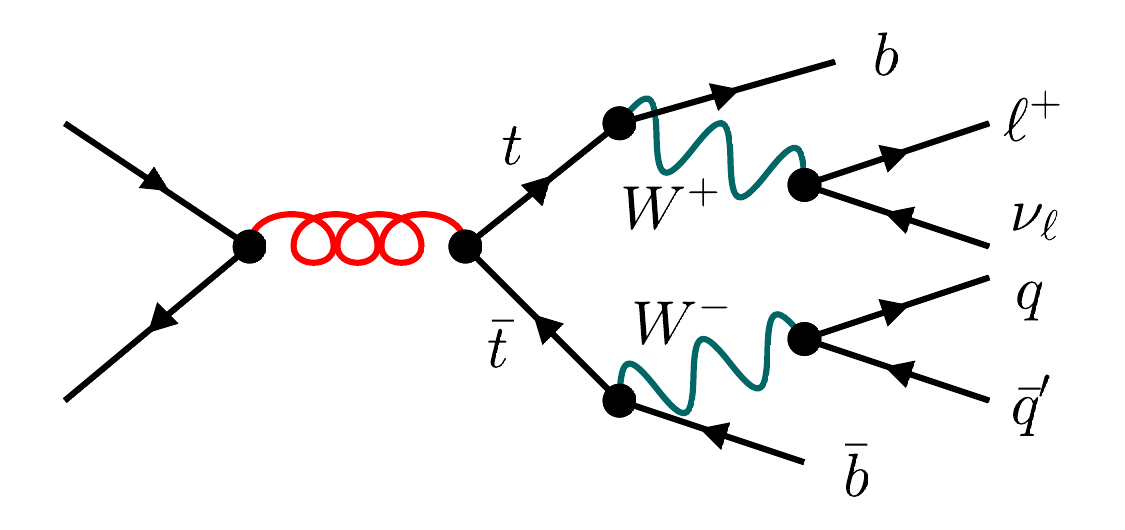}{$pp\to t\bar t$}
\bbfourlgraph{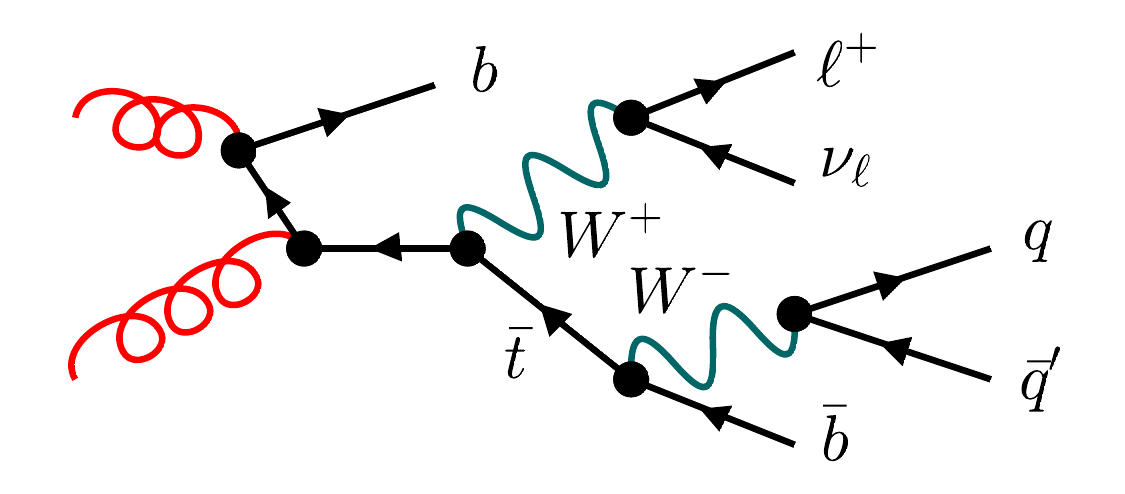}{$pp\to \bar t W^+b$}
\bbfourlgraph{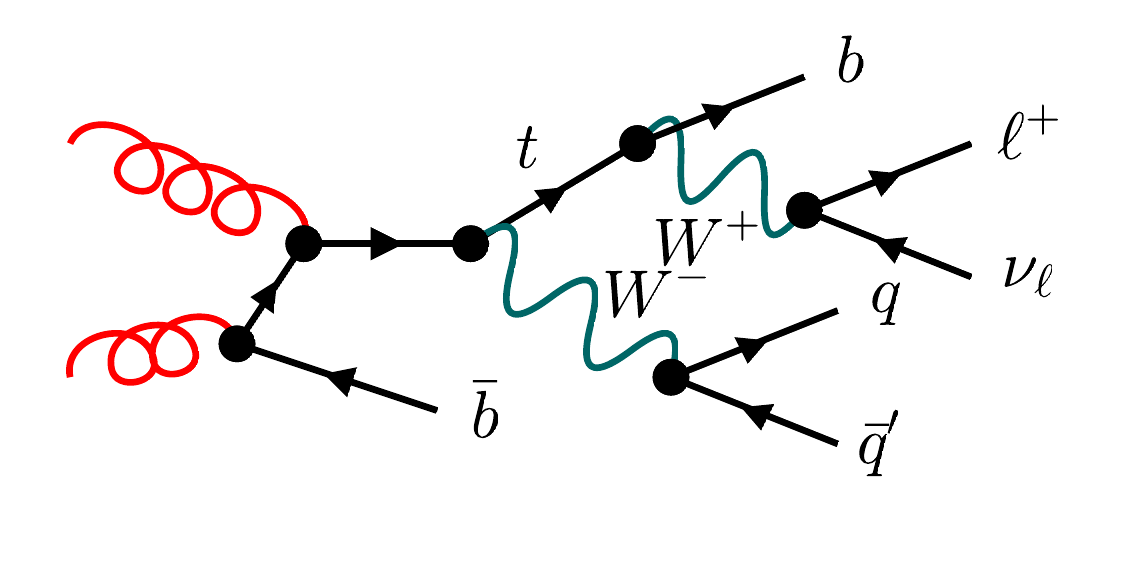}{$pp\to tW^-\bar b$}\\[3mm]
\bbfourlgraph{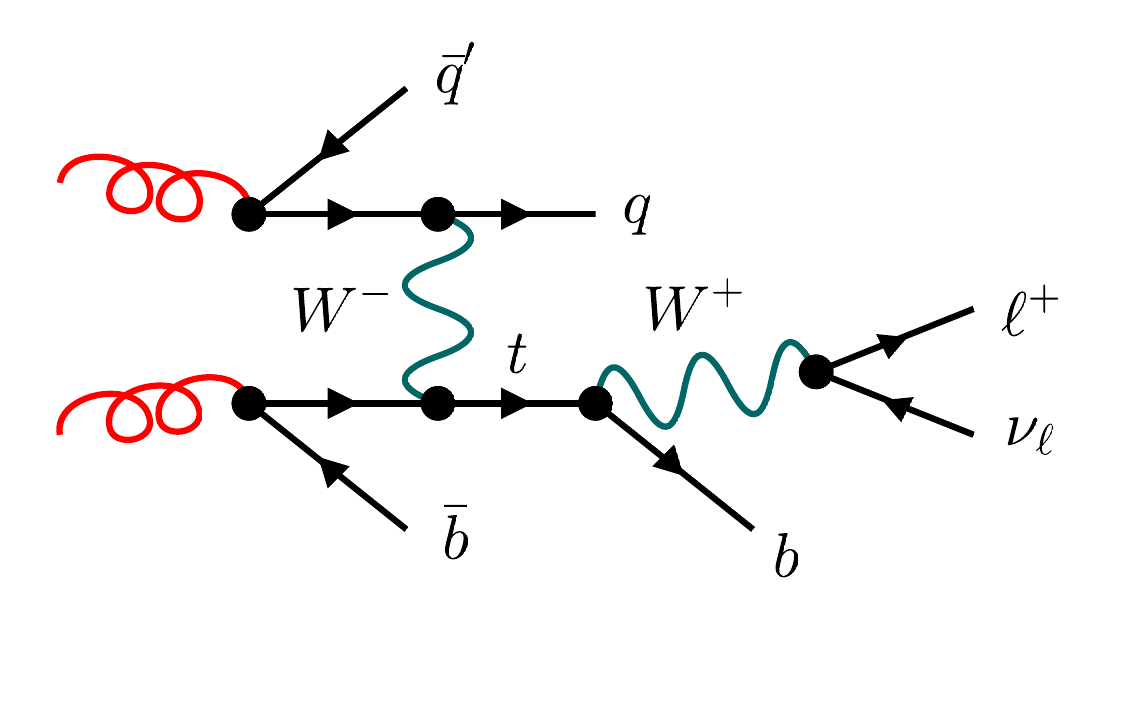}{$pp\to tj+2$\,jets }
\bbfourlgraph{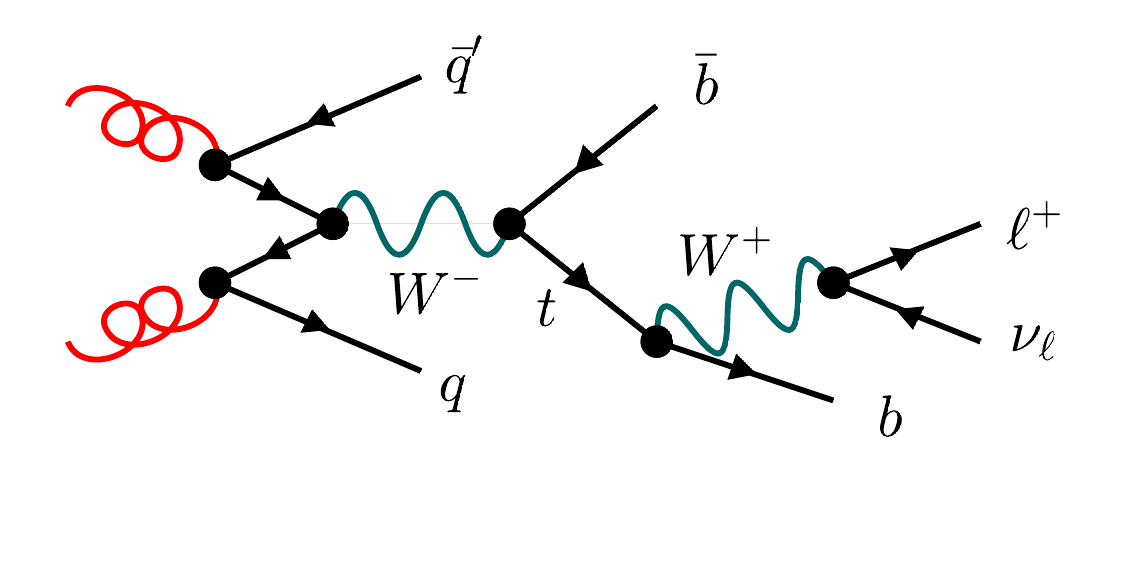}{$pp\to t{\bar b}+2$\,jets}
\bbfourlgraph{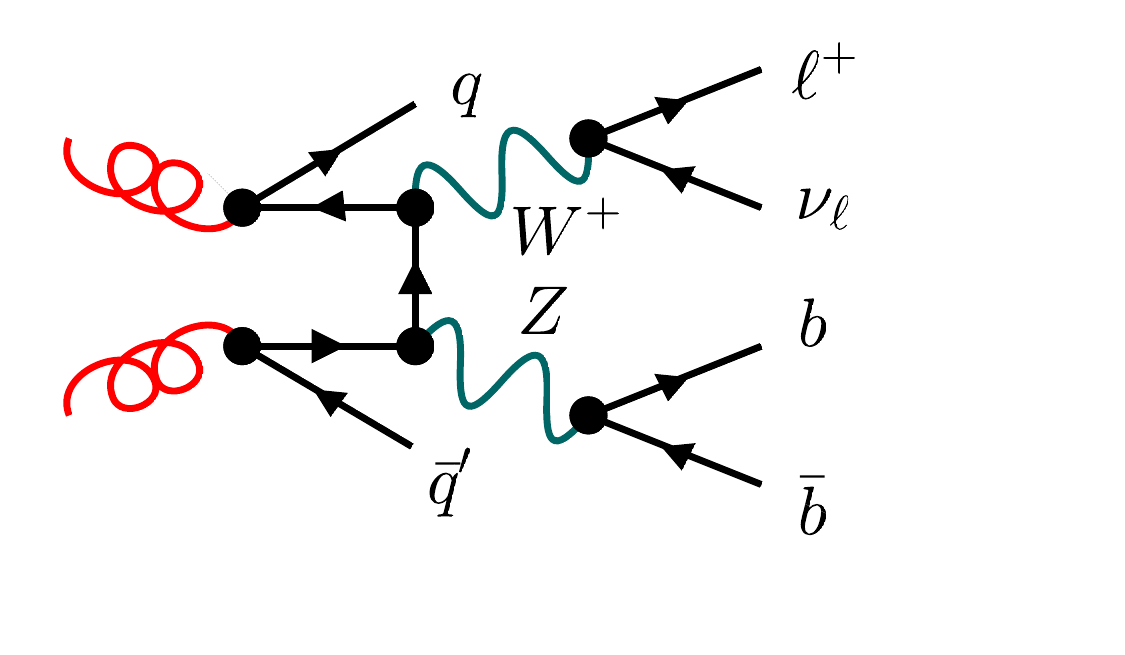}{$pp\to W^+Z(b\bar b)+2$\,jets}\\[3mm]
\bbfourlgraph{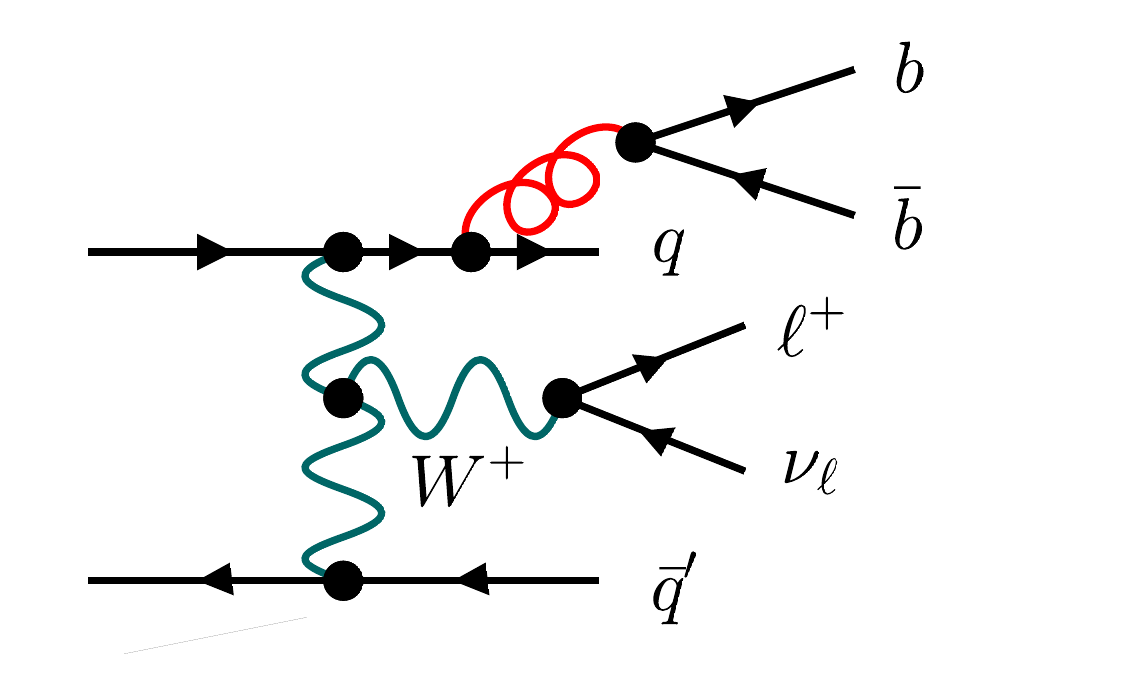}{$pp\;\xrightarrow[\mathrm{\hspace{-3ex}VBF\hspace{-3ex}}]{}\; W^+jj+2$\,$b$-jets}
	\end{center}
\caption{Representative tree diagrams for 
various $\ord(\as^2\aew^4)$  contributions 
to $pp\to \ell^{\pm} \nu_{\ell} q\bar q' b\bar b$.
See the main text for more details.
}
\label{fig:diagrams}
\end{figure}

Let us first consider the contributions that fulfill the criteria S1 and S2, i.e.~the
$\ord(\as^2\aew^4)$ and $\ord(\as^3\aew^4)$ contributions to
\beq
\label{eq:sl-qqp_signature}
pp \to \ell^{\pm} \nu_{\ell} q \bar q'  b\bar b\,,
\eeq
where in the case of a negatively (positively) charged lepton
the $q \bar q'$ pair must be consistent with the decay of a $W^+\,(W^-)$
boson, i.e.~$q \bar q'\,=\, u \bar d\, (d\bar u)$ or 
$c \bar s\, (s\bar c)$.
Note that the selection of this 
quark-flavour configuration is infrared safe at NLO.
At this order, the process~\refeq{eq:sl-qqp_signature}
involves all possible contributions 
and interference effects that can arise from 
$t\bar t$ or $\tW$ production 
with semileptonic decays.
Examples of the corresponding Born diagrams are 
depicted in Figs.~\ref{fig:diagrams}\subref{diag_tt}--\subref{diag_tw}, 
while the remaining diagrams in~\reffi{fig:diagrams},
see also~\refta{tab:pertorders},
correspond to various other physics processes 
that contribute to 
\refeq{eq:sl-qqp_signature} at $\ord(\as^2\aew^4)$.
These include $t$-channel single-top production in the 
4FNS at NLO, i.e.~with one extra jet (\reffi{fig:diagrams}\subref{diag_tchan}),
$s$-channel single-top production at NNLO, i.e.~with two extra jets 
(\reffi{fig:diagrams}\subref{diag_schan}), 
$pp\to WZ(b\bar b)$ at NNLO
(\reffi{fig:diagrams}\subref{diag_wz}),
and $Wjj$ production via vector-boson fusion
with an extra $b\bar b$ pair 
(\reffi{fig:diagrams}\subref{diag_vbf}). 
Formally all these different processes contribute to the 
same $\ord(\as^2\aew^4)$ and $\ord(\as^3\aew^4)$ cross section
interfering with each other.
Thus, in principle, they would have to be
collectively treated as a single off-shell process.
However, as shown in detail in \refapp{app:approximation},
the process \refeq{eq:sl-qqp_signature}  
can be well approximated as the incoherent sum
of two ingredients:
on the one side a process corresponding to off-shell $\ttbar + \tW$ production with interference 
and, on the other side,
all other processes.
In phase-space regions dominated by $\ttbar + \tW$ production
this approximation in fact holds at the permil level.

Based on this observation, in step S3 of our process definition 
we select all $\ttbar + \tW$ contributions and interferences
while discarding all other processes. 
Technically, at NLO this is achieved through the following two steps:
(a) selecting the subset of semileptonic 
Feynman diagrams that originate from 
the full set of diagrams for the 
dileptonic process~\refeq{eq:dileptprocdef}
by replacing a  lepton-neutrino 
pair with a $q\bar q'$ pair with the same 
weak-isospin quantum numbers;
(b) adding, as described below,
extra virtual and real-emission contributions 
associated with the QCD interactions of the $q\bar q'$ pair.
Schematically, this process definition can be written as
\beq
\label{eq:ll_signature}
pp \,\to\, \ell^{\pm} \nu_{\ell}  \ell'^{\mp} \nu_{\ell'} b\bar b\,\bigg|_{\ell'^{\mp} \nu_{\ell'}\;\rightarrow\; q\bar q'}
\;.
\eeq
Since it is based on the full set of Feynman diagrams for the dileptonic process,
this selection is guaranteed to be gauge invariant.
At the NLO, the same conversion~\refeq{eq:ll_signature} is applied also to the virtual and real corrections.

Regarding the additional NLO corrections that arise when the
lepton-neutrino pair is converted into a $q\bar q'$ pair, we note that the
$q\bar q'$ fermionic line, which was originally a leptonic line, couples
only to electroweak bosons and does not exchange any SU(3) colour with
the other QCD partons. For this reason, the NLO corrections that are inherited
from the dileptonic process via~\refeq{eq:ll_signature} do not interfere with the
additional QCD corrections that result from the interaction of virtual and
real gluons with the $q\bar q'$ fermionic line. 
Based on this observation,
we handle these extra QCD corrections as a separate contribution, 
which is directly implemented in the \POWHEGRES framework,
as explained in the next section. For efficiency reasons, for the
relevant matrix elements we use a $W^+W^-b\bar b$ 
double-pole approximation (DPA),
where we include only topologies 
that involve two resonant $W$ bosons.
This DPA embodies all possible $t\bar t+tW$
contributions to the full process~\refeq{eq:sl-inclusive_signature}
and provides also an accurate description of
the associated off-shell effects.
In fact, at LO the DPA description agrees at the permil level with the
\bbfourlsl{} description for all relevant inclusive and fiducial cross
sections and differential distributions.
Note that the $W^+W^-b\bar b$ DPA is used only for the 
QCD corrections associated with the
$q\bar q'$ pair, which arises only through the
$W\to q\bar q'$ decay in the DPA, while
all other NLO QCD ingredients 
are based on exact off-shell matrix elements 
for the dileptonic process.

\subsection[\POWHEGRES approach for $\ell^{\pm} \nu_{\ell} q \bar q'  b\bar b$ production]{\POWHEGRES approach for \boldmath$\ell^{\pm} \nu_{\ell} q \bar q'  b\bar b$ production}
\label{se:PWGRESsl}

Based on the above process definition,
the \POWHEGRES generator for the semileptonic
process~\refeq{eq:ll_signature}
can be implemented as an extension of the
original \bbfourldl{} generator.
The only missing ingredient 
that needs to be supplemented 
are the QCD corrections associated 
with the $q\bar q'$ pair.
This can be achieved by generating dileptonic
events with the \bbfourldl{} generator, 
and converting them into
semileptonic events according to the extended \POWHEGRES formula
\begin{eqnarray}
\label{eq:RESsla}
\mathd \sigma_{\bbfourlsl} & = & 
\mathd \sigma_{\bbfourldl}\,
K_{\Whad}
\left[
\Delta_{\Whad} (q_{\tmop{cut}}) \,+ 
\hspace{-.5em}
\sum_{\labcoll\,\in\, \setcoll(\Whad)} 
\hspace{-.5em}
\Delta_{\Whad} (k_{\rT,\labcoll})\, 
 \frac{R_{\DPA}(\Phi_{\rR,\labcoll})}{B_{\DPA}(\Phi_\rB)} 
\,\mathd 
\Phiradal{\labcoll} 
\right]\,.
\nonumber\\
\end{eqnarray}
Here $\mathd \sigma_{\bbfourldl}$ corresponds to 
dileptonic LHEs generated according
to the \POWHEGRES formula~\refeq{eq:RESe}
in the {\tt allrad} mode,
which gives rise to up to three \POWHEG~emissions.
Such dileptonic events should be reinterpreted 
as semileptonic ones as indicated in~\refeq{eq:ll_signature}.
The real radiation emitted by the resulting 
$q\bar q'$ pair is then generated 
as an extra \POWHEGRES+\,{\tt allrad}
emission, handling the $q\bar q'$ pair 
as the decay products of a $W$~resonance.
This extra emission is 
described by the expression between squared brackets in~\refeq{eq:RESsla},
which corresponds to the insertion of an extra 
$W\to q\bar q'(+g)$ decay
subprocess 
into the \POWHEGRES+\,{\tt allrad} formula~\refeq{eq:RESe}.
The sum over $c\in \calC(\Whad)$ accounts for the 
two collinear sectors in $W\to q\bar q' g$\,, and the
associated resonance-aware mappings ensure that 
the virtuality of the
intermediate $W$~boson is preserved.
Note that the $W\to q\bar q'(+g)$ 
subprocess is the same for all $t\bar t+tW$ histories, 
thus $\mathd \sigma_{\bbfourldl}$ is the only ingredient
that needs to be split into resonance histories.

The $R/B$ ratio in~\refeq{eq:RESsla}
is computed in the DPA as discussed above. 
More precisely, 
$R_{\DPA}(\Phi_{\rR,\labcoll})$ consists of all $2\to 7$ 
real-emission topologies of type
\begin{equation}
pp \to W^{\pm}(\to \ell^{\pm} \nu) W^{\mp}( \to  q \bar q' g) b\bar b\,, 
\end{equation}
where the extra gluon is emitted only within the
$W\to q \bar q'$ decay, 
while $B_{\DPA}(\Phi_{\rB})$ consists of all 
$2\to 6$ tree topologies of type
\begin{equation}
pp \,\to\, W^{\pm}(\to \ell^{\pm} \nu) W^{\mp} (\to  q \bar q') b\bar b\,,
\end{equation}
The argument of $B_{\DPA}(\Phi_{\rB})$ 
corresponds to the original underlying Born 
event of the dileptonic process.
The DPA is implemented only through a diagrammatic
filter that requires the presence of two $W$ resonances, while we refrain
from applying on-shell projections.
The Sudakov form factors $\Delta_{\Whad}$ are constructed as
in~\refeq{eq:RESs}, but using $R_{\DPA}/B_{\DPA}$ as emission probabilities.
This approach guarantees an accurate distribution of
QCD radiation in the $W$-decay phase space, 
including off-shell effects in the DPA
and with an exact treatment of spin correlations.

By construction, the total probability of 
the \POWHEG emission in~\refeq{eq:RESsla} is equal to one.
Thus, the 
Sudakov form factors effectively account for
the part of the virtual corrections to $W\to q\bar q'$ that cancels against
the real corrections.
The main effect of the remaining finite part of the virtual correction is a
relative shift of the differential cross section, which corresponds to the overall NLO
correction to the $W\to q\bar q'$ branching ratio, while we do not 
expect any other significant effect from the virtual corrections. 
Based on this observation,
in~\refeq{eq:RESsla} the finite part of the 
virtual corrections to $W\to q\bar q'$ is accounted for by the 
matching factor
\begin{eqnarray}
\label{eq:WBRmatching}
K_{\Whad} \,=\, 
\frac{\BR(W \to jj)}{\BR_{\bbfourl}(W \to l \nu)}\,,
\end{eqnarray}
which adapts the normalisation of the \bbfourlsl{}
cross section in a way that compensates for the different branching ratios
for hadronic and leptonic $W$ decays.
To this end, the denominator and numerator on the rhs of~\refeq{eq:WBRmatching}
should be chosen consistently with the
content of the \bbfourlsl generator, which corresponds to\footnote{Note that 
the normalisation of \bbfourldl{}
does not involve any sum over final-state lepton flavours,
while the \bbfourldl{} normalisation involves the sum over 
both generations of light quarks in the final state.
}
\begin{eqnarray}
\label{eq:WBRa}
\BR_{\bbfourl}(W \to l \nu)\,=\,
\left.\frac{\Gamma_{W\to\, \ell\nu}}{\Gamma_{W}}\,\right|_{\NLO}
\,=\,\left[9 + \frac{6}{\pi}\as(m_W)\right]^{-1}\,,
\end{eqnarray}
and
\begin{eqnarray}
\label{eq:WBRa}
\BR_{\bbfourl}(W \to jj)\,=\,1-3\times\BR_{\bbfourl}(W \to l \nu)\,.
\end{eqnarray}
The matching factor \refeq{eq:WBRmatching} guarantees a consistent treatment
of hadronic $W$ decays, without any expansion of $1/\Gamma_{W,\NLO}$.
As for the treatment of the top-quark width, 
the semileptonic generator automatically inherits
the inverse-width expansion~\refeq{eq:GEXPu}--\refeq{eq:GEXPv}
from the dileptonic generator through~\refeq{eq:RESsla}.

We note that the above procedure can be easily extended
to $t\bar t+tW$ production with fully hadronic final states.
To this end, one should simply handle both hadronic $W$ decays 
as described above.

\subsection[Implementation of the {\bbfourlsl{}} generator and
interface to \PythiaEight{}]{Implementation of the {\bbfourlslInTitle{}} generator and
interface to \PythiaEightInTitle{}}

\label{se:technicalimp}

The semileptonic extension of the \bbfourl{} generator is
implemented in the form of a \bbfourlsl{} plugin, which takes
dileptonic LHEs generated with \bbfourldl{} as input and transforms them into
semileptonic events according to \refeq{eq:RESsla}.
After reading in dileptonic
events, the \bbfourlsl plugin replaces the appropriate lepton and neutrino 
by a corresponding quark and anti-quark, 
adds the latter to the list of valid emitters,
and applies the normalisation factor~\refeq {eq:WBRmatching}.
Subsequently, up to one additional \POWHEG radiation is emitted from the
$W\to q\bar q'$ decay. The required 
Born and real-emission amplitudes in DPA
are evaluated using \OpenLoops{}~\cite{Cascioli:2011va,Buccioni:2017yxi, Buccioni:2019sur}
and its interface to \RES.

The LHEs that are generated by \bbfourldl and subsequently processed
by the \bbfourlsl plugin are stored in the standard LHE
format~\cite{Boos:2001cv} with the addition of 
non-standard information that is needed 
by the plugin to generate radiation
in the $W\to q\bar q'$ decays.
As in the original version of \bbfourl{},
each LHE involves 
multiple \POWHEG emissions that are generated by 
the various production and decay
subprocesses in the {\tt allrad} mode.
In practice, \bbfourldl
(\bbfourlsl) generates LHEs containing $6+n$
final-state particles with 
$0\le n \le 3\,(4)$.
Those \POWHEG emissions that originate from decay
suprocesses are linked to the corresponding 
resonances, whose momenta are also stored in the LHEs.
In addition, in each LHE we also store  
the kinematics of the associated underlying Born event $\Phi_\rB$.
For a reliable calculation of the DPA amplitudes, 
we also increased the number of printed 
digits for kinematic quantities 
to the maximum available in a
64-bit floating point type.
This is backward compatible with the original \POWHEGBOX{} Les Houches
reader.

The consistent matching of radiation generated by \bbfourlsl and 
\PythiaEight{} is guaranteed by a 
dedicated shower veto prescription within \PythiaEight{},
which is implemented in the language of {\tt UserHooks}, 
in {\tt PowhegHooks.h} and {\tt PowhegHooksBB4L.h}.
As usual, \Pythia{} is allowed to shower 
without restrictions, 
and each new emission is analysed by the
{\tt UserHooks} code, which decides whether to veto it or not,
based on the presence, the type and the hardness 
of \POWHEG emissions in the LHE.
In the first step {\tt PowhegHooksBB4L.h} identifies 
whether the new emission is being attached to the 
production subprocess or to the top or anti-top
decay.
These three different kinds of emissions are matched 
independently of each other, and only 
to \POWHEG emissions of the same kind.
Emissions stemming from the production subprocess are handled by the
standard {\tt PowhegHooks.h} code, while {\tt PowhegHooksBB4L.h} takes care
of radiation emitted from top decays.
In \bbfourlsl, hadronically decaying
top quarks may generate up to two \POWHEG emissions, one from the
$t\to Wb$ decay and one from the $W\to q\bar q'$ decay.
For what concerns the matching procedure, 
the hadronic $W$ decay needs to be handled as a third 
independent decay subprocess, on the same footing as the 
$t\to W^+b$ and $\bar t\to W^-\bar b$ decays.
This new feature has been implemented as a extension of the
original {\tt PowhegHooksBB4L.h} algorithm.

\section{Setup for numerical studies\label{sec:setup}}

In Sects.~\ref{sec:origVsMei}--\ref{sec:nlops} we investigate various
features of the original {\tt bb4l} generator~\cite{Jezo:2016ujg}, its new version with
matrix-element--improved resonance histories, and its extension to
semileptonic final states based on the {\tt bb4l-sl} plugin.
The required matrix elements are evaluated with 
\OpenLoops{}~\cite{Cascioli:2011va,Buccioni:2017yxi,Buccioni:2019sur}.

Quarks of the first two generations are  treated as
massless, and the Cabibbo--Kobayashi--Maskawa matrix is assumed to be
trivial. Bottom and top quarks are treated as massive quarks and,
similarly as in the 4FNS, they are excluded form the list of 
possible initial-state partons.
However, they are included in the 
loop corrections and are handled as active quarks 
in the renormalisation of the strong coupling, for which we use
\beq
\as(m_Z^2)\,=\,0.118\,.
\eeq
The consistent matching of this $b$-quark treatment 
with the PDFs is discussed below.

For the top quark and for $W$, $Z$ and Higgs bosons the complex-mass scheme~\cite{Denner:1999gp, Denner:2005fg} 
is used. In this approach,  
particle masses are replaced throughout by the
complex-valued parameters
\beq
\begin{split}
\label{eq:complexmasses}
\mu^2_i\,=&\;\;M_i^2-\ri\Gamma_i m_i\,.
\end{split}
\eeq
The electromagnetic coupling $\alpha$ and the weak mixing angle
$\theta_{\mathrm{w}}$
are derived from the gauge-boson masses and the
Fermi constant, 
\begin{equation}
\GF=1.16585\times10^{-5}~\GeV^{-2}\,, 
\end{equation}
in the $\GF$ scheme, via
\begin{equation}
\alpha=\sqrt{2}\, \frac{G_\mu}{\pi} \left|\mu_{\sss W}^2\(1-\frac{\mu_{\sss
    W}^2}{\mu_{\sss Z}^2}\)\right|=\frac{1}{132.50698}\,,
\end{equation}
and $\cos \theta_{\mathrm{w}}\,=\,{\mu_{\sss W}}/{\mu_{\sss Z}}$.
The employed input masses are 
\begin{align}
  m_{W} &= 80.419  \;\GeV\,,  &  m_{Z} &= 91.188 \;\GeV\;\GeV\,,\nonumber\\
  m_{t} &= 172.5 \;\GeV\,,   &   m_{b} &= 4.75 \;\GeV\,, \nonumber\\
  m_{H} &= 125 \;\GeV\,.  
\end{align}
For the gauge bosons we use the NLO QCD widths 
\begin{align}
  \Gamma^{\NLO}_{W} &= 2.10134 \;\GeV\,,  &  \Gamma_{Z}^{\NLO} &= 2.51080 \;\GeV\,,
\end{align}
and for the Higgs boson we use
\begin{align}
	  \Gamma_{H} &=4.03\times10^{-3}\;\GeV\,.
\end{align}
The value of the top-quark width is consistently calculated at NLO QCD from all other input
parameters 
at the level of the off-shell three-body decays $t\to f \bar f' b$
with light fermions $f, \bar f'$ and a massive $b$~quark. This yields
\begin{align}
  \Gamma_{t}^{\LO} &= 1.45258 \;\GeV\,, &  \Gamma_{t}^{\NLO} &= 1.32733 \;\GeV\,.
\end{align}
Here we also state the LO top width as required 
for the inverse-width expansion~\refeq{eq:GEXPu}--\refeq{eq:GEXPv}.
To compute the NLO QCD top-quark widths
we employ a numerical routine of the \MCFM implementation of~\citere{Campbell:2012uf}.

All numerical studies are performed for LHC collisions at 13\,TeV
using the acceptance cuts described in~\refses{sec:origVsMei}{sec:lhe}.
As PDFs we employ the
five-flavour NNPDF~3.1
set with $\as = 0.118$~\cite{NNPDF:2017mvq},
as implemented in the LHAPDF6 library~\cite{Buckley:2014ana}
with LHAPDF \mbox{id\,=\,303400}.
The usage of five-flavour PDFs is motivated by the fact that the typical
scales in $t\bar t$ production are far above the bottom mass.
However, this choice is not consistent with the fact that
the partonic cross sections are evaluated by treating $b$ quarks as
massive and excluding them from the initial state.
This different treatment of $b$ quarks in the PDFs and in the 
perturbative calculations can be easily compensated by appropriate $\ord(\as)$
matching factors~\cite{Cacciari:1998it}.
At the level of the $\bar B(\Phi_\rB)$ weights for the
$q\bar q$ and $gg$ channels, these matching factors can be written as
\begin{eqnarray}
\label{eq:45Fmatcha}
\bar B_{q\bar q}(\Phi_\rB) 
& \rightarrow & 
\bar B_{q\bar q}(\Phi_\rB) - 
\frac{4}{3} T_F \frac{\as}{2 \pi}
\ln\left( \frac{Q_\rR^2}{m_b^2} \right)\,
B_{q\bar q}(\Phi_\rB)\,,
\\
\label{eq:45Fmatchb}
\bar B_{gg}(\Phi_\rB) 
& \rightarrow & 
\bar B_{gg}(\Phi_\rB) -
\frac{4}{3} T_F \frac{\as}{2 \pi}
\left[\ln\left(\frac{Q_\rR^2}{m^2_b}\right)
+ \ln\left(\frac{m_b^2}{\mu^2_\rF} \right)\right] 
B_{gg}(\Phi_\rB)\,,  
\end{eqnarray}
where $\as$ is is the five-flavour strong coupling.
Here the logarithm of ${m_b^2}/{\mu^2_\rF}$
cancels the $\ord(\as)$ contribution of 
$b$-quark loops to the evolution of the five-flavour 
gluon density, while the 
logarithms of $Q_\rR^2/m_b^2$
cancel $b$-quark loop contributions to the running of 
$\as$ from $m_b$ to the scale $Q_\rR$.
In the case of a conventional 4FNS calculation, where 
$b$-quark loops are exlcuded or renormalised in the decoupling scheme,
one should set $Q_\rR=\mu_\rR$. However, in our case bottom loops are included 
in the matrix elements and handled as active contributions 
to the runnig of $\as$. Thus in \refeqs{eq:45Fmatcha}{eq:45Fmatchb}
we set $Q_\rR=m_b$. 

We note that in~\citere{Jezo:2016ujg} $Q_\rR$ was set equal to $\mu_\rR$ due
to the erroneous assumption of a decoupling of 
$b$-quark loops in the matrix elements.
The effect of replacing $Q_\rR=\mu_\rR$ by the correct setting 
$Q_\rR=m_b$ amounts to a shift of about $-6\%$ in the 
\bbfourl{} inclusive cross section.
However it turns out that, due to an accidental cancellation,
this shift is largely compensated by 
the inverse-width expansion~\refeq{eq:GEXPu}--\refeq{eq:GEXPv}
introduced in this paper.
Nevertheless, the correct implementation of the
matching factors~\refeqs{eq:45Fmatcha}{eq:45Fmatchb} 
and the inverse-width expansion are crucial 
for the consistency of the \bbfourl{}
cross section. 
In particular,
we note that the above mentioned accidental cancellation
can be spoiled by a different scale choice 
and/or in differential observables.

For the technical studies presented in this paper, the renormalization and
factorization scales are set to the fixed value\footnote{%
We note in passing
that the results of the original bb4l generator
presented in Ref.~\cite{Jezo:2016ujg} are based on the dynamic scale
\begin{equation} 
\label{eq:ttscale}
  \muR=\muF=\left[\(M_t^2+p_{{T},t}^2\)\(M_{\bar t}^2+p_{{T},{\bar
t}}^2\)\right]^{\frac{1}{4}}\;,
\end{equation}
where the (anti)top invariant masses and transverse momenta 
are defined in the 
underlying Born phase space
based on the particle identities and the full four-momenta of
the six (off-shell) decay products of the $t\bar t$ system
(alternatively, this scale choice can be applied at the level of 
physical momenta in the Born and real-emission phase spaces).
While this dynamic scale is not used in this paper, it is still 
available as default scale choice in the
improved version of the {\tt bb4l} generator. 
We also note that, in the new version of {\tt bb4l},
 the availability of resonance histories of 
$t\bar t$ and $\tW$ type makes it possible to use 
different scale choices for events of
$t\bar t$ and $\tW$ kind.
}
\begin{equation}
\label{eq:mtscale}
\muR = \muF = m_t\,.
\end{equation}
Conventional variations of the QCD scales and PDFs are not considered since
the focus of this paper is on 
the treatment of NLO radiation and its matching 
to parton showers in the presence of 
off-shell effects.

For the \POWHEGBOX{} parameter {\tt hdamp}, which defines the region of 
phase space where NLO radiation is resummed in the 
\POWHEG{} method~\cite{Alioli:2008tz}, 
we set
\begin{equation}
h_{\mathrm{damp}}\,=\, m_t\,. \nonumber
\end{equation}
This setting yields a transverse-momentum
distribution of the top pair that is more consistent with data at large transverse momenta. 
In~\refses{sec:origVsMei}{sec:nlops}
we always use 
the {\tt allrad} feature described in Sect.~\ref{sec:resmethod},
and the inverse-width expansion~\refeq{eq:GEXPu}--\refeq{eq:GEXPv} 
is applied throughout.
Regarding the treatment of resonance histories,
in \refse{sec:origVsMei} we compare results obtained with the original 
histories described in Sect.~\ref{se:bb4loriginal} 
and the matrix-element--based histories defined
in~\refeq{eq:MEresB}, while the latter are used throughout
in \refses{sec:lhe}{sec:nlops}.

All events are showered by \Pythia8.245 with the ATLAS A14 tune. For 
a consistent matching of radiation in top decays, 
we enable both the {\tt PowhegHooks.h} and {\tt PowhegHooksBB4L.h} hooks 
described in Sect.~\ref{se:technicalimp}.
For simplicity we switch off QED emissions, hadronisation, as well as multiparticle interactions.
Using the default procedure to unweight events before showering, we found a
significant fraction of events in which the {\tt btilde} upper bound was
violated, potentially leading to  unphysical distortions
of some of the kinematic spectra of the final-state particles 
To avoid this problem, we have enabled the {\tt ubexcess\_correct} feature,%
\footnote{The corresponding correction factors for {\tt btilde} and {\tt remnant} events
are: {\tt
ub\_btilde\_corr = 0.9983} and {\tt ub\_remn\_corr = 0.9969}.}
which ensures a fully consistent unweighting procedure.

All the event selections in this paper were implemented and
all the plots obtained using {\tt Rivet}~\cite{Buckley:2010ar}.

\section{Effects of resonance-history separation \label{sec:origVsMei}}

In the \POWHEGRES method,  the
consistent generation of QCD radiation in the presence of resonances is
guaranteed, as discussed in \refse{sec:PWGmethod},
by means of a splitting into contributions that are
associated with different resonance histories.
In this section, focussing on the
\bbfourl generator for the 
dileptonic process $pp \to e^+ \nu_e \mu^- \bar{\nu}_\mu b \bar{b}$, 
we compare predictions based on the
original definition of resonance histories (OrigH)
to the improved resonance histories based on
matrix elements (MeH).
As dicussed in \refse{se:bb4lgen},
the OrigH and MeH history definitions
are characterised by different probability distributions for the 
individual resonance histories, as well as 
different lists of resonance histories.
In particular, the MeH resonance histories 
include new histories
corresponding to $\tW$ production and decay.
The difference between results based on the MeH and OrigH resonance
histories can be regarded as an intrinsic 
uncertainty of the \POWHEGRES method, which 
is due to the ambiguity in the definition of resonance histories 
in the off-shell regions of phase space.
As we will see, this uncertainty turns out to be very small. Thus,
the observed agreement can also be regarded as 
a validation of our implementation of the new MeH 
resonance histories.

\subsection{Physics objects and event selection\label{sec:LLanalysis}}
Since our main focus is on technical aspects of the
\POWHEGRES method, in the following comparison we use physics objects
and selection cuts that are defined at the level of 
Monte Carlo~(MC) truth.
This makes it possible to identify the four-momenta of all partons, 
including the two neutrinos, and to reconstruct top resonances unambiguously.

\noindent{\it Physics objects} --- Before applying any cuts we define the four-momenta and the flavour of leptons, 
neutrinos and jets as follows.

\begin{itemize}
\item  Both for leptons and neutrinos we use the full information that is 
available at MC-truth level: we identify them according to their charge and
flavour, and we use their exact four-momenta at parton level.
Since QED radiation is switched off in \Pythia{} there is no need to recombine
collinear photon radiation off charged leptons.

\item Jets are built using the anti-$k_\rT$ algorithm~\cite{Cacciari:2008gp}
with $R=0.5$ and are
tagged according to their flavour content at MC-truth level.
Jets are categorised into $B$~jets and light jets depending on the presence
of $b$~quarks among their constituents. More precisely, 
jets containing at least one $b$ or $\bar{b}$~quark are classified as 
$B$~jets 
and labelled as $j_B$.
Jets of type $j_B$ are additionally labelled as $j_b$ and/or $j_{\bar b}$ 
if they contain at least one $b$ and/or $\bar b$~quark.
\end{itemize}

\noindent{\it Dilepton\,+\,$B$-jet selection (\twoLB)} --- We consider events that fulfill
the following acceptance cuts
\begin{itemize}
\item Two charged leptons with $p_\rT>25\,\GeV$ and $\vert\eta\vert<2.5$.

\item Two neutrinos with a combined transverse energy, 
$|\vec p_{\rT,\nu\bar\nu}|$,
larger than 25 GeV.

\item At least one $B$~jet with $p_\rT>25\, \GeV$ and $\vert\eta\vert<2.5$.

\end{itemize}
This event selection will in the remainder of the manuscript be labelled as
\twoLB~selection.

\noindent{\it Off-shell cut} --- In order to investigate off-shell effects
 we also define the following (optional) extra cut, which forces 
one of the two top quarks to be off-shell,
\newcommand{\Qoff}{Q_{\mathrm{off{\scriptscriptstyle-}shell}}}
\begin{equation}
\label{eq:offshellcut}
\Qoff = \max\Big\{|Q_t-m_t|,|Q_{\bar t}-m_t|\Big\} > 60\,\GeV\,.
\end{equation}
Here $Q_t$ and $Q_{\bar t}$ denote, respectively, the 
invariant masses of the reconstructed top
and anti-top quarks.
The top and anti-top quarks are reconstructed as follows.  We first assemble a
collection of all light jets, also outside of acceptance.  Then, for the top
quark, we start with the $e^+ \nu_e j_b$ system and add up to one light jet
from the collection to it if it brings its virtuality, $Q_t$, closer to $m_t$.
The anti-top quarks are reconstructed analogously but starting from the $\mu^-
\bar{\nu}_\mu j_{\bar{b}}$ system.
This reconstruction is done simultaneously for top and anti-top, minimising
$Q_t+Q_{\bar t} - 2 m_t$, and each light jet from the collection is only used once. 
If there is only one light jet it is only added once either to the top or the
anti-top systems.
The definition of $\Qoff$ involves two $B$~jets, one of which 
is selected by the \twoLB~cuts, while the other one is used to determine 
$\Qoff$ even if it lies outside the acceptance region.
When imposing the cut~\refeq{eq:offshellcut}
we only require the presence of at least two distinct  
$B$~jets of type $j_b$ and $j_{\bar{b}}$, one of which must lie 
within the acceptance cuts.
This event selection is not realistic and only serves the purpose of 
comparing our predictions in a region with increased fraction of 
$\tW$ resonance histories.

\subsection{Effect of matrix-element--based vs original resonance 
histories}

In the following we compare 
predictions for 
$pp \to e^+ \nu_e \mu^- \bar{\nu}_\mu b \bar{b}$
obtained with the 
\bbfourldl{} generator using the original 
resonance-history projectors (OrigH) described in 
\refse{se:bb4loriginal}
and, alternatively,  the matrix-element--improved 
histories (MeH) introduced in 
\refse{se:newreshist} and defined in~\refeq{eq:MEresB}.
Since we always employ the \allrad mode, which 
allows for up to one NLO emission in each production and decay 
subprocesses, when using OrigH histories 
nearly all events,\footnote{Except for events
with $Zb\bar{b}$ resonance history, which appear at a negligible rate.}
involve the emission of up to three \POWHEG{} partons,
two of which are gluons radiated by the two final-state 
$b$~quarks.
In contrast, in the case of MeH histories
only events of $t\bar t$ kind can give rise to three 
\POWHEG{} emissions, while events of $\tW$ kind radiate at
most two \POWHEG{} partons. In the latter case only the 
$b$~quark that originates from the top decay is 
guaranteed to emit \POWHEG{} radiation, while 
the first emission stemming from the other $b$~quark 
is typically generated at the stage of parton showering.
For this reason, the most significant differences between OrigH
and MeH histories are expected to originate from events of $\tW$ kind and
in observables that are sensitive to QCD radiation emitted by $b$~quarks.
In particular, significant differences may show up in 
observables that depend, either directly or through the acceptance 
cuts, on the $b$-jet momenta.
Such differences are mostly expected 
at the LHE level, where the number of \POWHEG{} emissions depends on the
history, while in complete NLOPS simulations, where 
the different number of \POWHEG{} emissions is 
compensated by the parton shower, possible 
differences between OrigH and MeH histories are expected to be mitigated.
Note also that the choice of resonance histories does not affect
observables that are insensitive to the kinematic distribution of QCD
radiation, and for those we anticipate
identical results when using OrigH or MeH histories.

In Tab.~\ref{tab:reshistXsec} and Figs.~\ref{fig:reshist}--\ref{fig:reshist_tw}
we compare predictions at the LHE and NLOPS levels 
with the setup of \refse{sec:setup}
and the cuts of~\refse{sec:LLanalysis}.
\begin{table}
  \centering
  \begin{tabular}{|l|l||c|c|c|}
    \hline
&&  inclusive &  \twoLB & \twoLB\,+\,off-shell \\
&&  phase space & cuts  & cuts\\
    \hline\hline
    LHE & OrigH  & 9.672(4) & {4.422(3)} & {0.1908(6)} \\
    LHE & MeH    & 9.653(3) & {4.411(2)} & {0.1912(4)} \\
    LHE &  $\tW$ fraction    & {4.31\%}  & {3.86\%} & {43.0\%} \\\hline
    NLOPS & OrigH& 9.672(4) & {4.419(3)} & {0.3515(8)} \\
    NLOPS & MeH  & 9.653(3) & {4.408(2)} & {0.3502(5)} \\
    NLOPS &  $\tW$ fraction    & {4.31\%}  & {3.86\%} & {23.3\%} \\
    \hline
  \end{tabular}
  \caption{Cross sections in picobarn for the inclusive phase space
  (third column), in the presence of dilepton+$B$-jet (\twoLB) cuts (fourth column), 
and with \twoLB~cuts in combination with the
off-shell cut $\Qoff>60\,\GeV$ (fifth column).
Predictions based on the 
original (OrigH) and matrix-element--based (MeH) resonance histories
are compared at the LHE and NLOPS levels.
The reported $\tW$ fractions are determined using the 
MeH resonance histories and correspond to the 
fraction of events with $W^-t$ or $W^+\bar t$ history in the respective
selections.
\label{tab:reshistXsec}}
\end{table}
In Tab.~\ref{tab:reshistXsec} 
we present integrated cross sections in three different phase spaces.
In the inclusive phase space (without cuts)
we find, as expected, an excellent agreement for 
LHE and NLOPS predictions based on the two different resonance histories.
This is a mandatory consistency check, which confirms that
resonance-history projectors as well as the operation of parton showering
are unitary.
In the presence of dilepton+$B$-jet cuts (\twoLB), 
due to the sensitivity of the $b$-jet cut to 
the radiation emitted by the parton shower,
the LHE and NLOPS cross sections are no longer
expected to be identical. However  
the different behaviour 
of QCD radiation induced by the OrigH and MeH resonance histories
gives rise to only sub-permil differences in the \twoLB~cross section.
Finally, we find that the additional off-shell cut, $\Qoff>60\,\GeV$, 
selects only about 8\% (4\%) of the \twoLB~cross section at the NLOPS (LHE) level, while increasing 
the fraction of events with $\tW$ resonance histories 
from 4.3\% to about 
23\% (43\%)
at NLOPS (LHE) level.
In this off-shell region, 
comparing LHE and NLOPS results 
we observe that the parton shower 
leads to a 84\%
enhancement of the cross section.
This can be attributed to the fact that
LHE events with $\Qoff<60\,\GeV$ migrate to the
$\Qoff>60\,\GeV$ region
as a result of shower induced $b$-jet fragmentation, \ie via 
radiation of a light jet that is emitted 
by the decay products of a top quark but is 
not included in the corresponding 
reconstructed invariant mass,
or $b$-jet contamination.
Given the strong sensitivity to QCD radiation and the
high fraction of $\tW$ events, in the off-shell \twoLB~region 
we may also expect
an enhanced sensitivity 
to the choice of resonance histories. Indeed, we observe a
difference between OrigH and MeH histories, 
which remains  however at the level of few permil.

In Figs.~\ref{fig:reshist}--\ref{fig:reshist_tw}
\newcommand{\reshistplot}[1]{
  \begin{minipage}{0.49\textwidth}
    \begin{center}
      \includegraphics[width=\textwidth,trim={0 27 0  0},clip]{figures/reshist/main/#1}\\
      \includegraphics[width=\textwidth,trim={0 27 0 19},clip]{figures/reshist/nlopsOverLhe/#1}\\
      \includegraphics[width=\textwidth,trim={0 27 0 19},clip]{figures/reshist/meiOverOrig/#1}\\
      \includegraphics[width=\textwidth,trim={0  0 0 19},clip]{figures/reshist/RHtW/#1}
    \end{center}
  \end{minipage}
}
\begin{figure}
  \reshistplot{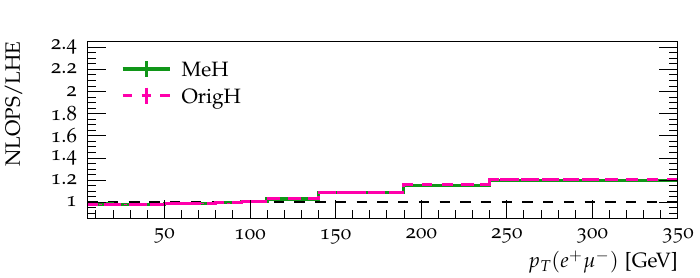}
  \reshistplot{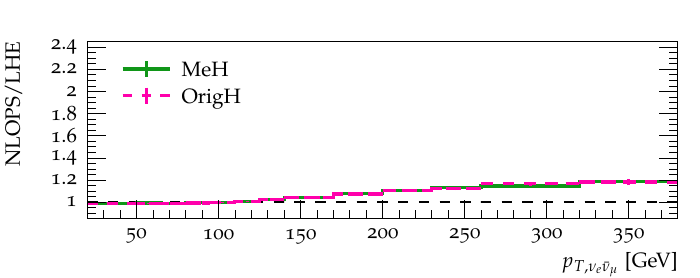} \\
  \reshistplot{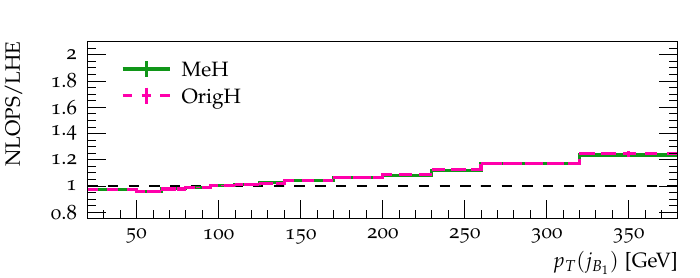}
  \reshistplot{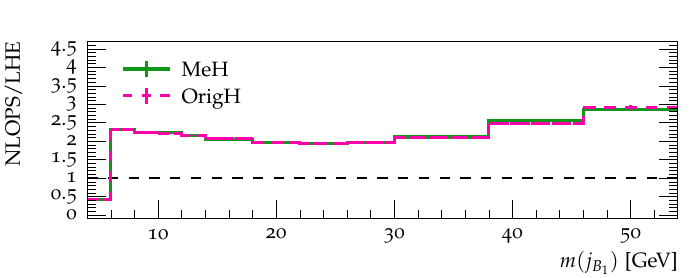}
  \caption{Differential distributions for 
$pp \to e^+ \nu_e \mu^- \bar{\nu}_\mu b \bar{b}$
with dilepton+$B$-jet cuts (\twoLB): comparison of LHE and NLOPS predictions
with original (OrigH) vs matrix-element--based (MeH)
resonance histories. The lowest frame shows the 
fraction of events of $\tW$ type.
See the main text for more details.
\label{fig:reshist}
}
\end{figure}
\begin{figure}
  \reshistplot{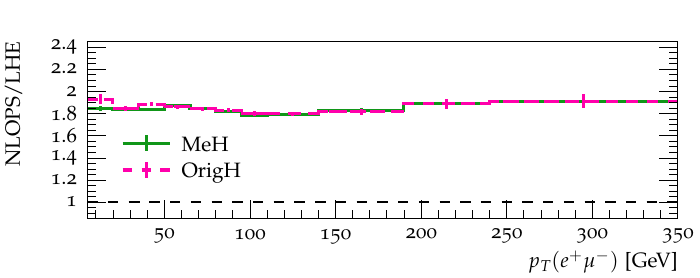}
  \reshistplot{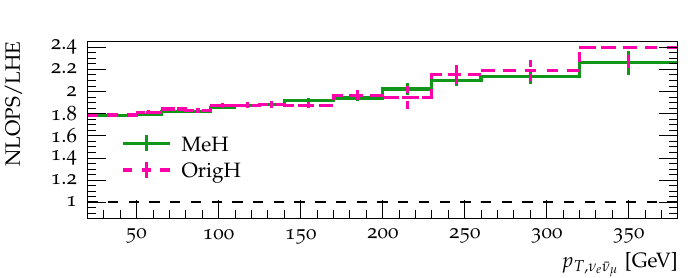}\\
  \reshistplot{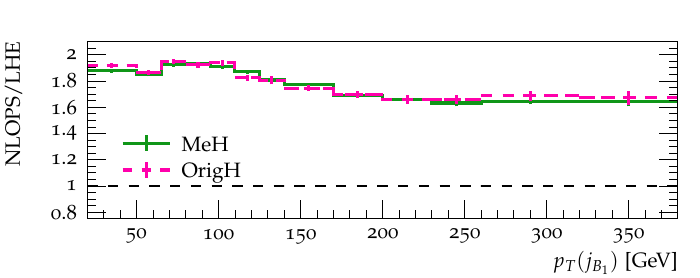}
  \reshistplot{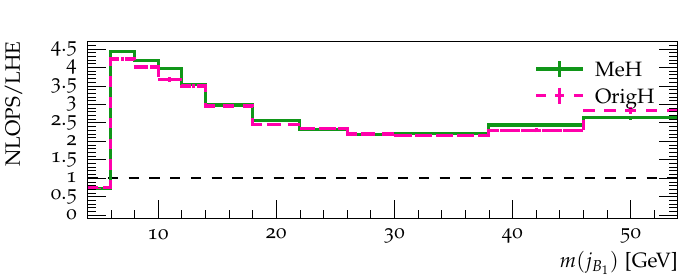}
  \caption{Differential distributions for 
$pp \to e^+ \nu_e \mu^- \bar{\nu}_\mu b \bar{b}$
with dilepton+$B$-jet cuts (\twoLB) and with the additional off-shell cut 
$\Qoff>60\,\GeV$.
Same observables and predictions as in \reffi{fig:reshist}.
\label{fig:reshist_tw}
}
\end{figure}
we 
investigate the sensitivity of differential distributions
to the different definitions of resonance histories.
In particular, we consider the distributions in the
transverse momentum of the lepton pair and 
in the missing $p_\rT$,
which feature a significant sensitivity to 
off-shell effects~\cite{Denner:2016jyo} and may thus 
be sensitive to the modification of 
resonance histories.
Because the prescription for resonance-history separation directly affects
the emission of POWHEG radiation off final-state $b$~partons,
depending on whether or not they originate from top decays, we also examine
the transverse-momentum and invariant-mass spectra of the hardest $b$~jet.

In \reffi{fig:reshist}
we compare LHE and NLOPS predictions with OrigH and MeH 
resonance histories in the presence of \twoLB~cuts.
The upper and the lower panels display 
absolute predictions and various ratios, respectively.
The NLOPS/LHE ratio, shown in the second panel, is 
independent of the choice of the resonance histories,
and in the dominant regions of phase space this ratio 
is close to one, as expected from the sub-percent NLOPS/LHE agreement 
of the integrated cross sections in Tab.~\ref{tab:reshistXsec}.
In the tails of the $p_\rT$ distributions, shower effects are more important
and can reach $+20\%$.
In the invariant mass of the leading $b$~jet the NLOPS/LHE ratio is 
around 250\%. This can be interpreted as a moderate invariant-mass shift of
order 5--10\,GeV, which is induced by shower radiation
and translates into a large NLOPS/LHE 
ratio due to the steepness of the 
distribution in $m_{j_{B_1}}$.

The ratio of distributions based on MeH vs OrigH resonance histories, 
shown in the third panel of \reffi{fig:reshist}, 
features an exceptional level of agreement, \ie very small sensitivity to the 
choice of resonance histories.
This holds both at the LHE and NLOPS level, and 
for all observables, even in the regions where 
shower effects are sizeable.
For each observable, 
in the fourth panel we show the fraction of events of $\tW$ 
type.
This fraction is determined within the MeH approach
as the ratio of events with $tW^-$ or $\bar tW^+$ 
resonance histories with respect to the total.
Its kinematic distribution in \reffi{fig:reshist}
shows that, in the \twoLB~phase space, events of $\tW$ type are most abundant 
in the region where the leading $b$~jet is soft.

In \reffi{fig:reshist_tw} we present the same distributions applying
\twoLB~cuts and the additional
cut $\Qoff>60\,\GeV$. As a result of this off-shell cut, 
the fraction of $\tW$ events increases by about a factor eleven~(six)
and  LHE~(NLOPS) level,
while its kinematic dependence is qualitatively similar as in the 
full \twoLB~phase space.
The off-shell cut enhances also shower effects,
and the NLOPS/LHE ratio can reach a factor two in the tails of the 
$p_\rT$ distributions. As discussed above, this enhancement can be
attributed to a shower-induced  migration of events across the off-shell cut
at $\Qoff=60\,\GeV$. 
This interpretation is supported by the fact
that the NLOPS/LHE ratio grows with the $p_\rT$ of the leading $b$~jet,
and in the 
$p_{\rT,j_{B_1}}$ 
tail the ratio of $t\bar t$ over $\tW$ events  
almost doubles when the parton shower is switched on.

For what concerns the effect of resonance histories, we observe that the
ratio of MeH/OrigH distributions is mostly consistent with one 
within statistics.  The only
significant exception is observed 
in the small invariant-mass region 
of the $m_{j_{B_1}}$ distribution, where MeH based predictions are up to 7\% below 
OrigH based ones. 
This effect can be attributed to the different OrigH/MeH
treatment of \POWHEG{} radiation in events of $\tW$ kind.
As discussed above, this OrigH/MeH difference is expected to 
be mitigated by the parton shower, which is indeed what we observe at
NLOPS level.
This interpretation is confirmed by the
fact that the effect at hand shows up in the $m_{j_{B_1}}$
region with the highest fraction of $\tW$ events
(see the fourth panel).
As expected, the observed differences at LHE level are strongly mitigated 
when the parton shower is switched on.
Indeed, at NLOPS level no statistically significant MeH/OrigH difference is
observed.

In general, the high level of agreement between MeH vs OrigH 
resonance histories 
demonstrates that the ambiguity 
associated with the choice of resonance-history projectors
represents a very small source of
uncertainty.
This finding inspires further confidence 
in the \POWHEGRES method. 
Moreover, the observed agreement 
can be regarded as a
validation of our implementation of the new matrix-element 
based separation of resonance histories.

\section{Comparison of dileptonic and semileptonic channels \label{sec:lhe}}
In this section we discuss predictions of the new \bbfourlsl{} generator
for the semileptonic (SL) process 
\begin{eqnarray}
\label{eq:SLprocess}
pp&\to& e^+ \nu_e d {\bar{u}} b \bar{b}\,,
\end{eqnarray}
and compare them to predictions of the \bbfourldl{} generator for 
the related dilepton (DL) process 
\begin{eqnarray}
\label{eq:DLprocess}
pp&\to& e^+ \nu_e \mu^- \bar{\nu}_\mu b \bar{b}\,,
\end{eqnarray}
using the new matrix-element--based resonance histories throughout.
The above processes are both dominated by a common
$pp\to W^+W^- b\bar b$
subprocess with a $W^+\,\to\, e^+ \nu_e$ leptonic decay,\footnote{$W$-boson
decays should always be understood as off-shell decays.}
and the only difference lies in the hadronic or leptonic nature 
of the $W^-$ decay, 
\begin{equation}
\label{eq:antitopdecays}
W^-\,\to\, \mu^- \bar{\nu}_\mu
\qquad\mbox{or}\qquad
W^-\,\to\, d\bar{u}\,.
\end{equation}
More precisely, upon identification of the leptonic and 
hadronic $W^-$ decay products,
the two processes are perfectly equivalent at Born level, while 
all differences arise from the breaking of the
trivial correspondence between hadronic and leptonic 
$W^-$ decays, which results from QCD radiation effects
in hadronic $W$ decays. 
Thus, depending on whether a given observable is
weakly or strongly sensitive to such QCD effects,
the comparison of \bbfourlsl{} and \bbfourldl{}
is expected to yield, respectively, 
good agreement or sizeable differences.
The former case provides a check of the mutual consistency of the two
generators, while the latter case can be exploited to gain
instructive insights into the 
origin and the manifestation of QCD radiation 
effects in various observables.
With these two objectives in mind, 
in the following we define physics objects and observables 
in a way that enables switching between the cases of 
weak and strong sensitivity
to QCD 
effects that originate from the fragmentation of jets 
and/or form their contamination through other 
sources of  QCD radiation.

\subsection[Physics objects, $W$ reconstruction and event selection]{Physics objects, $\boldsymbol{W}$ reconstruction and event selection}
\label{se:SLDLsetup}

In this section we 
define selection cuts and observables that
are designed such as to enable a consistent comparison of the SL and DL
processes~\refeq{eq:SLprocess}--\refeq{eq:DLprocess}.
In particular, we define optimised $W$-reconstruction
procedures that are 
based on MC truth and 
maximise the correspondence
between hadronic and leptonic $W$-boson decays
by minimising the sensitivity of the former to QCD radiation effects.
Such optimised $W$-boson reconstructions will be enabled and disabled
in order to 
maximise the consistency of 
\bbfourlsl{} and \bbfourldl{} predictions for validation purposes
and, alternatively, to 
gain insights into 
the origin and behaviour 
of QCD radiation effects.

\subsubsection*{Physics objects and raw reconstruction of \boldmath$W$ bosons}
\label{se:SLDLres}

Light jets are defined with the
anti-$k_\rT$ algorithm with $R=0.5$. 
The leptons, neutrinos and $B$ jets are
defined as in Sect.~\ref{sec:LLanalysis}, and in this Section 
the labels $j_b$ and $j_{\bar b}$ are used for the 
leading $b$ and $\bar b$ jets.

Leptonic and hadronic $W$-boson decays are reconstructed as follows.
In the DL and SL channels,  the leptonic 
$W^+\to e^+\nu_e$  and $W^-\to \mu^-\bar \nu_\mu$ decays are reconstructed based on the
particle identity and exact four-momenta of their decay products according
to MC truth. 
In the SL channel,
the hadronic $W^-\to d\bar u$ decays
are required to form 
two separate jets, which are labelled as $j_d^W$ and $j_{\bar u}^W$.
The associated $W^-$ bosons are reconstructed by identifying and combining 
such $j_d^W$ and $j_{\bar u}^W$ jets at MC truth level. 
In the following, this procedure will be referred to as 
raw $W$ reconstruction.

\subsubsection*{Recombination and decontamination of hadronic \boldmath$W$ bosons}

At the Born level, the reconstructed hadronic $W^-$ boson in the
SL channel is equivalent to the leptonic $W^-$ boson in the SL channel.
However, this simple correspondence can be largely obscured by 
the effect of QCD radiation and jet clustering.
In particular, the hadronic $W^-$ mass peak is expected to be smeared as compared to the
leptonic one.
This is because the $W$ virtuality can be either enlarged, 
if radiation from
outside of $W^-$ gets clustered together with $j_d^W$ and $j_{\bar u}^W$, 
or reduced,
if the radiation from within $W^-$ escapes the 
cones of $j_d^W$ and $j_{\bar u}^W$. 
In order to restore a high degree of similarity between $W^-$
bosons of leptonic and hadronic type, in the following we introduce a
recombination 
and decontamination 
procedure that minimises the effects of
QCD radiation and jet clustering in the reconstruction of hadronic $W^-$
bosons.

\noindent{\it Recombination of hadronic $W^-$ bosons} --- 
The fragmentation of $j_d^W$ and $j_{\bar u}^W$ jets 
can lead to a significant reduction of the invariant mass
of the reconstructed $W^-$ boson. To avoid this effect, 
based on MC truth we add to $j_d^W$ and $j_{\bar u}^W$ 
all missing partons that originate
form the $W^-\to d\bar u$ decay via
NLO radiation or parton showering.
Such partons are assigned each to the jet $j_d^W$ or $j_{\bar u}^W$  
with the smallest $\Delta R$ distance.
This $W$ recombination applies also to partons that are clustered into
$b$ jets, \ie together with $W$-recombination we apply a 
$B$-decontamination procedure, which consists of 
removing from $B$ jets all gluons and quarks that
originate from the $W^- \to d\bar u$ decay.
These corrections are expected to suppress 
DL--SL differences in the $W^-$-mass distribution 
below the peak as well as in the efficiency of cuts and in distributions
involving $j_d^W$, $j_{\bar u}^W$ and $B$ jets.

\noindent{\it Decontamination of hadronic $W^-$ bosons} ---
The clustering of QCD radiation originating from other parts of the process into  
$j_d^W$ and $j_{\bar u}^W$  
can significantly increase the invariant mass of the reconstructed 
$W^-$ boson. To avoid this effect, 
based on MC truth we remove from $j_d^W$ and $j_{\bar u}^W$ 
all partons that do not 
originate from the $W^-\to d\bar u$ decay.
This $W$-decontamination is combined with
a $B$-recombination procedure, where
all partons that were removed 
from $j_d^W$ and $j_{\bar u}^W$
are added to the $j_B$ jet that is closest in 
$\Delta R$ if $\Delta R<0.5$.
These corrections are expected to suppress 
DL--SL differences in the $W^-$ mass distribution 
above the peak, 
as well as
in the efficiency of cuts and in distributions
involving $j_d^W$, $j_{\bar u}^W$ and $B$ jets.

The combination of these recombination and decontamination procedures
amounts to an approximate 
reconstruction of the original hard $d$ and $\bar u$ quarks before the
emission of any radiation in the $W^-$ decay.  Thus it is expected to yield
a good correspondence between leptonic and hadronic $W$ decays.
Again, we note that the purpose of these idealised recombination and decontamination procedures
is to investigate 
QCD radiation patterns in a systematic way.

\subsubsection*{Acceptance cuts} 

After the above physics object identification and $W$-reconstruction
procedures, and before imposing acceptance cuts,
we require that the four objects stemming from $W$-boson decays,
\ie $e^+\nu_e$ and $\mu^-\bar \nu_\mu$ or $j^W_dj^W_{\bar u}$,
are pairwise separated by $\Delta R>0.5$.
Furthermore, we identify the leading $j_b$ and $j_{\bar b}$ jets,
which are required to be two individual jets, i.e.~separated by a distance $\Delta
R>0.5$.
In addition, we require that $j_b$ and $j_{\bar b}$ are at a distance
$\Delta R>1$ from the four objects stemming from the decay of the $W^+$ and $W^-$
bosons.

At this stage we count the number $N_B$ of jets of type $j_B$ with
$p_\rT>25\, \GeV$ and $\eta<2.5$, and we split the phase space in 
two complementary regions with 
$N_B=1$ and $N_B\ge 2$. 
Finally, we apply one of the following two selections, which
require the presence within acceptance cuts of three physics objects
corresponding either to a top or an anti-top decay.
\vspace{2mm}

\noindent{\it $W^+ b$ selection} ---
We require that the leading
$j_{b}$ and
the $W^+$ decay products, \ie $e^+$ and  $\nu_e$, fulfill $p_\rT>25\,\GeV$ and $\eta<2.5$.
These three physics objects represent a ``top candidate''. 
\vspace{2mm}

\noindent{\it $W^-\bar b$ selection} --- We require that the 
leading  $j_{\bar b}$ and the $W^-$ decay products, \ie
$\mu^-$ and $\bar \nu_\mu$ 
($j^W_d$ and $j^W_{\bar u}$) in the DL (SL) sample, 
fulfill $p_\rT>25\,\GeV$ and $\eta<2.5$.
These three physics objects represent an ``anti-top candidate''. 
\vspace{2mm}

\subsection{Results}

In the following  we present comparisons  of \bbfourl{} predictions in the
DL and SL channels for a $W^+ b$ and a $W^-\bar b$ selection
in the regions with $N_B=1$ and $N_B\ge 2$.  These comparisons are carried out at the
LHE level, \ie focussing on the effect of the hardest POWHEG emissions.

\begin{figure}
  \begin{center}
  \includegraphics[width=0.48\textwidth]{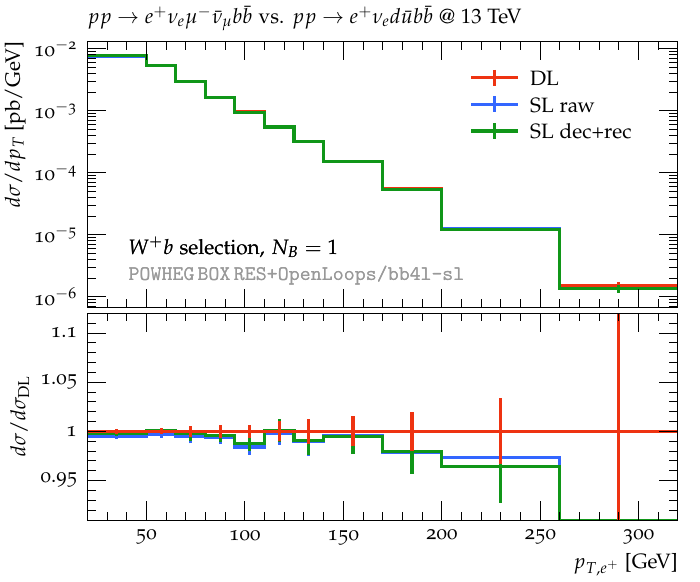}
  \includegraphics[width=0.48\textwidth]{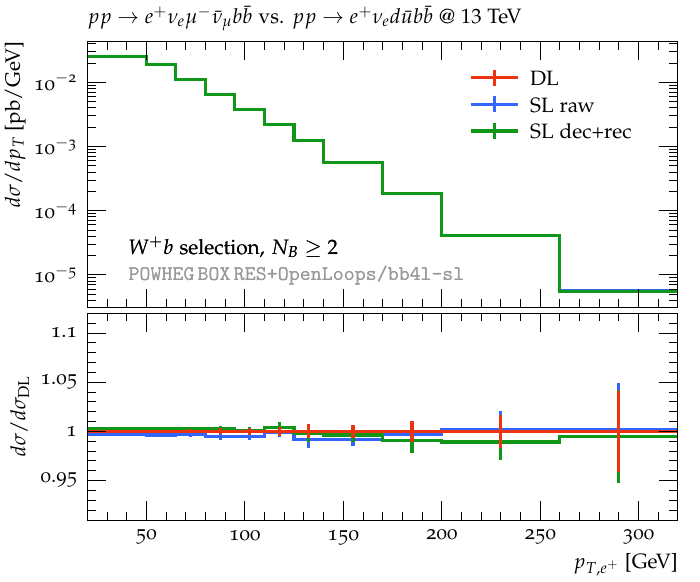}
\\
  \includegraphics[width=0.48\textwidth]{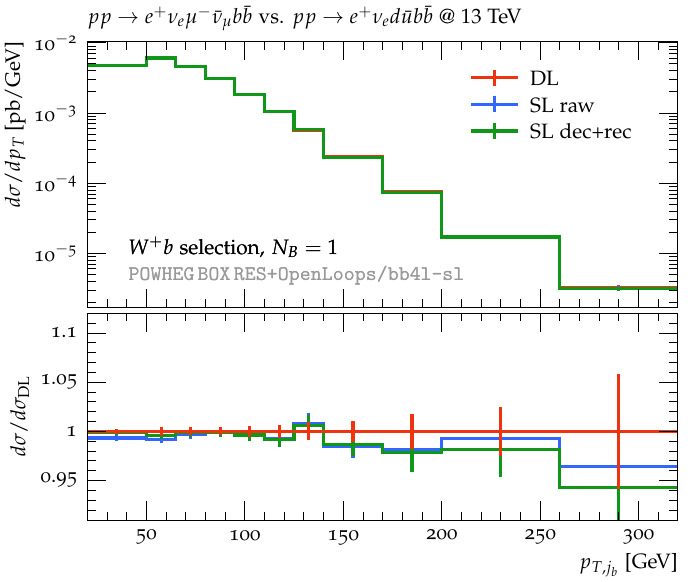} 
  \includegraphics[width=0.48\textwidth]{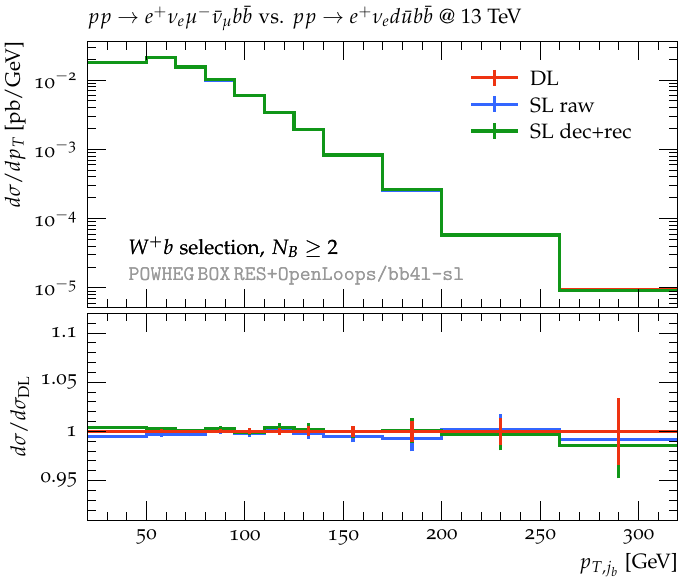}
\\
  \includegraphics[width=0.48\textwidth]{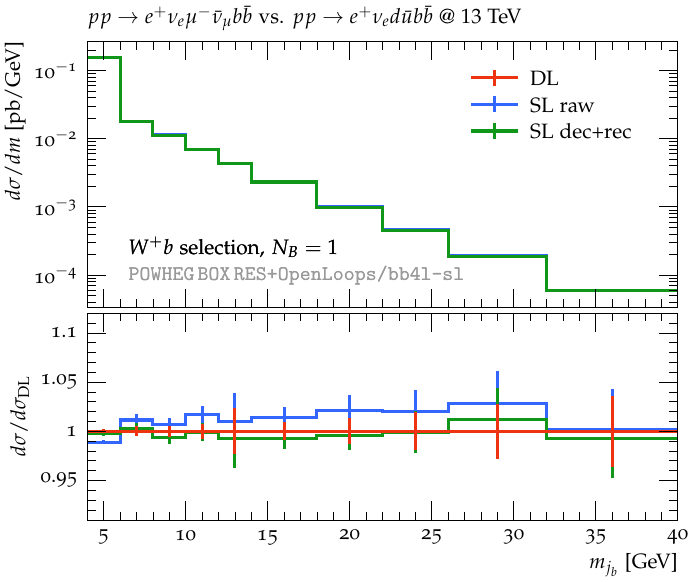} 
  \includegraphics[width=0.48\textwidth]{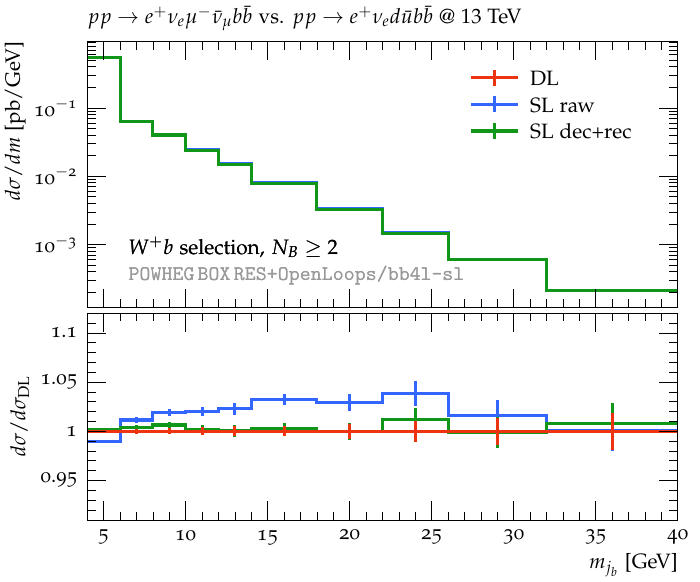} 
  \end{center}
  \caption{Comparison of {\tt bb4l} LHE predictions in the dilepton
channel (DL) and in the single-lepton channel with naive $W$-reconstruction 
(SL raw) or with
$W$-decontamination and -recombination corrections (SL dec+rec).
The distributions in the $p_\rT$ of the positron and in the
$p_\rT$ and the mass of the leading $b$-jet 
are compared in the $W^+$ phase space 
with $N_B=1$ (left column) and $N_B\ge 2$ (right column).  
See more details in the main text.
}
\label{fig:redecay_Wp_1_LHE}
\end{figure}

\begin{figure}
  \begin{center}
  \includegraphics[width=0.48\textwidth]{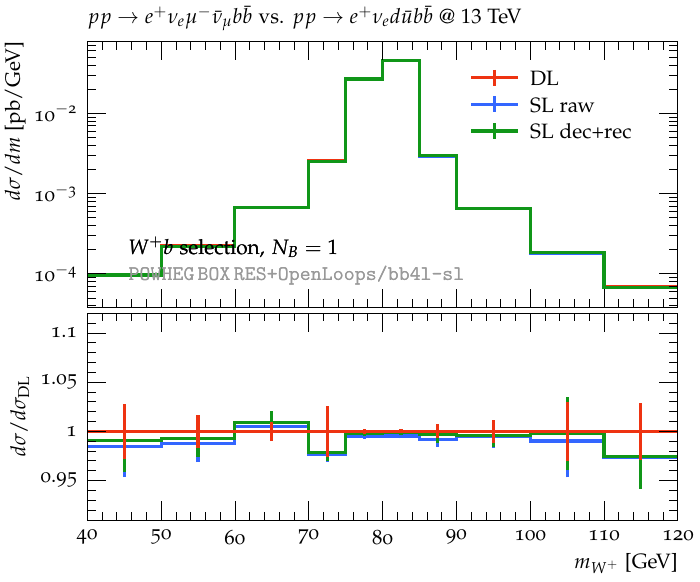} 
  \includegraphics[width=0.48\textwidth]{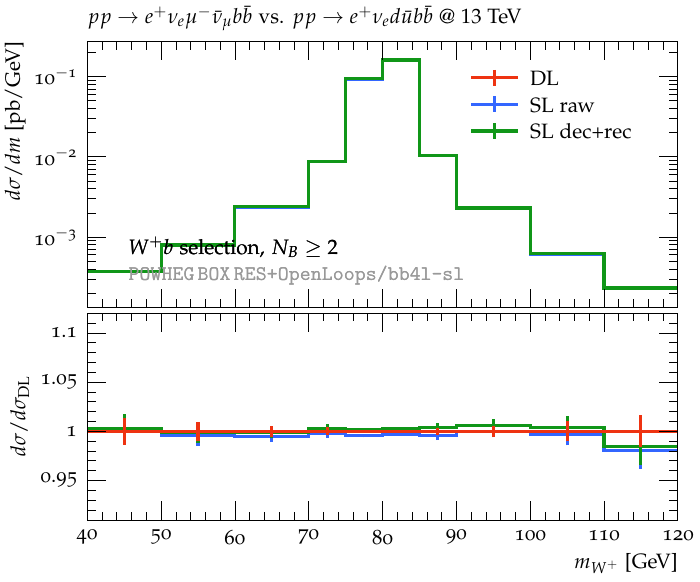}
\\
  \includegraphics[width=0.48\textwidth]{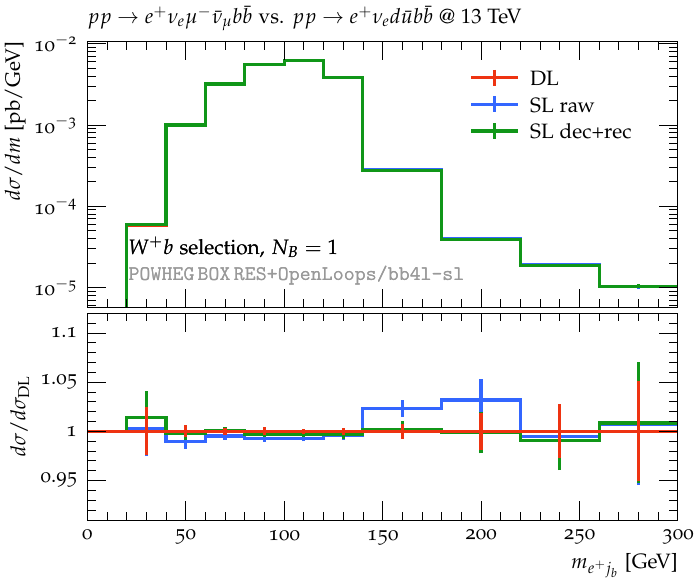}
  \includegraphics[width=0.48\textwidth]{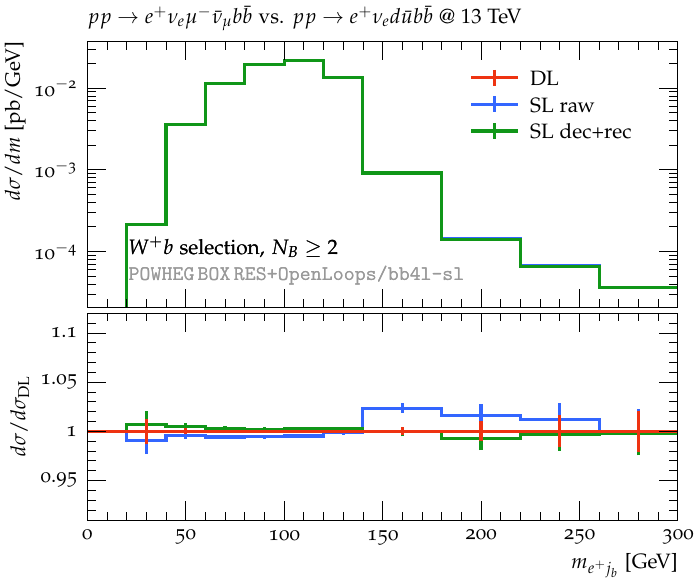}
\\
  \includegraphics[width=0.48\textwidth]{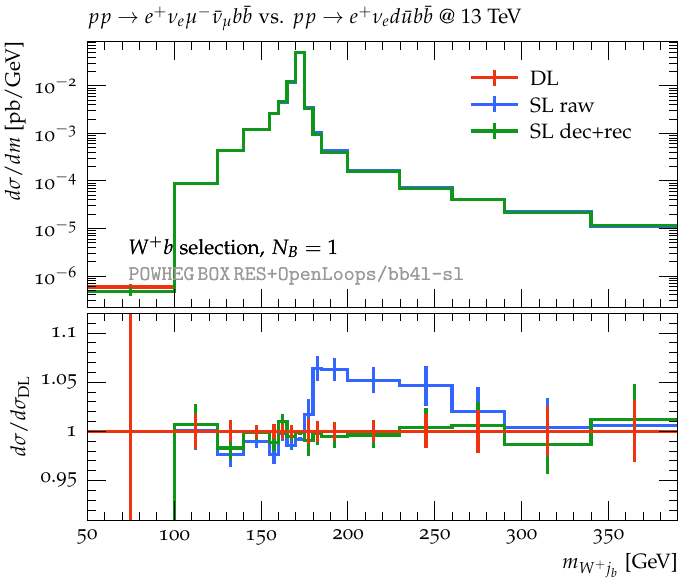}
  \includegraphics[width=0.48\textwidth]{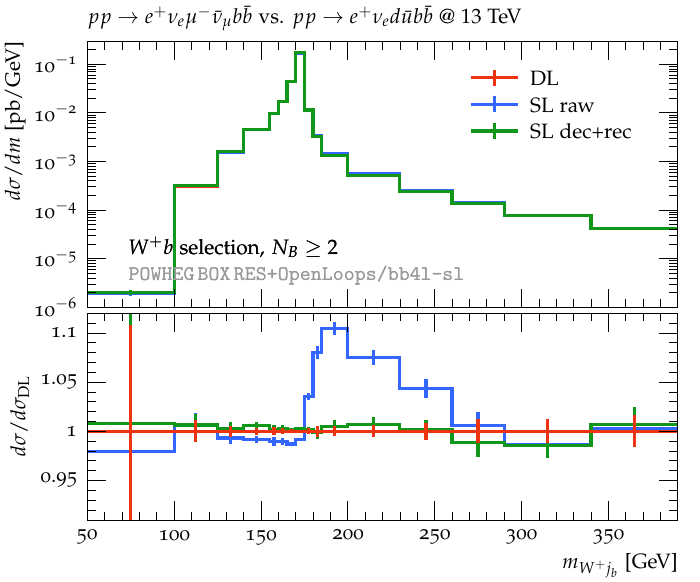}
  \end{center}

  \caption{Same comparison as in \reffi{fig:redecay_Wp_1_LHE} for
the invariant masses of the reconstructed $W^+$ boson, 
\ie $m_{W^+}=m_{e^+\nu_e}$,
and the invariant masses of the $e^+ j_{\bar b}$ and $W^+ j_{\bar b}$ systems.
See more details in the main text.
}
\label{fig:redecay_Wp_2_LHE}
\end{figure}

In Figs.~\ref{fig:redecay_Wp_1_LHE}--\ref{fig:redecay_Wp_2_LHE} we apply 
a $W^+b$ selection and we study various observables that depend on the
constituents of the ``leptonic top candidate''  $e^+ \nu_e j_{b}$. 
In general, since these three objects are treated in the same way in \bbfourldl{}
and \bbfourlsl{}, we expect to observe good agreement. 
In fact, in the first five observables of
Figs.~\ref{fig:redecay_Wp_1_LHE}--\ref{fig:redecay_Wp_2_LHE}
the difference between DL and SL predictions with
raw $W$ reconstruction (SL raw) never exceeds 2--3\%. This holds both in the
$N_B=1$ and $N_B\ge 2$ phase space.
Such small differences are due to the interplay of QCD radiation from 
hadronic $W^-$ decays with the $B$ jets that enter the acceptance cuts or 
the observables at hand. This is confirmed by the fact that 
the maximum DL--SL difference goes down to 1\% 
when the $W$-decontamination and -recombination corrections 
(DL dec+rec) are applied.
This reduction is clearly visible in the $j_b$ mass distribution,
where the 2--3\%  excess in the SL raw prediction 
is due to radiation stemming from the hadronic $W$-boson decay, which 
gets clustered into $j_b$ increasing its invariant mass.
A similar behaviour is observed
in the distribution in $m_{e^+j_b}$.

As for the distributions in $p_{\rT,e^+}$,\, $p_{\rT,j_b}$ and $m_{W^+}$,
we observe that 
SL raw predictions are already in
excellent agreement with DL ones. Applying the dec+rec corrections
this agreement persists.
Finally, in the invariant mass of the reconstructed top quark
in \reffi{fig:redecay_Wp_2_LHE} we see that SL raw predictions
exceed DL ones by up to 
10\% in the resonance
region. This excess reflects a migration of events from the Breit--Wigner peak
towards higher invariant mass, which is  due to 
radiation from the
hadronic $W^-$ decay that gets clustered into $j_b$.
Similarly as for the $m_{j_b}$ distribution, this effect 
disappears when the dec+rec corrections are applied.

\begin{figure}
  \begin{center}
  \includegraphics[width=0.48\textwidth]{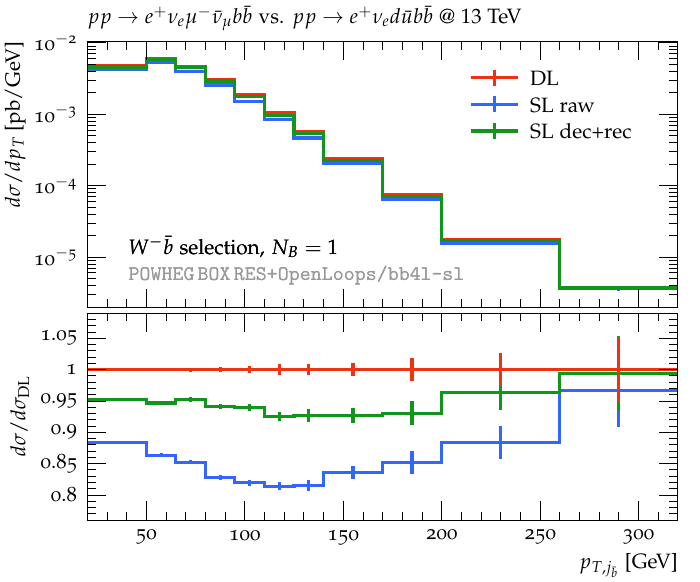} 
  \includegraphics[width=0.48\textwidth]{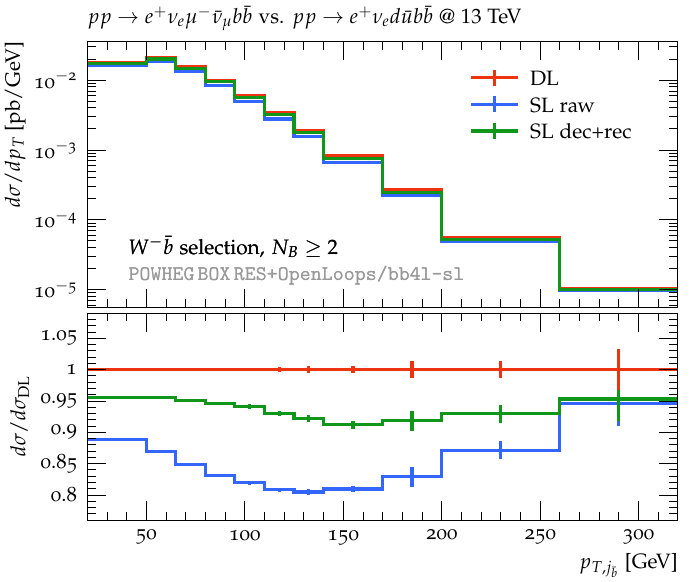}
  \includegraphics[width=0.48\textwidth]{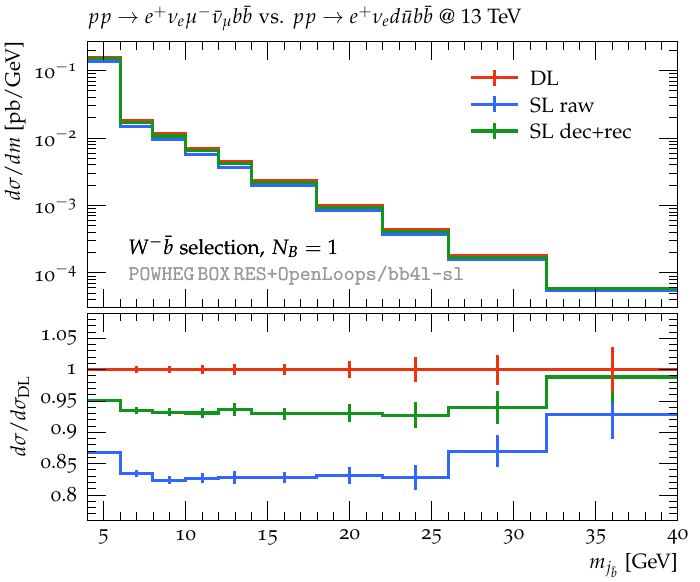} 
  \includegraphics[width=0.48\textwidth]{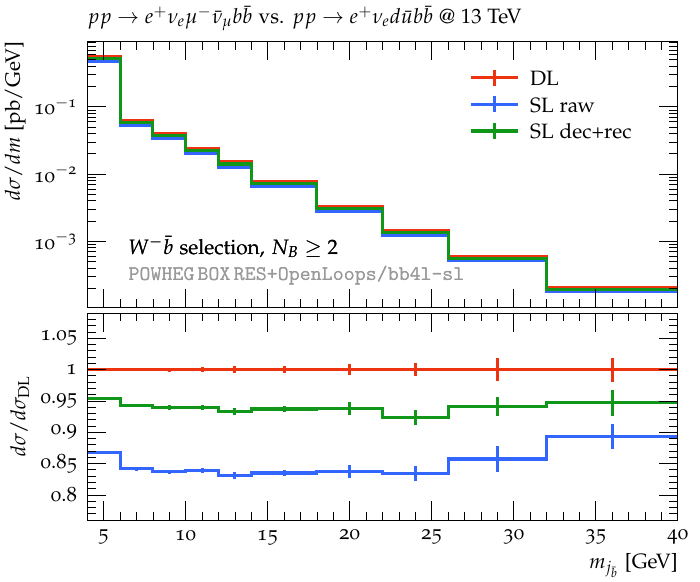} 
\\
  \includegraphics[width=0.48\textwidth]{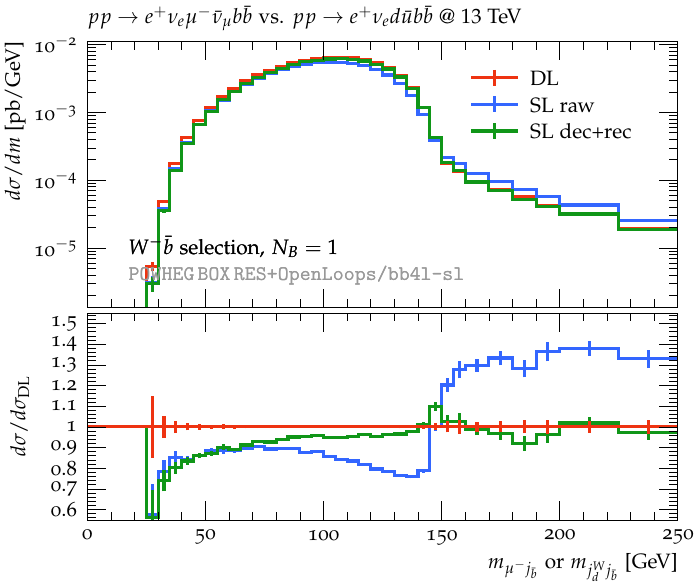}
  \includegraphics[width=0.48\textwidth]{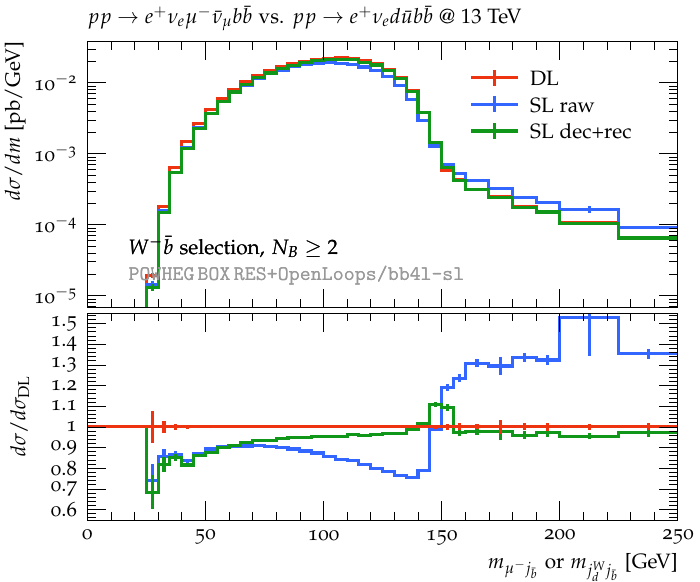}
  \end{center}
  \caption{%
Comparison of {\tt bb4l} LHE predictions in the the dilepton
channel (DL) and in the single-lepton channel with raw $W$-reconstruction  (SL raw) or with
$W$-decontamination and -recombination corrections (SL dec+rec).
The distributions in the $p_\rT$ and the mass of the leading $B$-jet
and in the mass of the $\mu^- j_b$ (DL channel) or $j_dj_b$ (SL channel)
systems are compared in the $W^+$ phase space 
with $N_B=1$ (left column) and $N_B\ge 2$ (right column).  
}
\label{fig:redecay_Wm_1_LHE} 
\end{figure}

\begin{figure}
  \begin{center}
  \includegraphics[width=72mm,trim= 0 28 0 0, clip]{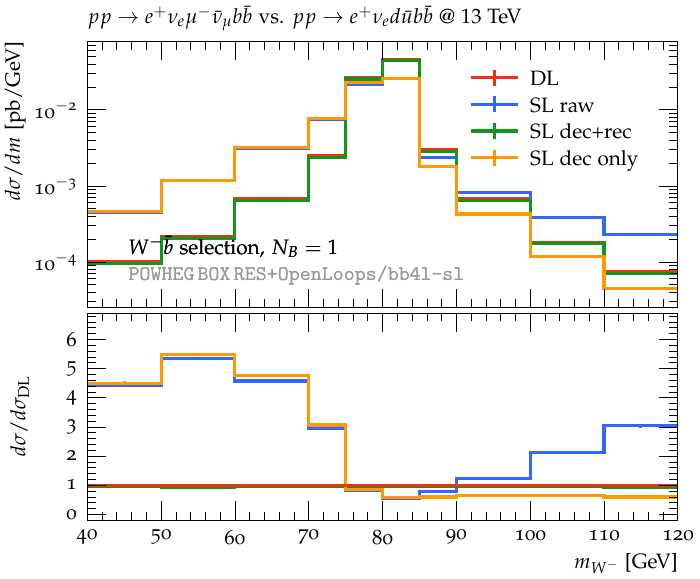} 
  \includegraphics[width=72mm,trim= 0 28 0 0, clip]{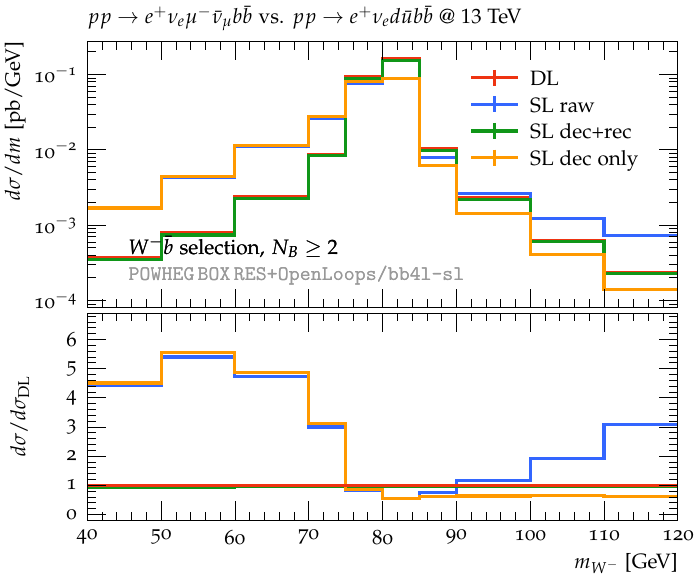} 
\\
  \includegraphics[width=72mm,trim= 0 0 0 19.0, clip]{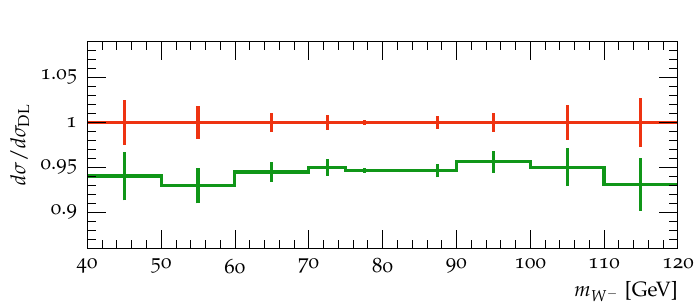}
  \includegraphics[width=72mm,trim= 0 0 0 19.0, clip]{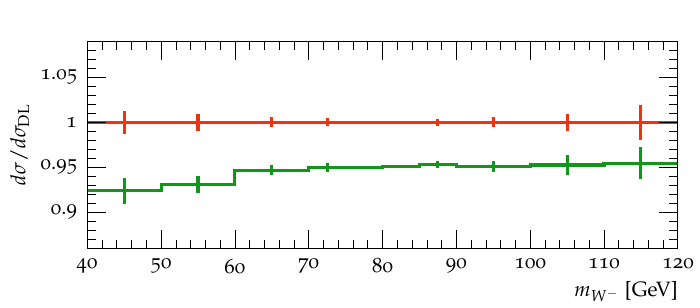}
\\
  \includegraphics[width=72mm,trim= 0 28 0 0, clip]{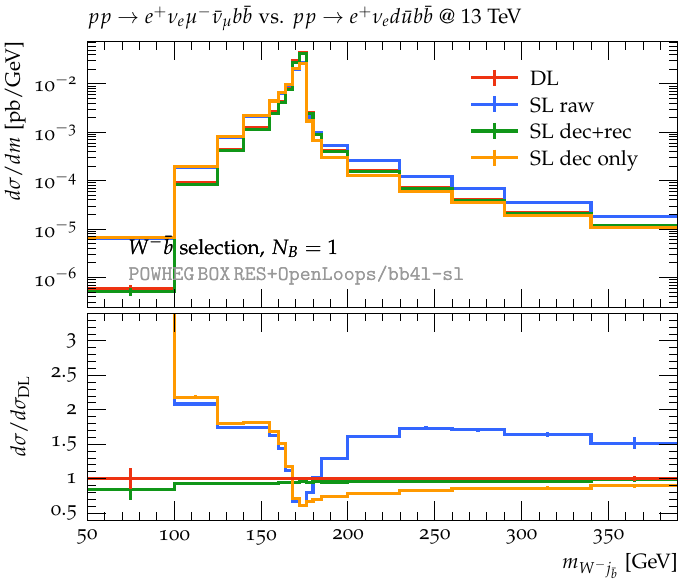} 
  \includegraphics[width=72mm,trim= 0 28 0 0, clip]{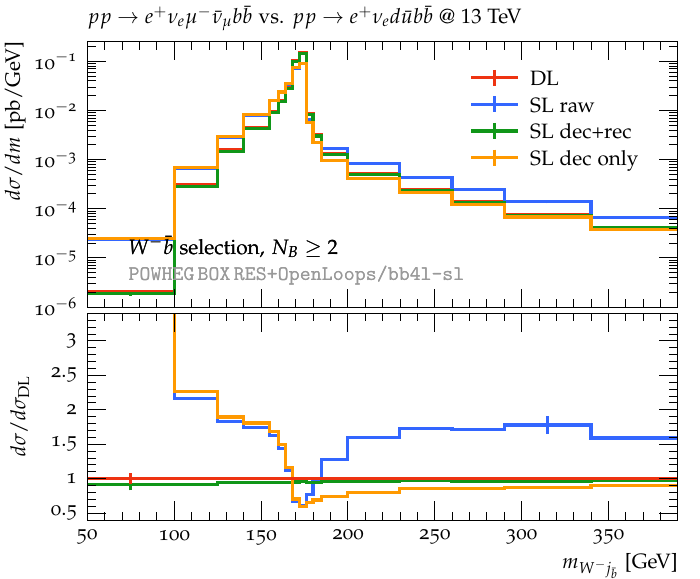}
\\
  \includegraphics[width=72mm,trim= 0 0 0 19.0, clip]{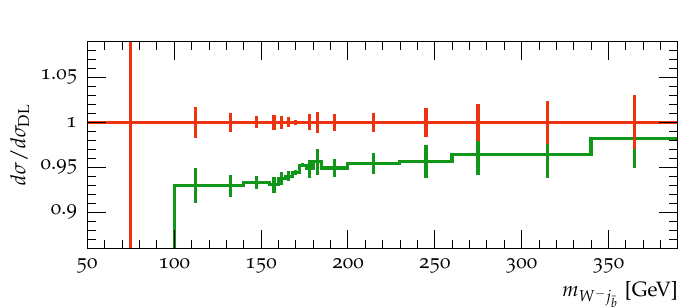}
  \includegraphics[width=72mm,trim= 0 0 0 19.0, clip]{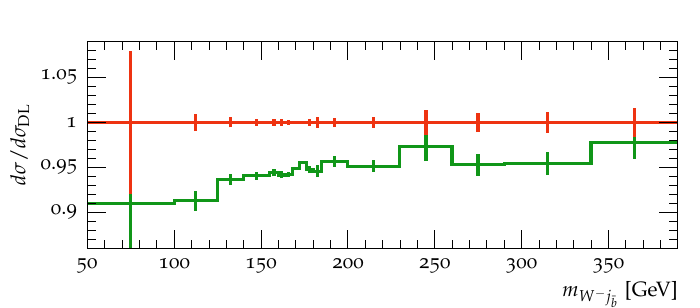}
  \end{center}
\caption{Similar comparison as in \reffi{fig:redecay_Wm_1_LHE} 
for the distributions in the invariant masses of the
reconstructed $W^-$ bosons, 
which correspond either to the
$\mu^-\bar\nu_\mu$ or the $j^W_dj^W_{\bar u}$ systems, and 
of the associated
 $W^-j_b$ systems.
Here in addition to DL, SL raw and SL dec+rec we also show 
predictions in the single-lepton channel with only $W$-decontamination  (SL dec only).
The central frame shows the various DL/SL ratios, while the 
lower frame shows the ratio of SL~dec+rec wrt DL with higher resolution.
}
\label{fig:redecay_Wm_0_LHE}
\end{figure}

In Figs.~\ref{fig:redecay_Wm_1_LHE}--\ref{fig:redecay_Wm_0_LHE} we apply a 
$W^-\bar b$ selection and we study various observables that depend on the
constituents of the $j^W_d j^W_{\bar u} j_{\bar b}$  hadronic ``anti-top candidate''
in the SL channel
and its 
$\mu^+\bar \nu_\mu j_{\bar b}$ leptonic counterpart in the DL channel.
In \reffi{fig:redecay_Wm_1_LHE} we show the distributions in the 
$p_\rT$ and the mass of $j_{\bar b}$ as well as the invariant mass of the 
$\mu^-j_{\bar b}$ ($j^W_{d} j_{\bar b}$) pair in the DL (SL) channel.
In general, due to the presence of hadronic $W^-$ decays,
in the $W^-\bar b$ selection we observe 
much more pronounced 
DL--SL differences 
as compared to the $W^+ b$ selection.
In the bulk of all distributions, 
SL raw predictions feature a deficit of about 10-20\% wrt the DL channel.
This effect can be attributed to the fragmentation of light jets from 
$W^-$ decays, which leads to reduced $p_\rT$ 
as compared to the decay products of the leptonic $W^-$ decay, and thus to 
fewer events passing the acceptance cuts in the SL channel.
In the tail of the $j^W_{d} j_{\bar b}$ mass distribution, above the edge 
that is located around 150\,GeV,
we observe the opposite behaviour, \ie an excess of SL raw events as compared to DL ones. 
This can be attributed to the migration of SL events across the edge 
as a result of large-angle QCD radiation that is emitted by the $W^-$ decay 
and absorbed by $j_{\bar b}$.
When applying dec+rec corrections, all DL--SL differences are largely suppressed 
as expected. More precisely, with the only exception of the suppressed 
region of small $j^W_{d} j_{\bar b}$ invariant mass,
the remaining DL--SL 
difference is at the level of 
5--8\%.
This residual 
mismatch can be attributed to imperfections of the dec+red corrections, such as
the ambiguity in the assignment of large-angle QCD radiation to the individual 
$j^W_d$ or $j^W_{\bar u}$ jets.

In \reffi{fig:redecay_Wm_0_LHE} we present the same comparison for the 
distributions in the invariant masses of the reconstructed 
$W^-$ boson and of the anti-top quark. 
In these observables,
the DL--SL differences due to 
the kinematic effects of QCD radiation and jet clustering
are strongly amplified by the presence of 
narrow Breit--Wigner peaks.
In the
$N_B=1$ and $N_B\ge 2$ regions, and for both invariant-mass distributions, 
the SL raw predictions feature huge enhancements in the off-shell regions on both sides of the
resonances. Such enhancements are due to 
two opposite mechanisms, where resonant events lose invariant mass
due to the fragmentation of the $j^W_{d}$ or $j^W_{\bar u}$ jets
or, alternatively, gain invariant mass due to QCD radiation 
that does not originate from the hadronic $W^-$ but is 
clustered into the $j^W_{d}$ or $j^W_{\bar u}$ jets.
The light-jet clustering mechanism can be reversed
by applying the $W$-decontamination correction (see ``SL dec only'' curves).
As a result, in the
high-mass tails the SL excesses of up to 100--200\% 
disappear, and the SL cross section goes down to
around 
10--30\%
below the DL
one. This deficit can be attributed to the light-jet 
fragmentation mechanism, which is also responsible for the 
huge DL excesses below the Breit--Wigner peaks.
These remaining DL--SL differences are strongly suppressed when
the $W$-decontamination is supplemented by the
$W$-recombination correction (SL dec+rec). In this case, in the $W$-mass
distribution the shape of DL and SL predictions become almost identical, and
in a broad region around the Breit--Wigner peaks we observe a rather
constant difference around 5\%, similarly as in
\reffi{fig:redecay_Wm_1_LHE}.
An equally good shape agreement and 5\% difference is observed also in the 
anti-top mass distributions.

In summary, the results of Figs.~\ref{fig:redecay_Wp_1_LHE}--\ref{fig:redecay_Wm_0_LHE}
demonstrate the mutual consistency of the \bbfourldl{} and \bbfourlsl{}
versions of the \bbfourl{} generator,
providing also a qualitative and quantitative picture of 
the jet-fragmentation and jet-clustering effects that 
are responsible for the expected differences between DL and SL channels.

\section{Shower effects and comparison against on-shell \boldmath$\ttbar+tW$ generators
\label{sec:nlops}}

In this section we turn our interest in 
the investigation of 
$\ttbar+tW$ production 
with semileptonic decays
to the parton shower,
and we quantify its effect by comparing \bbfourlsl{}
predictions at the LHE and NLO+PS levels.
Moreover we compare the 
\bbfourlsl{} generator against the well established
\POWHEG{} generators that describe $\ttbar$ and $\tW$ production
in the on-shell approximation, namely the \hvq~\cite{Frixione:2007vw} and 
the \ST~\cite{Re:2010bp} generators.\footnote{These generators are named 
as the corresponding
directories in the \POWHEGBOX{} package~\cite{Alioli:2010xd}.}
This comparison provides quantitative information on 
various physics ingredients---such as off-shell and interference 
effects---that are 
included at NLO+PS accuracy in \bbfourlsl{},
while in \hvq{}+\ST{} they are either absent, or only included at LO+PS accuracy.

\subsection{Differences between on-shell and off-shell generators}
\label{se:hvqSTdeficit}

As a basis for the comparison of on-shell against
off-shell \POWHEG generators for $\ttbar+\tW$ production with semileptonic decays,
in the following we list the main deficits of \hvq{}+\ST{} wrt 
\bbfourlsl{} in terms of physics ingredients and perturbative accuracy.

\begin{itemize}

\item In \bbfourlsl{}, all $t\to Wb$ and $W\to q\bar q'$ decays are 
NLO+PS accurate, i.e.~the hardest emission is based on matrix elements, 
while in \hvq{}+\ST{} such decays are only LO+PS accurate,
i.e.~QCD radiation is entirely generated by the parton shower.
The resulting differences can be quite sizeable at the LHE level. However, after parton showering 
they tend to be rather small. This is due to fact that 
shower emissions in $t\to Wb$ and $W\to q\bar q'$ decays 
are systematically improved through 
matrix-element corrections (MEC) in \Pythia.
 
\item The \ST{} generator describes $\tW$ production 
as a $gb\to tW$ hard process with massless $b$~quarks.
This implies that 
$\tW$ contributions to the phase space with two resolved $b$~jets
arise via $pp\to tWb$ real-emission channels and are 
only LO+PS accurate.
In contrast, in \bbfourlsl{}
$b$-quark mass effects 
are included throughout, and
$\tW$ contributions are NLO+PS accurate 
in the whole phase space with up to two resolved $b$ jets.

\item In \bbfourlsl the $\ttbar$--$\tW$ interference is entirely included
at NLO+PS accuracy, while the \ST{} generator supports two different treatments of the interference.
In the diagram-subtraction approach (\STDS), 
the interference is included in the 
$pp\to tWb$ real-emission processes, i.e.~with LO+PS accuracy,
while  in the diagram-removal approach (\STDR) it is omitted.
Thus, the difference between \STDR{} and \STDS{} can be regarded as a
LO+PS estimate of the $\ttbar$--$\tW$ interference.
 
\item In \bbfourlsl{} non-resonant effects, i.e.~contributions from diagrams that are free from 
$t\to bW$ sub-topologies, are included throughout at NLO+PS accuracy, while in 
\hvq{}+\ST{} they are entirely absent.

\item Off-shell effects are exact in \bbfourlsl{}, while  the
\hvq{}+\ST{} generators employ matrix elements with on-shell
top quarks together with an approximate treatment of 
off-shell effects, which consists of re-distributing the invariant masses of top quarks (and their decay products)
according to a Breit--Wigner distribution.

\end{itemize}

\begin{table}
  \centering
  \begin{tabular}{l||c|c|c}
                            & \bbfourlsl{}  & \hvq{}+\STDS{}  & \hvq{}+\STDR{} \\
\hline
$t\to Wb$ and $W\to q\bar q'$ decays 
          & NLO+PS          & LO+PS   & LO+PS    \\
$tWb$ production            & NLO+PS       & LO+PS   & LO+PS    \\
$\ttbar$--$\tW$ interference & NLO+PS       & -       & LO+PS    \\
off-shell effects           & NLO+PS       & approx. & approx.  \\
non-resonant contributions  & NLO+PS       & -       & -        \\
 \end{tabular}
  \caption{Summary of the main differences---in terms of 
physics ingredients and their formal accuracy---between \bbfourl and the \hvq+\ST{}
generators for the cases where \ST{}
is applied in diagram-subtraction (\STDS) or diagram-removal (\STDR) mode
(see the discussion in \refse{se:hvqSTdeficit}).
The indicated formal accuracy, NLO+PS or LO+PS, refers
to the phase space with two resolved $b$~jets.
\label{tab:gencomp}
}
\end{table}
The above differences between \hvq+\ST{} and \bbfourlsl{} are summarised in a
schematic way in~\refta{tab:gencomp}.

\subsection{Physics objects and event selection\label{sec:RealisticAnalysis}}
\label{se:hvqstanalysis}

The comparisons presented in the following subsections are based on a semi-realistic 
event selection that is inspired by experimental analyses of $\ttbar$ production.
In particular, as described in the following, we focus on semileptonic $\ttbar$ signatures with 
two resolved $b$~jets.

\vspace{2mm}
\noindent{\it Physics objects and cuts} --- The event selection consists of the following
steps.
\begin{itemize}
\item  Based on MC truth we identify the charged lepton and the neutrino
stemming from the decay of the $W^+$ boson.\footnote{In the analysis 
we use the exact
four-momenta of the $W^+$ decay products.
As before, since QED
radiation is switched off in \Pythia{} there is no need to dress
charged leptons.} For both we
require $p_\rT>25\,\GeV$, and in the case of the charged lepton also 
$|\eta|<2.5$. 

\item Jets are built using the anti-$k_\rT$ algorithm with $R=0.5$ and are
categorised
into $b$~jets, $\bar{b}$~jets and light jets 
depending on the presence of
$b$~quarks among their constituents
at MC-truth level.

\item We require (at least) two light jets, plus one $b$~jet and 
one $\bar b$~jet with $p_\rT>25\,\GeV$ and $|\eta|<2.5$.
In the following the label  
$j_{b}$ ($j_{\bar b}$) is used to denote 
the leading $b$~($\bar b$)~jet within acceptance.

\end{itemize}
\noindent{\it Reconstruction of resonances}---Certain observables are defined in terms 
of the $W$-boson and top-quark momenta, which are reconstructed as follows.

\begin{itemize}

\item The momentum of the leptonically decaying $W$ boson is defined as the
sum of the four-momenta of the corresponding lepton and neutrino (also when
$|\eta_\nu|>2.5$).

\item To reconstruct the hadronic $W$ boson 
we consider the four hardest light jets 
within acceptance and we select the
combination of 2 or 3 light jets 
whose invariant mass is closest to $m_W$.

\item The (anti-)top-quark momentum is reconstructed as the sum of the
momenta of the reconstructed ($W^-$) $W^+$ boson, 
plus the ($j_{\bar b}$) $j_{b}$ jet and up
to one of the remaining light jets within acceptance.
The latter is included in case it yields an invariant mass 
closer to $m_t$.

\end{itemize}

\subsection{Integrated cross section}

\def\bbfourlratio{\frac{\sigma}{\phantom{\big|}\sigma_{\bbfourlsl}^{\NLOPS}}}

\begin{table}
  \centering
  \begin{tabular}{|l|l||c|c||c|c||c|c|}
    \hline
&&  \multicolumn{2}{c||}{inclusive phase space} &\multicolumn{4}{c|}{$e^+\nu_ebbjj$ fiducial phase space} \\\hline
&&  \multicolumn{2}{c||}{}& \multicolumn{2}{c||}{$R=0.5$} &\multicolumn{2}{c|}{$R=0.2$} \\\hline
&& $\sigma$[pb] & $\bbfourlratio$ & 
   $\sigma$[pb] & $\bbfourlratio$ & 
   $\sigma$[pb] & $\bbfourlratio$ \\\hline
    \hline
    {\footnotesize \bbfourlsl}     & {\footnotesize NLOPS}     & 57.56(2)  & 1     & 16.30(1)  & 1     & 14.639(9) & 1     \\
    {\footnotesize \bbfourlsl}     & {\footnotesize LHE}       & 57.56(2)  & 1     & 16.33(1)  & 1.002 & 17.17(1)  & 1.173 \\\hline\hline
    {\footnotesize \hvq }          & {\footnotesize NLOPS}      & 54.340(9) & 0.944 & 15.910(6) & 0.976 & 14.244(6) & 0.973 \\
    {\footnotesize \hvq }          & {\footnotesize LHE}        & 54.339(9) & 0.944 & 17.446(7) & 1.070 & 18.614(8) & 1.272 \\\hline
    {\footnotesize \STDR }         & {\footnotesize NLOPS}      & 2.5524(3) & 0.044 & 0.5249(2) & 0.032 & 0.4683(2) & 0.032 \\
    {\footnotesize \STDR }         & {\footnotesize LHE}        & 2.5524(3) & 0.044 & 0.3082(1) & 0.019 & 0.3470(1) & 0.024 \\\hline
    {\footnotesize \STDS }         & {\footnotesize NLOPS}      & 2.5194(3) & 0.044 & 0.4877(2) & 0.030 & 0.4232(2) & 0.029 \\ 
    {\footnotesize \STDS }         & {\footnotesize LHE}        & 2.5194(3) & 0.044 & 0.3002(1) & 0.018 & 0.3280(1) & 0.022 \\\hline\hline
    {\footnotesize \hvq\,+\,\STDR} & {\footnotesize NLOPS}      & 56.892(8) & 0.988 & 16.475(6) & 1.011 & 14.780(6) & 1.010 \\
    {\footnotesize \hvq\,+\,\STDR} & {\footnotesize LHE}        & 56.893(8) & 0.988 & 17.754(7) & 1.089 & 18.961(7) & 1.295 \\\hline
    {\footnotesize \hvq\,+\,\STDS} & {\footnotesize NLOPS}      & 56.859(8) & 0.988 & 16.398(6) & 1.006 & 14.667(6) & 1.002 \\
    {\footnotesize \hvq\,+\,\STDS} & {\footnotesize LHE}        & 56.858(8) & 0.988 & 17.746(7) & 1.089 & 18.942(7) & 1.294 \\\hline
  \end{tabular}
  \caption{Cross sections for 
$pp \to e^+\nu_e jj b \bar{b}$ at 13\,TeV
in the fully inclusive phase space
(columns 3--4) and in the fiducial phase space 
with 
two $b$~jets plus two light jets, a lepton and a neutrino,
as defined in \refse{se:hvqstanalysis}
(columns 5--8). In the latter case we compare results 
based on the anti-$k_\rT$ algorithm with $R=0.5$
(columns 5--6)
and $R=0.2$ (columns 7--8).
The various rows report \bbfourlsl{} predictions
at the LHE and NLO+PS level, as well as  
corresponding \hvq+\STDR{} and \hvq+\STDS{} predictions.
In the case of \bbfourlsl{}, by default all LHE and NLOPS results include 
the 
inverse-width expansion
discussed in \refse{se:topwidthcorr}.
In this comparison all input parameters, QCD scales, PDFs and branching ratios are
chosen consistently with the setup described in \refse{sec:setup}.
\label{tab:bb4lhvqfidXS}
}
\end{table}

Cross sections obtained with the \bbfourlsl{} and \hvq{}+\ST{} generators 
at LHE and NLOPS levels are reported in 
\refta{tab:bb4lhvqfidXS}. 
Let us first focus on fully inclusive cross sections.
In this case, as expected from the unitarity of NLO+PS matching,
LHE and NLOPS predictions coincide with each other and correspond exactly to 
fixed-order NLO.
As compared to \bbfourlsl{}, the \hvq{} inclusive cross section
features a deficit of~5.6\%, which is mainly due to 
the missing $\tW$ contribution. The latter amounts 
to 4.4\% of the \bbfourlsl{} cross section, both in the 
DR and DS approach, which points to a very small
$\ttbar$--$\tW$ interference. The combined 
\hvq{}+\ST{} prediction lies only 1.2\% below 
\bbfourlsl{}. This small difference can be 
attributed to bottom-mass effects,  
off-shell and 
non-resonant contributions of $\ord(\Gamma_t/m_t)$,
or to effects that originate from $\ord(\as^2)$ 
differences in the 
treatment of bottom-loop contributions to the gluon-PDF,
and in the perturbative expansion of 
$1/\Gamma_t$. In this respect we note that 
the consistent expansion of $1/\Gamma_t$ terms
up to $\ord(\as)$,
as described in~\refse{se:topwidthcorr},
is mandatory in order to achieve percent-level agreement 
between \hvq{}+\ST{} and \bbfourlsl{}. 
In particular, we have checked that
omitting the 
inverse-width expansion~\refeq{eq:GEXPu}--\refeq{eq:GEXPv}
in~\bbfourlsl{} yields
\beq
\frac{\sigma_{\tt bb4l-sl}}{\sigma_{\tt hvq+ST}}\Bigg|_{\mbox{\scriptsize 
no $1/\Gamma_t$ expansion
}}
=\;
1.074\,,
\eeq
which corresponds to a difference bigger than 
the entire $\tW$ contribution. 
Based on the analysis of~\refse{se:topwidthcorr},
it is clear that
a similarly large 
mismatch would show up also in fiducial cross sections 
and differential observables.

Moving to the $e^+\nu_ebbjj$
fiducial phase space,
we observe that the \bbfourlsl{} cross section remains 
largely independent of shower effects. More precisely,
when jet cuts are applied using the standard jet resolution, $R=0.5$,
the difference between NLOPS and LHE predictions are at the 
permil level. 
This tiny sensitivity 
to parton showering is mainly due to the fact that,
in \bbfourlsl{},
the hardest QCD emissions from all
top-production and -decay subprocesses are entirely 
controlled by matrix elements.
This is not the case for \hvq{} and \ST{}, where 
QCD radiation in top decays is entirely generated by the parton shower.
Moreover, in the case of \ST{} also the hardest radiation on top of the
$pp\to tWb$ production process is entirely due to the shower.
In the fiducial \hvq{} cross section such shower effects 
are around $-10\%$ and are most likely dominated by negative 
jet-$p_\rT$ shifts 
resulting from jet fragmentation.
As for the \ST{} fiducial cross section,
we observe that parton showering shifts the LHE result by about $+65\%$.
This large positive effect can be attributed to positive 
jet-$p_\rT$ shifts resulting from ISR contamination of hard jets.
This interpretation is supported by the fact that the
shower sensitivity of the \ST{} fiducial cross section is
strongly attenuated when the 
jet resolution is reduced to $R=0.2$, while in the case of
\hvq{} its is strongly enhanced.
Note that, for $R=0.2$, shower effects become quite significant
also in the \bbfourlsl{} fiducial cross section, where they amount to 
$-15\%$, while their impact in \hvq{}+\ST{} 
is around $-23\%$.

Comparing \bbfourlsl{} to \hvq+\ST{} at the NLOPS level,
we find that the percent-level agreement observed 
in the inclusive cross section
persists also in the fiducial phase space,
both for $R=0.5$ and $R=0.2$. 
More precisely, when cuts are applied 
the relative difference wrt \bbfourlsl{}
moves form about $-1\%$ to $+1\%$.
This level of agreement is remarkable,
since the fiducial cross section is 
sensitive to the modelling of QCD radiation in $\tW$ production and 
top decays, which is entirely controlled by \Pythia in \hvq+\ST.
In this respect, we note that \Pythia's  
matrix-element corrections play a 
significant
role. We have checked 
that disabling such corrections 
shifts the fiducial \hvq+\ST{}
cross section (for $R=0.5$)
by about $-2.5\%$,
while in \bbfourlsl{} the effect of \Pythia's  
matrix-element corrections is at the few-permil level.
Finally we note that, in the fiducial region,
the {\tt DS--DR} difference points to interference effects
of the order to 10\% of the $\tW$ contribution, which amounts to only 2--3 
permil of the total fiducial cross section.

\subsection{Shower effects in differential distributions}
\label{se:showerdist}

\newcommand{\showerplot}[1]{
  \begin{minipage}[t]{.45\linewidth}
  \includegraphics[width=72mm,trim= 0 28 0 0, clip]{figures/nlops/mainLHE/#1} 
  \\
  \includegraphics[width=72mm,trim= 0 32 0 19, clip]{figures/nlops/ratio05/#1}
  \\
  \includegraphics[width=72mm,trim= 0 0 0 19, clip]{figures/nlops/ratio02/#1}
  \end{minipage}
}

\newcommand{\showerplotL}[2]{
\begin{subfigure}{0.49\linewidth}
\showerplot{#1}
\caption{\phantom{\hspace{-12mm}}}
\label{#2}
\end{subfigure}}

\newcommand{\showerplotR}[2]{
\begin{subfigure}{0.49\linewidth}
\showerplot{#1}
\caption{\phantom{\hspace{-12mm}}}
\label{#2}
\end{subfigure}}

\begin{figure}
\centering
\showerplotL{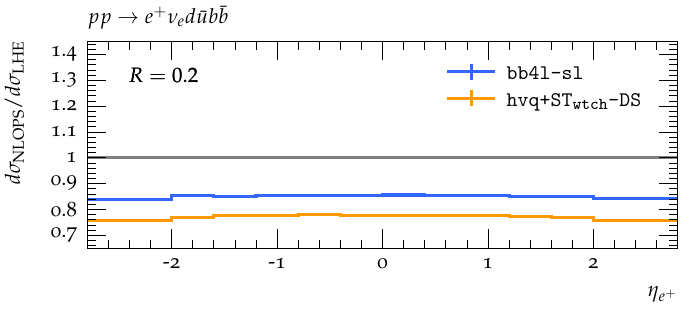}{figPS:2_W+_etal}
\showerplotR{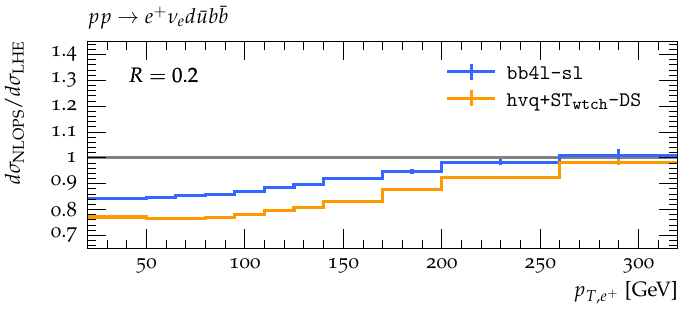}{figPS:2_W+_ptl}
\\[5mm]
\showerplotL{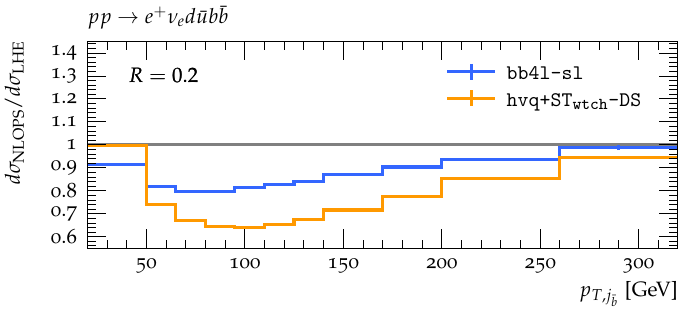}{figPS:2_W-_ptb}
\showerplotR{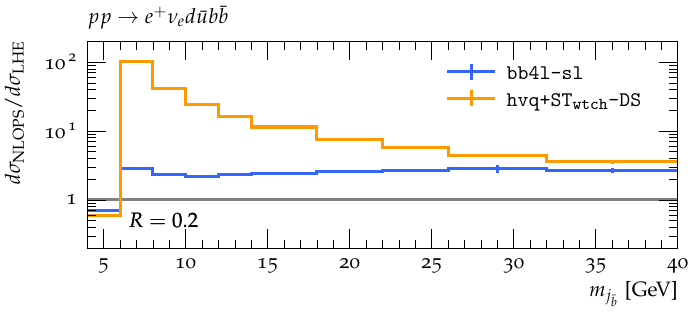}{figPS:2_W-_mb}
\caption{Impact of the parton shower 
for $pp\to e^{+} \nu_{e} j j b\bar b\,$ at 13\,TeV
in the fiducial phase space  with 
two $b$-jets plus two light jets, a positron and a neutrino,
as defined in \refse{se:hvqstanalysis}:
distributions in the rapidity of the positron (\subref{figPS:2_W+_etal}),
its transverse momentum (\subref{figPS:2_W+_ptl}), as well as the transverse momentum
of the hardest $\bar b$-jet (\subref{figPS:2_W-_ptb}) and its invariant mass
(\subref{figPS:2_W-_mb}).
The upper frame shows LHE and NLOPS  
predictions of \bbfourlsl for $R=0.5$,
while the middle (lower) frames compare the 
NLOPS/LHE ratios of the \bbfourlsl{}
and \hvq+\STDS{} generators
for $R=0.5$ ($R=0.2$). 
}
\label{figPS:1}
\end{figure}

In \reffis{figPS:1}{figPS:3}
we investigate the impact of shower effects
in the \bbfourlsl{} and \hvq+\ST{} generators 
by comparing LHE and NLOPS predictions
for various differential 
distributions that describe the behaviour of the
positron, the 
leading $b$ and $\bar b$~jet, the 
leptonically decaying $W^+$ boson and $t$~quark,
as well as the hadronically decaying
$W^-$~boson and $\bar t$~quark.
These comparisons are carried out in the 
$e^+\nu_ebbjj$ fiducial region, and the 
decaying $W$~bosons and (anti-)top quarks are reconstructed as
detailed in \refse{se:hvqstanalysis}.

In the positron-rapidity distribution (\reffi{figPS:2_W+_etal})
we observe that the
NLOPS/LHE ratios for $R=0.5$ and $R=0.2$ are nearly independent of 
$\eta_{e^+}$ and behave as in the
case of the fiducial cross section.
This holds both for \bbfourlsl{} and \hvq+\ST.
A similarly stable behaviour is observed also 
in the positron-$\pT$ distribution (\reffi{figPS:2_W+_ptl}),
but only for moderate $p_{\rT,e^+}$, while 
above 100\,GeV shower effects become increasingly 
sensitive to $p_{\rT,e^+}$.
This behaviour is quite similar in 
\bbfourlsl{} and \hvq+\ST{}. Thus it is probably due 
to the interplay of ISR with the acceptance cuts,
since ISR is handled in a similar way in the 
different generators.

The above interpretation is supported by \reffi{figPS:2_W-_ptb},
where we observe that the NLOPS/LHE
ratios are rather sensitive 
to the $p_\rT$ of the hardest $\bar b$~jet.
In particular, the effect of the shower 
grows significantly at large $p_\rT$. 
We have checked that the 
corresponding distribution for the case of the hardest $b$~jet 
behaves in a very similar way.

\begin{figure}
\centering
\showerplotL{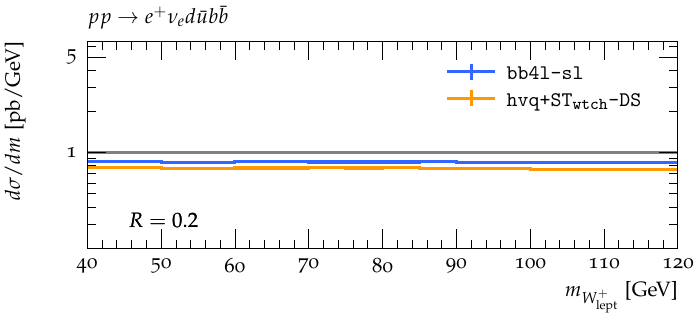}{figPS:2_W+_mW_zoom}
\showerplotR{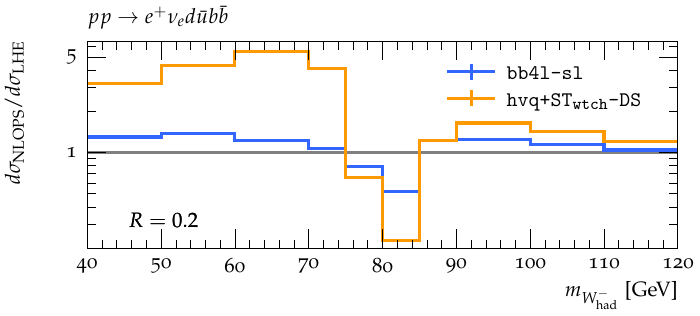}{figPS:2_W-_mW_zoom}
\\[5mm]
\showerplotL{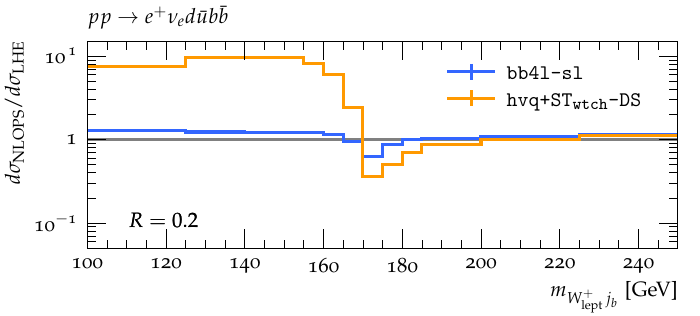}{figPS:2_W+_mWb}
\showerplotR{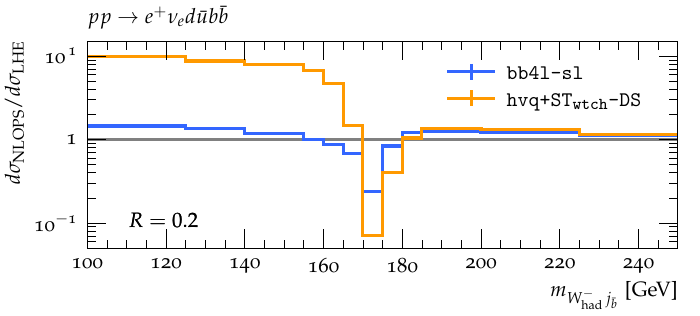}{figPS:2_W-_mWb}
\caption{Distributions in
the reconstructed invariant masses of 
the leptonically decaying $W^+$ (\subref{figPS:2_W+_mW_zoom})
and hadronically decaying $W^-$ boson (\subref{figPS:2_W-_mW_zoom}),
as well as of the leptonically decaying  
top (\subref{figPS:2_W+_mWb})
and hadronically decaying anti-top quark (\subref{figPS:2_W-_mWb}).
Same acceptance cuts, predictions and ratios as in \reffi{figPS:1}.
}
\label{figPS:2}
\end{figure}

Also the distributions in the masses of the hardest 
$b$ and $\bar b$~jets behave in a similar way, and in
\reffi{figPS:2_W-_mb} we present the one for the 
$\bar b$~jet.
This observable is highly sensitive to the parton shower,
and NLOPS/LHE corrections are completely different in 
\bbfourlsl{} and \hvq+\ST. The reason is that 
all bins with $m_{j_{\bar b}}>m_b$ are mostly populated by QCD radiation
off $\bar b$~quarks, which is modelled in a different way in the
different generators. 
In the case of \hvq+\ST, where QCD radiation 
off $\bar b$~quarks is 
entirely stemming from the parton shower,
NLOPS/LHE corrections have a huge impact, which varies between a factor four and a hundred 
in the plotted region.
In contrast, in \bbfourlsl{}, where the hardest emission is controlled by the
matrix elements,
NLOPS/LHE corrections are around a factor 2--3 and roughly constant.

The distributions in the invariant masses of the 
reconstructed $W^+$ and $W^-$ bosons,
$m_{W^+_{\mathrm{lept}}}$ and 
$m_{W^-_{\mathrm{had}}}$,
are shown in
Figs.~\ref{figPS:2_W+_mW_zoom}--\subref{figPS:2_W-_mW_zoom}.
In the case of the leptonically decaying $W^-$,
apart from normalisation effects
at the level of the fiducial cross section,
the invariant-mass distribution 
is completely insensitive to QCD radiation.
In contrast, the modelling of QCD radiation 
has an important impact on the invariant mass of the
hadronically decaying $W^-$. 
Below resonance, shower radiation tends to increase the cross section
due to jet-fragmentation processes, where QCD partons escape from the 
reconstructed $W$~boson. 
In \bbfourlsl{} such effects are around $20\%$, while 
in \hvq+\ST{} they reach a factor four. 
The shower tends to increase the cross section also 
above the $W^-$ resonance. This 
can be attributed to 
the contamination of the reconstructed $W^-$ boson
through ISR or QCD radiation stemming from $b$~quarks. 
Here the differences between \bbfourlsl{} and 
\hvq+\ST{} are less dramatic since 
in both cases the hardest initial-state emission is generated by \POWHEG.
The above interpretations of shower effects
are consistent with the observed dependence on the 
jet radius: when $R$ is reduced from 0.5 to 0.2,
jet-fragmentation effects (below resonance)
tend to increase, while 
jet-contamination effects (above resonance) tend to decrease.

The distributions in the invariant masses of the 
reconstructed $t\to W^+ j_b$ and $\bar t\to W^-j_{\bar b}$ 
resonances,
$m_{W^+_{\mathrm{lept}}\,j_b}$ and 
$m_{W^-_{\mathrm{had}}\,j_{\bar b}}$,
are shown in
Figs.~\ref{figPS:2_W+_mWb}--\subref{figPS:2_W-_mWb}.
In the case of the $t\to W^+ j_b$ resonance (\reffi{figPS:2_W+_mWb}),
in spite of the leptonic nature of the $W^+$ decay,
we observe qualitatively similar 
shower effects as for the hadronically 
decaying $W^-$ (\reffi{figPS:2_W-_mW_zoom}).
In particular, in \hvq+\ST{}
we find sizeable positive shower corrections
that tend to increase (decrease)
below (above) resonance 
when $R$ is reduced 
from 0.5 to 0.2.
Since the involved $W^+$ boson decays leptonically,
this behaviour must be due to 
fragmentation and contamination processes 
that involve the associated $b$~jet.
We note also that, as compared to \hvq+\ST,
the shower sensitivity of  
\bbfourlsl{} is much smaller below resonance,
while above resonance it is similar.
This suggests that $b$-jet contamination is 
dominated by ISR, which is handled 
in a similar way in the 
different generators.
Finally, we observe that shower effects in the
invariant-mass distribution of the 
hadronically decaying anti-top quark
(\reffi{figPS:2_W-_mWb})
are quite similar as in the case of the
leptonically decaying top (\reffi{figPS:2_W+_mWb}).
Thus, the sizeable shower corrections to the invariant mass of 
the hadronic $W^-$ (\reffi{figPS:2_W-_mW_zoom})
seem to have little impact on the reconstructed mass of the
hadronic anti-top.
This is most likely due to the fact that, inside a hadronic 
$t\to Wb \to jjb$ decay, the radiation that is emitted from the 
$b$~quark which contaminates the reconstructed hadronic 
$W$~boson (or vice versa) does not have any impact on the 
reconstructed top-quark momentum.

\begin{figure}
\centering
\showerplot{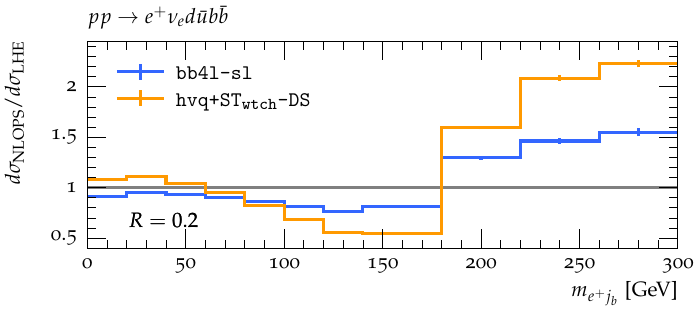}
\caption{Distribution in 
the invariant mass $m_{e^+j_b}$
of the positron and the 
hardest $b$ jet.
Same acceptance cuts, predictions and ratios as in \reffi{figPS:1}.
}
\label{figPS:3}
\end{figure}
Finally, in \reffi{figPS:3} we present the distribution 
in the invariant mass of the positron and the $b$~jet,
i.e.~the visible decay products of the leptonically decaying top quark.
At LO, this observable features a kinematic edge   
at $m_{e^+\,j_b}^2 = m_t^2-m_W^2\simeq (152\,\GeV)^2$,
which can be exploited for top-quark mass measurements
or in order to design cuts that suppress backgrounds due to 
on-shell top-quark production.
In the region above the edge, which is entirely populated by events with QCD
radiation, we observe that \bbfourlsl{} and \hvq+\ST{} are both strongly sensitive 
to shower radiation.
This suggests that the observed NLOPS/LHE corrections are dominated by ISR,
which is handled in a similar way in the 
different generators.
In the region below the edge, the impact of the 
shower is much less pronounced. For $R=0.5$,
in \hvq+\ST{} it can reach 20\%, while \bbfourlsl{} 
is largely insensitive to the parton shower.

\subsection{Off-shell vs on-shell generators}
\label{se:offshelldist}

In \reffis{figOFF:1}{figOFF:3} we compare 
predictions of the off-shell \bbfourlsl{} generator 
and its on-shell counterparts, \hvq+\ST{},
for the same set of observables investigated 
in \refse{se:showerdist}. 
In addition, in order to quantify
the relative importance of the $\ttbar$ and $\tW$ production modes
we also present pure $\hvq$ results, and
to assess $\ttbar$--$\tW$ interference effects we compare \ST{} predictions
in diagram-subtraction (DS) and diagram-removal (DR) mode. 
Moreover, both for \bbfourlsl{} and \hvq+\ST{},
we present extra ratios that illustrate the
effect of the matrix-element corrections that are applied within 
\PythiaEight~when showering top-quark and
$W$-boson decays.

\newcommand{\offshellplot}[1]{
  \begin{minipage}[t]{.45\linewidth}
  \includegraphics[width=72mm,trim= 0 28 0 0, clip]{figures/nlops/mainNLOPS/#1} 
  \\
  \includegraphics[width=72mm,trim= 0 32 0 19, clip]{figures/nlops/ratioCombined/#1}
  \\
  \includegraphics[width=72mm,trim= 0 0 0 19, clip]{figures/nlops/ratioNomec/#1}

  \end{minipage}
}

\newcommand{\offshellplotL}[2]{
\begin{subfigure}{0.49\linewidth}
\offshellplot{#1}
\caption{\phantom{\hspace{-12mm}}}
\label{#2}
\end{subfigure}}

\newcommand{\offshellplotR}[2]{
\begin{subfigure}{0.49\linewidth}
\offshellplot{#1}
\caption{\phantom{\hspace{-12mm}}}
\label{#2}
\end{subfigure}}

\begin{figure}
\centering
\offshellplotL{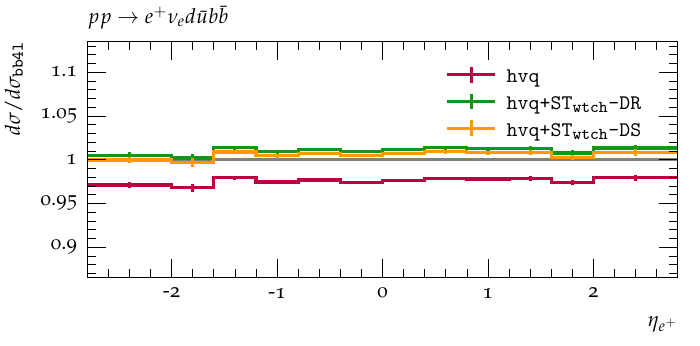}{figOFF:2_W+_etal}
\offshellplotR{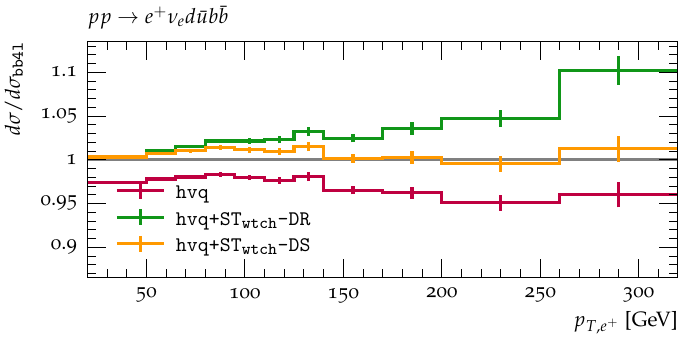}{figOFF:2_W+_ptl}
\\[5mm]
\offshellplotL{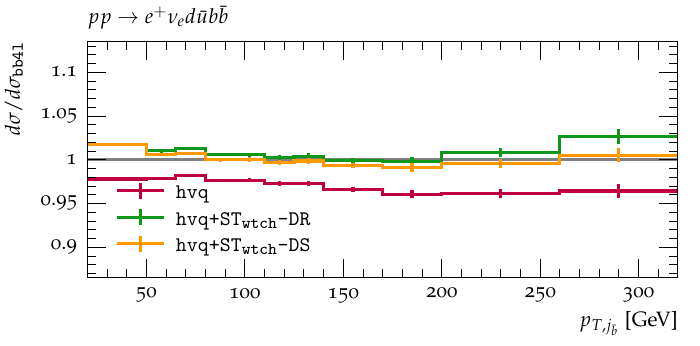}{figOFF:2_W-_ptb}
\offshellplotR{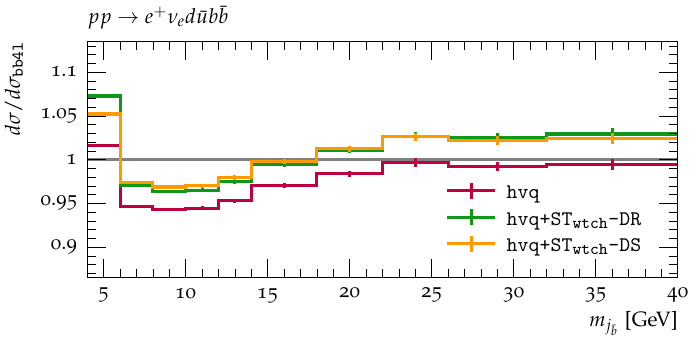}{figOFF:2_W-_mb}
\caption{Comparison of NLOPS predictions of
on-shell and off-shell 
 generators for $pp\to e^{+} \nu_{e} j j b\bar b\,$ at 13\,TeV
in the fiducial phase space  with 
two $b$~jets plus two light jets, a positron and a neutrino,
as defined in \refse{se:hvqstanalysis}:
distributions in the rapidity of the positron (\subref{figOFF:2_W+_etal}),
its transverse momentum (\subref{figOFF:2_W+_ptl}), as well as the transverse momentum
of the hardest $\bar b$~jet (\subref{figOFF:2_W-_ptb}) and its invariant mass
(\subref{figPS:2_W-_mb}).
The upper frame shows NLOPS 
predictions of \bbfourl for $R=0.5$.
The middle frame shows
ratios wrt \bbfourlsl{} for the case of  
\hvq, \hvq+\ST{} in DS and DR mode, 
while the lower frame shows the ratio of 
\hvq+\STDS{} and \bbfourlsl{}
distributions obtained with default \Pythia{} 
settings
($\rd\sigma_{\mathrm{def}}$) and disabling 
\Pythia{}'s 
matrix-element corrections
($\rd\sigma_{
{\mathrm{ME\,off}}}$).
}
\label{figOFF:1}
\end{figure}

In the lepton-rapidity distribution (\reffi{figOFF:2_W+_etal}) 
all ratios between predictions of \bbfourlsl{}, \hvq{},
\hvq+\STDR{} and \hvq+\STDS{}
are nearly constant, and their values are consistent with those observed 
at the level of the fiducial cross section. This holds also for the
distribution in the positron's transverse momentum
(\reffi{figOFF:2_W+_ptl}) in the region of moderate $p_\rT$,
while above 100\,GeV the 
DS and DR prescriptions yield increasingly different cross sections 
pointing to a sizeable $\ttbar$--$\tW$ interference.
For $p_{\rT, e^+}$ around 300\,GeV, the pure \hvq{} prediction lies $4\%$ below 
\bbfourlsl{}, and adding \ST{} in DR mode, \ie the pure $\tW$ contribution, 
shifts the result by $+15\%$, while 
switching from DR to DS mode, \ie including the 
interference, results in a shift of $-9\%$, which brings 
\hvq+\ST{} in very good agreement with \bbfourlsl.
When using the DS prescription, 
the difference between \hvq+\ST{} and \bbfourlsl{}
is below 1--2\% in the entire plotted range.
As demonstrated in the second ratio plot, this 
excellent consistency is guaranteed also by \Pythia's matrix-element corrections, 
whose effect at high $p_{\rT,e^+}$ is around 
$+6\%$ in \hvq+\ST{} and only $+2\%$ in \bbfourlsl{}.
In general, in all considered distributions 
the sensitivity of \bbfourlsl{} to 
\Pythia's matrix-element corrections is strongly suppressed 
as compared to \hvq+\ST{}. This is expected, since
in \bbfourlsl{} such shower corrections affect only 
the second and subsequent emissions.

Also for the distribution in the
$p_\rT$ of the hardest $\bar b$~jet, shown in \reffi{figOFF:2_W-_ptb},
we find agreement at the few-percent level between \hvq+\STDS{} 
and \bbfourlsl{}.
Here the most significant deviation shows up in the first bin
and amounts to only~2\%.
The relative weight of $\tW$ grows from 2\% at low $p_\rT$
to about 4\% in the tail, while \Pythia's matrix-element corrections
in \hvq+\ST{}
vary between $+2\%$ and -$5\%$ depending on the $p_{\rT}$.

The distribution in the invariant mass of the
hardest $\bar b$~jet, shown in \reffi{figOFF:2_W-_mb},
features 
slightly larger 
differences, ranging from $-3\%$  to $+5\%$,
between \hvq+\STDS{}
and \bbfourlsl{}.
The most significant deviation is observed 
in the shape of the $m_{j_{\bar b}}$ distribution 
between 5 and 25\,GeV, 
and is most likely due to the different treatment of 
parton showering in the different generators.
In fact, as discussed in \refse{se:showerdist}, 
$b$-mass distributions are extremely sensitive to 
the parton shower. 
We also note that 
the observed difference 
between the generators is 
smaller as compared to the impact of
\Pythia's matrix-element corrections
in \hvq+\ST{}.

In the distributions in the invariant masses of the $W^+$ and $W^-$ bosons 
(Figs.~\ref{figOFF:2_W+_mW_zoom}--\subref{figOFF:2_W-_mW_zoom})
the relative weights
of the \hvq{} and \ST{} contributions are nearly constant.
In the case of the leptonically decaying $W^+$ boson
also the relative effect of matrix-element corrections on
\hvq+\ST is nearly constant, 
while the differences between \hvq+\ST\,\DS{}\,
and \bbfourlsl{} vary from zero to 5\%
depending on $m_{W^+}$.
In contrast, for the hadronically decaying $W^-$ boson
we find that matrix-element corrections to
\hvq+\ST{}
are sizeable and
depend on $m_{W^-}$ in a way that is consistent with the
behaviour of shower effects in the on-shell generators.
Also the difference between \bbfourlsl{} and
\hvq+\ST{}, which ranges between $-2\%$ and $+2\%$,
features a similar kinematic dependence, 
which suggests that such difference originates from 
shower uncertainties.
Here it should be stressed that,
in the light
of the large magnitude of shower effects in \hvq+\ST{},
the percent-level agreement 
with \bbfourlsl{} is quite remarkable. 

The distributions in the invariant masses of the leptonically decaying  $t$ and the hadronically
decaying $\bar t$~quark (Figs.~\ref{figOFF:2_W+_mWb}--\subref{figOFF:2_W-_mWb}) 
provide interesting insights into the
different treatment of top resonances in \bbfourlsl{} and \hvq+\ST.
For both distributions, comparing \hvq{} alone to the combined \hvq+\ST{} prediction, 
we observe irrespective of the DR or DS modes that the relative weight of $\tW$ production, which amounts to
$3\%$ in the fiducial cross section,
is quite sensitive to the top and anti-top invariant masses:
in the vicinity of the (anti-)top resonance it goes down to 2\%, while in the off-shell region
it increases up to 10\% and beyond.
This behaviour is due to the fact that in the regions 
where $t$\,($\bar t$) is, respectively,
on-shell or off-shell, 
the relative weight of the corresponding resonance-free 
single-top channel, i.e~$pp\to \bar t W^- b$\;($t W^+ b$),
is strongly suppressed or enhanced.
Note that in the case of the leptonic $t$-mass distribution
the single-top contribution to the off-shell region is much more 
pronounced as compared to the hadronic $\bar t$-mass distribution.
This is most likely due to the fact that 
QCD radiation effects associated with the 
hadronic $W^-$ decay lead to a strong dilution of the 
reconstructed $\bar t$~resonance, which implies a
significant migration of on-shell $\ttbar$ events towards the off-shell regions.

\begin{figure}
\centering
\offshellplotL{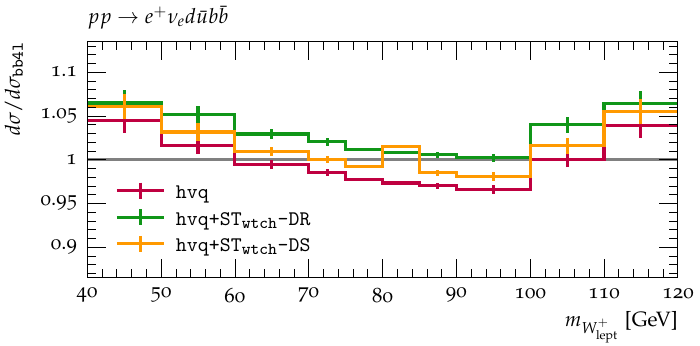}{figOFF:2_W+_mW_zoom}
\offshellplotR{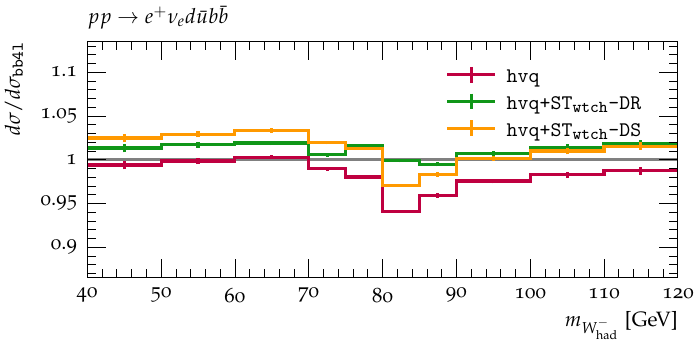}{figOFF:2_W-_mW_zoom}
\\[5mm]
\offshellplotL{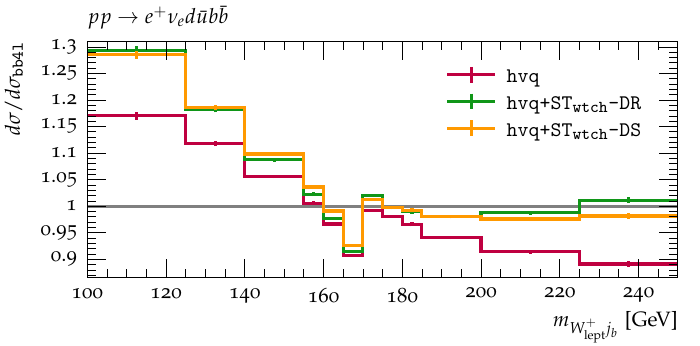}{figOFF:2_W+_mWb}
\offshellplotR{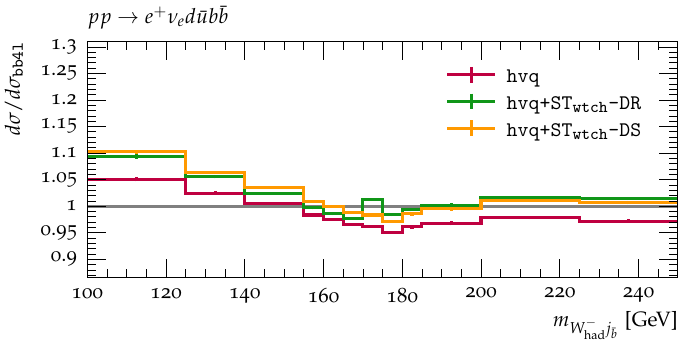}{figOFF:2_W-_mWb}
\caption{Distributions in 
the reconstructed invariant masses of 
the leptonically decaying $W^+$ (\subref{figOFF:2_W+_mW_zoom})
and hadronically decaying $W^-$ boson (\subref{figOFF:2_W-_mW_zoom}),
as well as of the leptonically decaying  
top (\subref{figOFF:2_W+_mWb})
and hadronically decaying anti-top quark (\subref{figOFF:2_W-_mWb}).
Same acceptance cuts, predictions and ratios as in \reffi{figOFF:1}.
}
\label{figOFF:2}
\end{figure}

Comparing \bbfourlsl{} to \hvq+\ST{} in Figs.~\ref{figOFF:2_W+_mWb}--\subref{figOFF:2_W-_mWb}, 
in the case of 
the hadronically decaying $\bar t$~resonance we observe a qualitatively similar 
behaviour as for the associated  $W^-$ resonance: in the on-shell region
\hvq+\ST{} features a deficit, while  
in the off-shell regions  it exceeds \bbfourlsl{} by up to 10\%
As discussed above for the $W^-$-mass distribution,
these deviations can be attributed to parton-shower uncertainties, and the 
fact that in the $\bar t$-mass distribution they
are much more sizeable is consistent with the fact that 
shower effects are much bigger than in the 
$W^-$-mass distribution (see \reffi{figPS:2_W-_mW_zoom}).
Note that the same holds also for \Pythia's
matrix-element corrections to \hvq{}+\ST, 
which behave in a significantly different way on
the two sides of the peak.
As for the leptonic $t$~resonance,
in the off-shell region below the peak 
the relative difference between \hvq+\ST{} and
\bbfourlsl{} turns out to be much more pronounced 
than in the hadronic $\bar t$~resonance.
This is related to the fact 
that the leptonic resonance is much steeper 
due to the absence of dilution effects stemming from the 
hadronic $W^-$ decay.
In particular, moving form the peak to the neighbouring bin below
the peak, the cross section goes down by 
a factor four, while in the difference between 
\hvq+\ST{} and \bbfourlsl{} we observe 
an abrupt change of about $-10\%$,
which can be interpreted as a difference of 
$-2.5\%$ in the number of events that migrate from one bin to the other, 
as a result of  QCD radiation.
We also note that this abrupt change is 
correlated with a similarly sharp variation of 
\Pythia's matrix-element corrections
in \hvq+\ST{}.  This suggests that 
the significant shape differences between
\hvq+\ST{} and \bbfourlsl{} 
may be due to parton-shower uncertainties 
in \hvq+\ST{} and,
given the huge size of shower effects in
\hvq+\ST, the observed agreement with 
\bbfourlsl{} is better than one may expect.
In this respect, one should also keep in mind that on-shell generators
like \hvq+\ST are not expected to provide an accurate description of
off-shell effects.

\begin{figure}
\centering
\offshellplot{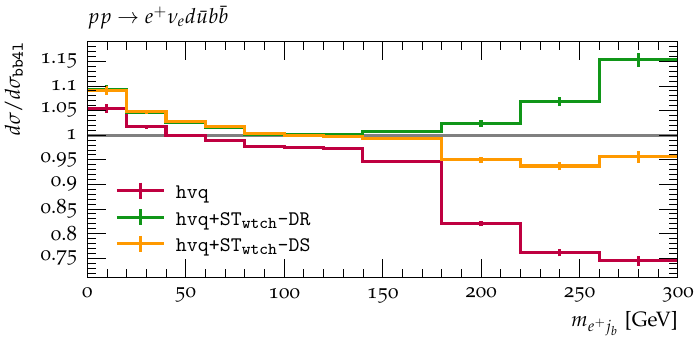}
\caption{Distribution in 
the invariant mass $m_{e^+j_b}$
of the positron and the 
hardest $b$~jet.
Same acceptance cuts, predictions and ratios as in \reffi{figOFF:1}.
}
\label{figOFF:3}
\end{figure}

Finally, the distribution in the invariant mass of the positron and 
the $b$~jet,
shown in \reffi{figOFF:3}, features an interesting pattern of 
off-shell effects. In the region above the
kinetic edge at $m_{e^+\,j_b}\simeq 152\,\GeV$,
the difference between \hvq{} and \bbfourlsl{} grows quite rapidly, reaching
$-25\%$ at 300\,\GeV. At the same time, the difference between
\ST{} predictions in DS and DR mode indicates large  
$t\bar t$--$\tW$ interference effects that grow up to 
20\% in the tail. In spite of these large off-shell effects,
at high $m_{e^+\,j_b}$ the
difference between \bbfourlsl{} and \hvq+\STDS{} never exceeds 
5\%. 
This is quite remarkable given that shower effects are around a factor two 
in the tail (see \reffi{figPS:3}).
In the region of small $m_{e^+\,j_b}$,
the DR and DS modes of \hvq+\ST{} agree, which indicates negligible $t\bar t$--$\tW$ 
interference effects. Nevertheless,
we observe a deviation between \bbfourlsl{} and \hvq+\ST{}, 
which approaches 10\% in the first bin.
This may be related 
to the very strong sensitivity of the $b$-jet mass to shower radiation in 
\hvq+\ST{} (see Figs.~\ref{figPS:2_W-_mb} and~\ref{figOFF:2_W-_mb}).

\section{Summary and conclusions}
\label{se:conclusions}

We have presented methodological improvements, new results and tools for the
NLOPS description of  off-shell $\ttbar+\tW$ production with dileptonic or semileptonic decays at the LHC.
In the dileptonic case, NLOPS predictions of this kind are
available through the \bbfourl{} generator in~\citere{Jezo:2016ujg}, which is
based on exact NLO matrix elements for the full $\ell^+ \nu_\ell
\,\ell^{\prime\,-} \bar{\nu}_{\ell'} b \bar{b}$ production process. This
provides, among others, a unified NLO description of off-shell $\ttbar$ and
$\tW$ production including their interference, as well as a NLO accurate
description of all involved decay subprocesses. 
Within the \bbfourl{} generator, the consistent matching of off-shell matrix
elements to the parton shower is guaranteed by the 
\POWHEGRES method, i.e.~the
resonance-aware extension of the \POWHEG method.
This technique is based on so-called resonance histories, which
provide a probabilistic decomposition of the full off-shell process into
separate production and decay subprocesses. 
Such resonance histories 
play a key role in order to guarantees the correct NLO shapes of 
resonances. Moreover, they
make it possible to associate NLO radiation to specific subprocesses
and to generate one \POWHEG emissions from each subprocess, i.e.~multiple
\POWHEG emissions per event.

In this paper we have presented a new version of the dileptonic \bbfourl
generator together with a new generator that provides the same kind of NLOPS
accuracy for the case of off-shell $\ttbar+\tW$ production with semileptonic
decays. These new dileptonic and semileptonic generators are both implemented in
the \bbfourl framework and are called \bbfourldl and \bbfourlsl,
respectively. Technically, semileptonic Les Houches events can be generated 
in an efficient way by re-processing pre-existing dileptonic events.

The definition of the physics content of the \bbfourlsl generator involves
nontrivial subtleties, which are due to the presence of various irreducible
backrounds that contribute to the same signature and 
interfere with $\ttbar+\tW$ production. 
Based on a detailed analysis of such irreducible backgrounds, we have
implemented a theoretical process definition, which guaratees that
\bbfourlsl includes all possible
$\ttbar+\tW$ contributions and the $\ttbar$--$\tW$ interference, while all 
relevant backgrounds---such as $t$- and $s$-channel single-top
production processes with extra jet radiation---can be separated and simluated with
independent tools.
In practice, we have adopted a process definition based on the same
LO and NLO Feynman diagrams that contribute to the dileptonic 
process, plus additional QCD correction effects associated with the
hadronic $W\to q\bar q'$ decay at NLO. For the latter we have adopted a
double-pole approximation that retains all Feynman diagrams with 
subtopologies of type $pp\to W^+W^-b\bar b$ plus extra radiation.

In addition to the new \bbfourlsl generator, we have presented two
significant methodological improvements that are implemented both 
within \bbfourldl and \bbfourlsl.

The first improvement deals with spurious effects of $\ord(\as^2)$ that
arise from the inconsistent perturbative treatment of decay widths at NLO.
In the context of the NWA, this problem is well known and can be solved by
means of a systematic perturbative expansion of terms of the form
$1/\Gamma_\NLO$. 
Off-shell NLO calculations, whose zero-width limit coincides with the NWA,
suffer from the same problem. However, a direct expansion of $1/\Gamma_\NLO$
terms is not possible in the off-shell case, since such terms are not
explicitly present. Nevertheless, as we have shown, the inverse-width
expansion can be generalised to NLO calculations for arbitrary off-shell 
processes and their matching
to parton showers in the \POWHEGRES framework.
As we pointed out, in the case of off-shell $\ttbar+\tW$ production
this inverse-width expansion plays an important role for the
consistent normalisation of total and differential NLO 
cross section. In fact, if left uncancelled, spurious terms can 
shift the cross section by up to $7\%$, which exceeds
the size of the overall off-shell and $\tW$ contributions,
and is similarly large as the complete NLO corrections to 
top decays.

The second methodological improvement concerns the definition of 
the \POWHEGRES resonance histories for 
off-shell $\ttbar+\tW$ production and decays.
While the original \bbfourl{} generator was effectively 
restricted to resonance histories of $\ttbar$ type, 
the new \bbfourlsl and \bbfourlsl generators 
implement also new histories of $\tW$ type. In addition, for both types of
resonance histories we have implemented realistic history probabilities
based on Born matrix elements.
This guarantees a reliable separation of events that are dominated by
$\ttbar$ or $\tW$ production, and 
a more consistent treatment of the hardest QCD radiation
for these different types of events. 
In the case of \bbfourldl, this improved \POWHEGRES implementation was
compared to the one of~\citere{Jezo:2016ujg} for several distributions,
finding only minor numerical differences. This suggests that \bbfourl
predictions  are largely free from uncertainties related to technical choices in
the definition of resonance histories.

In another technical study we have compared predictions of the \bbfourlsl
and \bbfourldl generators in a way that highlights those differences that
originate from QCD effects associated with the hadronic $W$~decay in
\bbfourlsl. In particular, we have isolated QCD effects stemming from jet
fragmentation and jet contamination, and we have analysed their impact on
various differential observables.

Finally, we have presented a tuned comparison of \bbfourlsl against 
the on-shell
\POWHEG generators of $\ttbar$ and $\tW$ production, i.e.~\hvq{} and~\ST.
In this context, in
spite of the deficits of the on-shell generators in terms of physical content
and perturbative accuracy, for several observables we have observed an
unexpected level of agreement. 
As we have demonstrated, this good agreement is due also to \Pythia's
built-in matrix-element corrections, which promote the formal accuracy for
top and $W$~decays within on-shell generators from LO to approximate NLO.
The most significant deviations between on-shell and off-shell generators
arise, among others, from the
$\ttbar$--$\tW$ interference, which is described with NLO 
accuracy within \bbfourl, while in the on-shell 
case it can only be estimated at LO by comparing different
version of the \ST{} generator.

The presented results deal only with selected kinematic distributions that
are typically studied in the context of $\ttbar$ production, 
while we did not investigate more exclusive phase-space regions
where off-shell and $\tW$ contributions can be strongly 
enhanced. In that case, and in general, \bbfourldl and \bbfourlsl
are expected to provide more reliable predictions as compared to 
the on-shell NLOPS generators.

The methodology developed in this paper 
is also applicable to off-shell $\ttbar+\tW$ production
with fully hadronic decays, and will make it possible
to generate fully inclusive event samples including dileptonic, semileptonic, and hadronic decays.

\FloatBarrier

\section*{Acknowledgments}
S.~P.~is grateful to Kirill Melnikov for 
useful discussions on the consistent perturbative treatment of
the top-quark width. 
T.~J.~would like to thank Silvia Ferrario Ravasio for numerous enlightening discussions.
This research was supported in part by the Swiss National Science Foundation~(SNF) under contracts BSCGI0-157722, PP00P2-153027, and CRSII2-160814, as well as from the Research Executive Agency of the European Union under the Grant Agreement PITN--GA--2012--316704~({\it HiggsTools}). J.~L.~was supported by the Science and Technology Research Council (STFC) under the Consolidated Grant ST/T00102X/1 and the STFC 
Ernest Rutherford Fellowship ST/S005048/1. 
The work of T.~J.~was supported by the SFB 1225 ``Isoquant,'' {project\nobreakdash-id} 273811115. 
We acknowledge the use of the DiRAC Cumulus HPC facility under Grant No. PPSP226. 

\appendix

\section{Real-to-Born kinematic mappings}
\label{app:mappings}

In this appendix we provide the explicit form of the kinematic
mappings that enter the new
matrix-element--based resonance histories presented in~\refse{se:newreshist}.
The definition of such histories is based on the weights~\refeq{eq:MEresB},
and their alternative versions~\refeq{eq:MEresC}--\refeq{eq:MEresD}. Such 
weights are defined in the Born phase space and are extended to the real-emission
phase space through \rtb kinematic mappings as specified 
in~\refeq{eq:RESdd}--\refeq{eq:RESde}.
The goal of these \rtb mappings, which transform real-emission events into Born events,
\begin{equation}
\label{eq:mappingform}
\phi_\rR \to \widetilde{\Phi}_{\rB,\labcoll}\,,
\end{equation}
is to ensure that the accuracy 
of the $\ttbar$ and $\tW$ Born probabilities, which 
is guaranteed by the usage of Born matrix elements,
is not spoiled by their extension to the real-emission phase space.
To this end, as discussed in~\refse{se:newreshist},
the mappings~\refeq{eq:mappingform}
are designed such as to preserve 
the key quantities that control the relative probabilities of 
$\ttbar$ and $\tW$ histories, namely the virtualities of the $W^+ b$ and $W^- \bar b$ systems,
as well as the transverse energies of the 
$b$~or $\bar b$~quarks that are potentially 
involved into initial-state 
$g\to b\bar b$ splittings in the case of $\tW$ production.

The real-to-Born mappings~\refeq{eq:mappingform}
are independent of the resonance history
and depend only on the collinear sector $\labcoll$.
More precisely, they only depend on 
whether the emitter is an initial-state (IS)
particle or a final-state (FS) $b$ or $\bar b$ quark. These different cases are 
discussed in \refapp{app:FSR} and \ref{app:ISR}.

\subsection{The FS case}
\label{app:FSR}

In the case of final-state radiation there are only two possible collinear sectors,
which correspond to $b\to b g$ or $\bar b \to \bar b g$ splittings
 and can be associated, respectively,
to an off-shell $t\to \ell^+ \nu_l b g$ or $\bar t \to \ell^-\bar\nu_l \bar b g$ decay.
In the following we refer to this decay as the radiative top decay, 
for which we use the charge-independent notation
\begin{equation}
t\to \ell\nu b g\,.
\end{equation}
In the FS case, the \rtb mapping~\refeq{eq:mappingform} acts only on the subset of momenta 
that arise from the
radiative top decay, which is identified according to the actual collinear sector.
Such momenta are labelled as
\begin{equation}
\label{eq:primemomA}
\{p^\prime_l, p^\prime_\nu, p^\prime_b, p_g^\prime\}
\,\subset\,
\Phi_\rR 
\,,
\end{equation}
while for the related momenta of the $bg$ system,
the off-shell $W$~boson and the off-shell top quark,
we use the symbols
\begin{equation}
\label{eq:primemomB}
p^\prime_W \,=\, p^\prime_l+p^\prime_\nu\,,
\qquad
p^\prime_{bg} \,=\, p^\prime_b+p^\prime_g\,,
\qquad
p^\prime_{t} \,=\, p^\prime_W+p^\prime_{bg}\,.
\end{equation}
The \rtb FS mapping turns the real-emission momenta~\refeq{eq:primemomA}--\refeq{eq:primemomB}
into the Born momenta
\begin{equation}
\label{eq:radmomA}\{p_l, p_\nu, p_b\}
\,\subset\,
\widetilde{\Phi}_{\rB,\labcoll}\,, 
\end{equation}
and
\begin{equation}
\label{eq:radmomB}
p_W \,=\, p_l+p_\nu\,,
\qquad
p_{t} \,=\, p_W+p_{b}\,.
\end{equation}
All external momenta that do not belong to the radiative top system are kept 
unchanged. Thus, also the off-shell momentum of the radiating top
itself remains unchanged, i.e.
\begin{equation}
\label{eq:topmomcons}
p_t \,=\, p^\prime_t\,.
\end{equation}
Within the radiating-top system, the \rtb mapping can be defined as a transformation 
\begin{equation}
\label{eq:redmapping}
p_{bg}^\prime \to p_b\,,
\end{equation}
which turns the off-shell real momenum 
of the $bg$ system into an on-shell $b$-quark Born momentum.
The three d.o.f.~of~$\vec p_b$
can be parametrised by its components 
in the beam direction, $p_{\parallel,b}$,
its magnitude in the transverse plane,
$|\vec p_{\perp,b}|$, and its angle $\phi_{bt}$ 
w.r.t.~the
parent top momentum in the transverse plane, which obeys
\begin{equation}
\cos\phi_{bt} \,=\, \frac{{\vec p}_{\perp,b}\cdot {\vec p}^{\,\prime}_{\perp,t}}{|{\vec p}_{\perp,b}|\,|{\vec p}^{\,\prime}_{\perp,t}|}\,.
\end{equation}
Once $\vec p_b$ is fixed, the $b$-quark energy is given by 
$E_b=\sqrt{|\vec p_b|^2+m_b^2}$, and for the
$W$-boson momentum in the Born phase space we have
\begin{equation}
\label{eq:Wmomentum}
p_W \,=\, p_t - p_b\,=\, p^\prime_t - p_b\,.
\end{equation}

As discussed in the following, the three d.o.f.~of $\vec p_b$ are fixed in a
way that preserves the $W$-boson virtuality, and, if possible, also the
transverse energy of the $bg$ system and its direction.
For the $W$~virtuality we always impose
\begin{eqnarray}
  p_W^{2} \,=\, p_W^{\prime 2}\,,
\label{eq:Wvirteq}
    \end{eqnarray}
by using
    \begin{eqnarray}
p_W^{2}\,=\, 
(p_t-p_b)^2\,=\, p_t^{2} + p_b^2 - 2 p_t\cdot p_b\,=\, p^{\prime 2}_t + m_b^2 - 2 p^\prime_t\cdot p_b\,.
    \end{eqnarray}
In terms of $|p_{\perp,b}|$, $p_{\parallel,b}$, and
$\cos\phi_{bt}$, this yields
    \begin{eqnarray}
p_t^{\prime 2} + m_b^2 - 2 \left[ E^\prime_t \sqrt{p^2_{\parallel,b}+|\vec p_{\perp,b}|^2+m_b^2} 
- |{\vec p}^{\,\prime}_{\perp,t}|| \vec p_{\perp,b}| 
\cos\phi_{bt} 
- p^\prime_{\parallel,t}\, p_{\parallel,b} \right]
&=&   p_W^{\prime 2}\,,
    \end{eqnarray}
which corresponds to a quadratic equation in $p_{\parallel,b}$, 
with solutions
    \begin{equation}
\label{eq:pbparall} 
p_{\parallel,b}
\,=\,      \frac{p^\prime_{\parallel,t}\,a\pm E^\prime_t\sqrt{\Delta
}}{c}\,, 
    \end{equation}
where $E^\prime_t=\sqrt{|{\vec p}^{\,\prime}_t|^2+m_t^2}$,
    \begin{eqnarray}
a &=&
p_t^{\prime 2} + m_b^2 - p_W^{\prime 2}
+ 2 |{\vec p}^{\,\prime}_{\perp,t}| |\vec p_{\perp,b}| 
\cos\phi_{bt}\,,\qquad
c \,=\, 2 E^2_t - p^2_{\parallel,t}\,,
\end{eqnarray}
and
\begin{eqnarray}
\label{eq:Deltadet}
\Delta &=&
a^2  - 2\,(|\vec p_{\perp,b}|^2+m_b^2)\,c\,.
\end{eqnarray}
Between the two solutions in~\refeq{eq:pbparall} 
we always pick the one where $p_{\parallel,b}$ is closer to $p^\prime_{\parallel,bg}$. 

Before fixing $p_{\parallel,b}$ according to~\refeq{eq:pbparall}, i.e.~in a way that preserves the $W$-boson
virtuality, 
when possible we fix the values of $|\vec p_{\perp,b}|$ and
$\cos\phi_{bt}$ in a way that preserves the transverse energy and the direction
of the $b$~quark. This is not always possible, since in some cases 
the latter two conditions lead to a negative determinant, $\Delta<0$,
in~\refeq{eq:pbparall}. For this reason we set 
$|\vec p_{\perp,b}|$ and $\cos\phi_{bt}$ according to different criteria in the 
following three cases.

\begin{itemize}

\item[(i)] By default we choose $|\vec p_{\perp,b}|$ in such a way that the 
resulting $b$-quark
transverse energy,
$E_{\perp,b}=\sqrt{|{\vec p}_{\perp,b}|^2 +m_b^2}$,
is equal to the one of the original bg system, i.e.
\begin{equation}
\label{eq:ETb}
E_{\perp,b}\,=\,E^\prime_{\perp,bg} \,=\, 
\sqrt{ |{\vec p}^{\,\prime}_{\perp,bg}|^2+(p^{\prime}_{bg})^2  }\,.
\end{equation}
Moreover, we also align the direction of the $b$~quark to the one of 
the original $bg$ system by imposing
\begin{equation}
\cos\phi_{bt} \,=\, \cos\phi_{bt}^\prime \,=\, \frac{{\vec p}^{\,\prime}_{\perp,bg}\cdot {\vec p}^{\,\prime}_{\perp,t}
}{|{\vec p}^{\,\prime}_{\perp,bg}|\,|{\vec p}^{\,\prime}_{\perp,t}|}\,.
\label{eq:bdirec}
\end{equation}
These choices are applied to all events for which they yield $\Delta>0$. Otherwise
we switch to case (ii) or (iii).

\item[(ii)] If $\Delta< 0$ in case~(i), then we try to impose only~\refeq{eq:ETb} 
and to fix $\cos\phi_{bt}$ in such a way that $\Delta=0$, which 
corresponds to a linear equation in $\cos\phi_{bt}$.
These choices are applied to all events for which 
they yield a physical direction, i.e.~$|\cos \phi_{bt}|\le 1$.
Otherwise we switch to case (iii).
 
\item[(iii)] If $\Delta=0$ yields unphysical solutions with $|\cos\phi_{bt}|>1$
in case~(ii), then we set $\cos\phi_{bt} = \pm 1$, picking the same sign as
the one of the unphysical solution,  
and we set $|\vec p_{\perp,b}|$ such that $\Delta=0$, which is always possible.

\end{itemize} 

Once the $b$-quark momentum is fixed as described above, the
$W$-momentum is determined as indicated in~\refeq{eq:Wmomentum}.
Finally, the momenta $\{p_\ell, p_\nu\}$
of the $W$-decay products, 
are obtained from the related momenta $\{p^\prime_\ell, p^\prime_\nu\}$
by applying a sequence of two boosts that transform
$p^\prime_W$ to its rest frame and subsequently turn it into 
$p_W$.

Note that in the strict soft limit the result of this mapping
coincides with the underlying Born event.

\subsection{The IS case}
\label{app:ISR}

\newcommand{\momqr}[2]{{q}^{\prime #1}_{#2}}
\newcommand{\momqb}[2]{{q}^{ks
#1}_{#2}}
\newcommand{\momqbT}[2]{{k}^{\prime #1}_{#2}}
\newcommand{\mombtil}[2]{{\tilde{p}}^{#1}_{#2}}
\newcommand{\vecmomr}[2]{{\vec p}^{\,\prime #1}_{#2}}
\newcommand{\momr}[2]{p^{\prime #1}_{#2}}
\newcommand{\momb}[2]{p^{#1}_{#2}}

In the IS case we have implemented \rtb mappings that 
preserve the virtuality of all relevant $W$~bosons and $Wb$~pairs 
together with the invariant mass of the virtual $b$~quark 
that 
is involved in the initial-state
$g\to b\bar b$ splitting in the case of
$pp\to tWb$ production. To this end, as detailed below,
we have determined the most likely parton-level topology of type
\begin{equation}
\label{eq:ISmapsA}
IJ\rightarrow t W b + X\,,
\end{equation}
where $I$ and $J$ are the initial-state partons,
$X$ is the parton emitted via ISR,
$t$ denotes a generic (off-shell) top or anti-top quark
that undergoes a three-body decay,
and $Wb$ is the off-shell  $W^+b$ or $W^-\bar b$
system with the highest probability of 
being the non-resonant $Wb$ system 
resulting from $tWb$ production.
In the following,  the bottom quark or anti-quark 
$b$ that belongs to the ``non-resonant'' $Wb$ system is 
referred to as the $b$~spectator, since it corresponds to the
$b$~(anti-)quark with the highest probability of stemming from
the initial-state
$g\to b\bar b$ splitting within 
a $Wtb$ production subprocess.

The general form of our \rtb mappings is
\begin{equation}
\label{eq:ISmapsB}
\momr{}{I},\,
\momr{}{J},\,
\momr{}{t},\,
\momr{}{X}\,
\;\rightarrow\;
\momb{}{I},\,
\momb{}{J},\,
\momb{}{t}\,,
\end{equation}
i.e.~the momenta of the non-resonant $Wb$ system are kept unchanged.
In order to determine the most likely topology of type
\refeq{eq:ISmapsA} we consider eight different 
cases---and corresponding 
probabilities---that
 arise from all possible combinations of the following options.

\begin{itemize}
\item[(i)] The initial-state leg that emits ISR\footnote{Note that 
in the POWHEG method the two initial state singular regions are treated
together.}
 is 
$E=I$ or $E=J$. The corresponding probability is chosen as
\begin{equation}
P^{(\mathrm{em})}_E \,=\, \frac{1}{(1-\cos\phi_E)^2}
\qquad\mbox{with}\qquad
\cos\phi_E\,=\,
\frac{\vecmomr{}{E}\cdot\vecmomr{}{X}}{|\vecmomr{}{E}||\vecmomr{}{X}|}\,.
\end{equation}
 
\item[(ii)] The initial-state leg that undergoes a
$g\to b\bar b$ splitting is $S=I$ or $S=J$. The corresponding 
probability is chosen as
\begin{equation}
P^{(g\to \bbbar)}_S \,=\, \frac{1}{\momqr{2}{S} - m_b^2}\,,
\end{equation}
where $q_{S}$ is the momentum of the 
virtual $b$ quark that results from the $g\to b\bar b$ splitting, i.e.
\begin{equation}
\momqr{}{S}  \,=\, 
\begin{cases}
\momr{}{S} - \momr{}{b} &\; \mbox{for}\;S\neq E\,,\\
\momr{}{S} - \momr{}{b} - \momr{}{X} &\; \mbox{for}\;S = E\,.\\
\end{cases}
\end{equation}

\item[(iii)] The non-resonant $Wb$ system corresponds either to the
$W^+b$ or to the $W^-\bar b$ system. In order to discriminate between these
two options we make use of the probability
\begin{equation}
P^{(\mathrm{top})}_{Wb} \,=\, \frac{m_t^2}{
\left[(\momr{}{W}+\momr{}{b})^2-m_t^2\right]^2
+\Gamma_t^2m_t^2
}\,.
\end{equation}
\end{itemize}
The most likely topology~\refeq{eq:ISmapsB} is selected as the one that 
maximises the combined ``probability''
\begin{equation}
P\,=\, \frac{P^{(\mathrm{em})}_E\,P^{(g\to \bbbar)}_S}{P^{(\mathrm{top})}_{Wb}}.
\end{equation}
 
Before we proceed with the definition of our kinematic mapping let us 
introduce the auxiliary off-shell momentum
\begin{eqnarray}
\momqbT{}{E}  \,&=&\,  \momr{}{E} - \momr{}{X}\,,
\end{eqnarray}
where $\momr{}{E}$ is the momentum of the initial-state leg that radiates,
and the auxiliary on-shell momentum
\begin{equation} 
\mombtil{}{E} = x \momr{}{!E} + \momqbT{}{E} \qquad \text{ with } \qquad 
  x = - \frac{\momqbT{2}{E}}{(\momr{}{!E} + \momqbT{}{E})^2 -\momqbT{2}{E}}\,,
\end{equation}
where $!E$ corresponds to the initial-state leg that does not radiate, 
i.e.~$!E=J(I)$ for $E=I(J)$. 
By construction $\mombtil{2}{E} = 0$. 
For the definition of the real-to-Born mappings we consider the following two cases.

\begin{itemize}
\item[(A)] If the ISR emitter and the leg that undergoes the 
$g\to b\bar b$ splitting are different, i.e.~$E\neq S$,
than the form~\refeq{eq:ISmapsB}
of the mapping automatically guarantees that
all external particles that are connected to the
$g \to b\bar{b}$ splitting are kept unchanged.
In this case, 
$\momr{}{E}$ and $\momr{}{t}$ are turned into the Born momenta 
\begin{eqnarray}
\momb{}{E} &=& y_A \mombtil{}{E} \,,\qquad
\momb{}{t} \,=\, \momr{}{t} + y_A\mombtil{}{E} - \momqbT{}{E} \,,
\end{eqnarray}
where 
\begin{equation}
  y_A = - \frac{\momqbT{}{E} \cdot \momqbT{}{E} - 2 \momr{}{t} \cdot \momqbT{}{E}}{2 \left(
  \momr{}{t} \cdot \mombtil{}{E} - \momqbT{}{E} \cdot \mombtil{}{E} \right)}
\end{equation}
guarantees that the virtuality of the resonant top quark is preserved,
i.e.~$\momb{2}{t} = \momr{2}{t}$.

\item[(B)] If the ISR emitter is instead the same leg that undergoes the 
$g\to b\bar b$ splitting, i.e.~$E = S$,
we again manipulate the initial-state leg $E$ and the 
momentum of the resonant top quark, which 
are turned into  
\begin{eqnarray}
\momb{}{E} &=& y_B \mombtil{}{E} \,,\qquad
\momb{}{t} \,=\,  \momr{}{t} + (y_B-1) \momqbT{}{E} - z \momr{}{!E} \,.
\end{eqnarray}
Here, in order to preserve the invariant mass of the virtual $b$ quark 
in the  $g\to b\bar b$ splitting, i.e.~$q^2_S=(p_S-p_b)^2=\momqr{2}{S}$,
and the virtuality of the resonant top quark, i.e.~$\momb{2}{t} = \momr{2}{t}$, we choose
\begin{eqnarray}
  y_B & = & -\frac{ \momqbT{2}{E} - 2 \momr{}{!b} \cdot \momqbT{}{E} }{2 \momr{}{!b} \cdot
  \momqbT{}{E}} \ ,\qquad
  z \, = \, \frac{(y_B - 1) \momqbT{}{E} \cdot \left(2 \momr{}{t} + (y_B - 1) \momqbT{}{E}
  \right)}{2 \momr{}{!E} \cdot \left(  \momr{}{t} + ( y_B - 1 ) \momqbT{}{E} \right) }\ ,  
\end{eqnarray}
where $p_{!b}$ is the momentum of the $b$~quark stemming form the decay of the 
resonant top quark.
\end{itemize}
Finally, in both cases we boost the decay products of the resonant top quark 
according to the $\momr{}{t}\to \momb{}{t}$ transformation.

In the soft limit this mapping  reproduces
the underlying Born kinematics with the soft radiation removed, and in the
collinear limit it reproduces it with the momentum of the collinear radiation
subtracted from the momentum of the emitter.

\section{The \bbfourlslInTitle{} definition of semileptonic 
\boldmath$t\bar t+tW$  production and decay}
\label{app:approximation}

As discussed in \refse{see:bbfourlsl}, the physics content of the
\bbfourlsl{} generator corresponds to the ingredients of the full
process~\refeq{eq:sl-inclusive_signature} that satisfy the selection
criteria S1--S3 (see \refse{sect:bb4lsldef}).
This selection captures all possible contributions associated with
off-shell $\ttbar+tW$ production and decays to semileptonic 
final states. Moreover, it is in one-to-one
correspondence with the ingredients of the dileptonic 
\bbfourldl{} generator, with the addition
of QCD effects in $W\to q\bar q$ decays.
In the following, we present technical studies that 
justify this process definition at 
LO and NLO. In doing so, we focus on the contributions that satisfy
S1, i.e.~the contributions of order $\as^2\alpha^4$ and
$\as^3\alpha^4$,
and we analyse the effect of the selection steps S2 and S3.

The studies presented in this appendix were done
with {\tt Sherpa}~\cite{Sherpa:2019gpd} for the case of
$\ell^{-} \bar \nu_\ell j j b\bar b$ production, while the 
charge-conjugated process yields 
qualitative identical results.
All predictions were obtained 
with the setup of~\refse{sec:setup}, and requiring two light jets
plus either one or more than one 
$B$~jet with $R=0.5$, $\pT > 25\,$GeV and
$\vert\eta\vert < 2.5$, 
as well as 
a charged lepton with $\pT > 25\,$GeV 
and $\vert\eta\vert < 2.5$, and no cuts on the neutrino.

\subsection{Effect of the S2 and S3 selections at LO}
\label{sec:LOapproximation}

\newcommand{\appBplot}[2]{
\begin{subfigure}{0.45\linewidth}
\includegraphics[width=\textwidth]{./figures/loplots/#1}
\caption{\phantom{\hspace{-12mm}}}
\label{#2}
\end{subfigure}}

\begin{figure}[t]
	\begin{center}
\appBplot{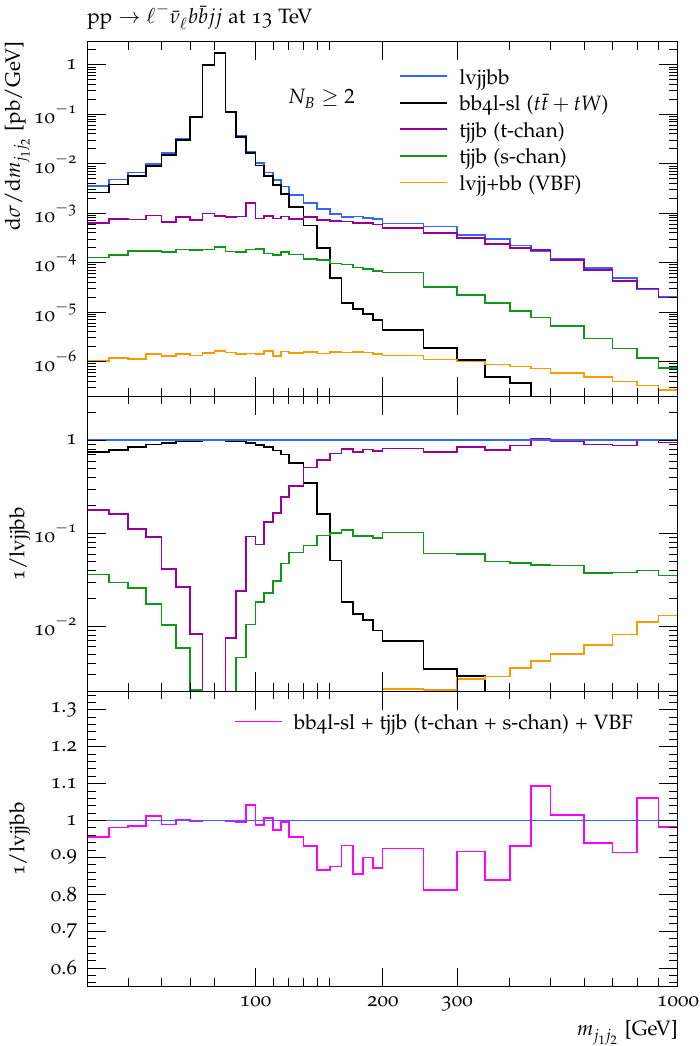}{mjj_BB_full}
\appBplot{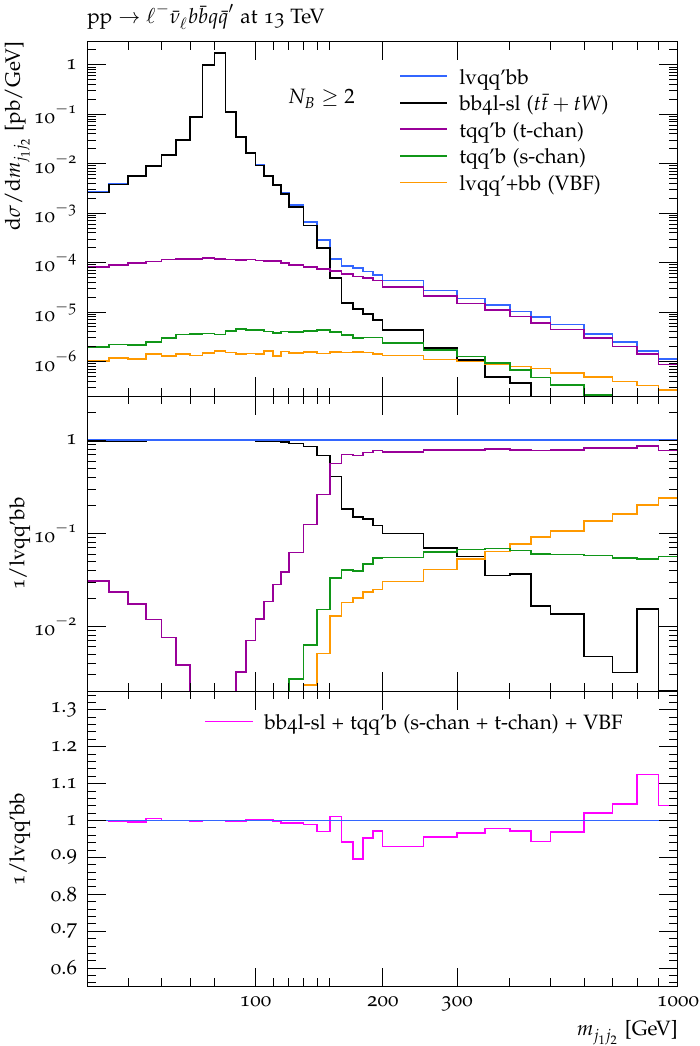}{mjj_BB}
	\end{center}
	\caption{Contributions of $\ord(\as^2\alpha^4)$ to the
distribution in the invariant mass of the two light jets 
for off-shell 
$\ell^{-} \bar\nu_\ell j j b\bar b$ (\subref{mjj_BB_full})
and 
$\ell^{-} \bar\nu_\ell q \bar q' b\bar b$ (\subref{mjj_BB})
production
in the phase space with $N_B\ge 2$ \mbox{$B$~jets}.
In both cases, exact off-shell predictions are compared to the
individual contributions from $t\bar t +tW$ production 
(corresponding to \bbfourlsl{} at LO), 
$t$- and $s$-channel single-top production, and
$Wjj+b\bar b$ production via VBF.
The mid panel shows ratios of individual contributions wrt 
the full off-shell process at hand,
while the lower panel shows the same ratio for the
additive combination of the $t\bar t+tW$ plus
single-top plus VBF production subprocesses.
}
\label{fig:lo_comparison_2b_mjj}
\end{figure}

In \reffi{fig:lo_comparison_2b_mjj} we present the distribution in the
invariant mass of the two light jets in the phase space with 
two or more $B$-jets. The plotted $m_{j_1j_2}$ range
includes a wide region around the 
$W\to jj$ resonance and
extends up to dijet masses of 1\,TeV.

The predictions in \reffi{fig:lo_comparison_2b_mjj}\subref{mjj_BB_full}
correspond to the outcome of the S1~selection, 
i.e.~the off-shell $\ell^{-} \bar \nu_\ell j j b\bar b$ 
production process at 
$\ord(\as^2\alpha^4)$, while in 
\reffi{fig:lo_comparison_2b_mjj}\subref{mjj_BB} we show the
outcome of the S1+S2 selection steps, which corresponds to
off-shell $\ell^{-} \bar \nu_\ell q \bar q' b\bar b$ production
at $\ord(\as^2\alpha^4)$ with
a $q \bar q'$-pair consistent with a $W^+$ decay, 
i.e.~$u \bar d$ or $c \bar s$. 
In Figs.~\ref{fig:lo_comparison_2b_mjj}\subref{mjj_BB_full}--\subref{mjj_BB} 
we also show the 
contributions of the most significant subprocesses,
namely
off-shell $t\bar t+tW$ production, $s$-~and $t$-channel 
single-top production, and $Wjj$ production via VBF 
in association with a $b\bar b$ pair 
(see \refta{tab:pertorders} and the discussion in \refse{sect:bb4lsldef}).
Here the $t\bar t+tW$ contribution should be understood 
as the  physics content of \bbfourlsl{} at LO, and is equivalent to the
outcome of the S1+S2+S3 selection steps.
This contribution is the same in  
\reffi{fig:lo_comparison_2b_mjj}\subref{mjj_BB_full}
and~\ref{fig:lo_comparison_2b_mjj}\subref{mjj_BB}, 
while in~\reffi{fig:lo_comparison_2b_mjj}\subref{mjj_BB}
the single-top and VBF contributions have a smaller cross section
due to the requirement of a $q\bar q'$ pair in the final state.

The absolute predictions in the upper panels of
Figs.~\ref{fig:lo_comparison_2b_mjj}\subref{mjj_BB_full}--\subref{mjj_BB} 
demonstrate that the 
total cross section is strongly dominated by the $W\to jj$ resonance
and is very well approximated by off-shell $t\bar t+tW$ production
and decay, i.e.~\bbfourlsl, which is the only subprocess 
with a $W\to jj$ resonance.
When a $q\bar q'$ pair is required
(Fig.~\ref{fig:lo_comparison_2b_mjj}\subref{mjj_BB})
the \bbfourlsl contribution provides an almost perfect approximation in 
a broad region around the resonance, 
with sub-percent precision for 
$\vert m_{j_1j_2}- m_W \vert < 30\,$GeV.

Extra contributions become significant only 
for $m_{j_1j_2} \gtrsim 130\,$GeV 
and can exceed \bbfourlsl by more than two 
orders of magnitude in the off-shell tail. 
In this highly off-shell regime the
$pp\to \ell^{-} \bar \nu_\ell q \bar q' b\bar b$ 
process is
dominated by $t$-channel single-top production, 
and receives sizeable contributions also from $s$-channel single-top
production and, at very large dijet masses,
also from the VBF subprocess.
As demonstrated in the lower panel 
of~\reffi{fig:lo_comparison_2b_mjj}\subref{mjj_BB},
the incoherent sum of the \bbfourlsl{}
plus the single-top and the VBF subprocesses
agrees with the exact description based
on off-shell $\ell^{-} \bar \nu_\ell q \bar q' b\bar b$ 
matrix elements
at the sub-percent level for $m_{j_1j_2} <
150\,$GeV, and at the level of $5\%$ in the entire 
plotted range above 150\,GeV. 
Extending this analysis of the $m_{j_1j_2}$ distribution 
to the phase space with $N_B=1$ $B$-jets
(not show here)  
we found a qualitatively and quantitatively 
similar behaviour 
as in the $N_B\ge 2$ phase space.

\begin{figure}
	\begin{center}
\appBplot{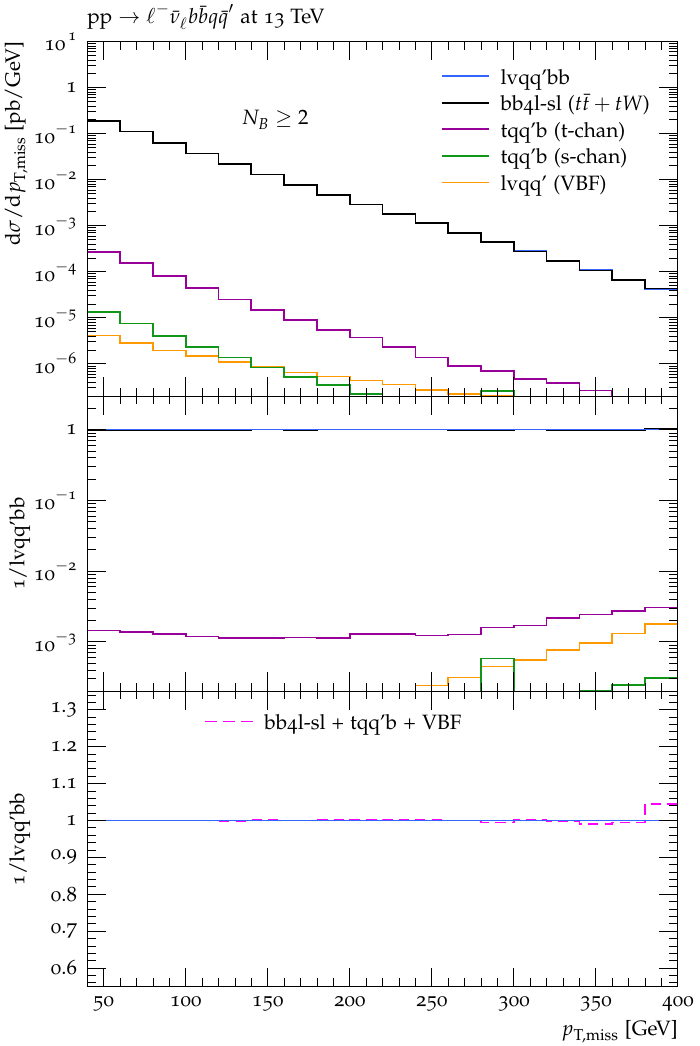}{MET_BB}
\appBplot{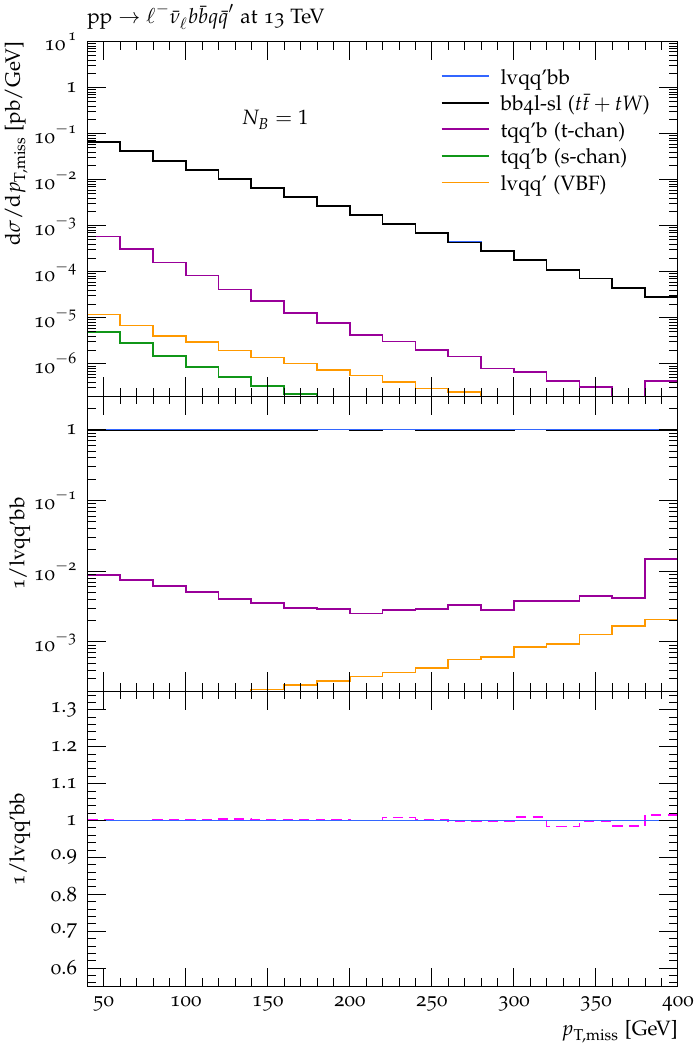}{MET_B}\\
\appBplot{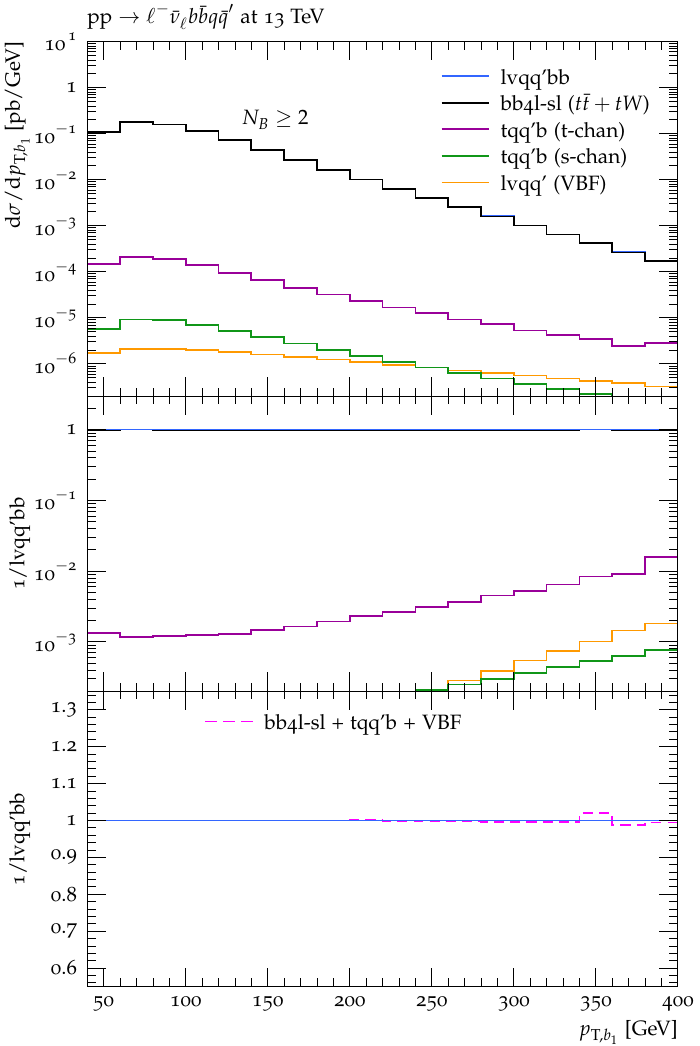}{PTB1_BB}
\appBplot{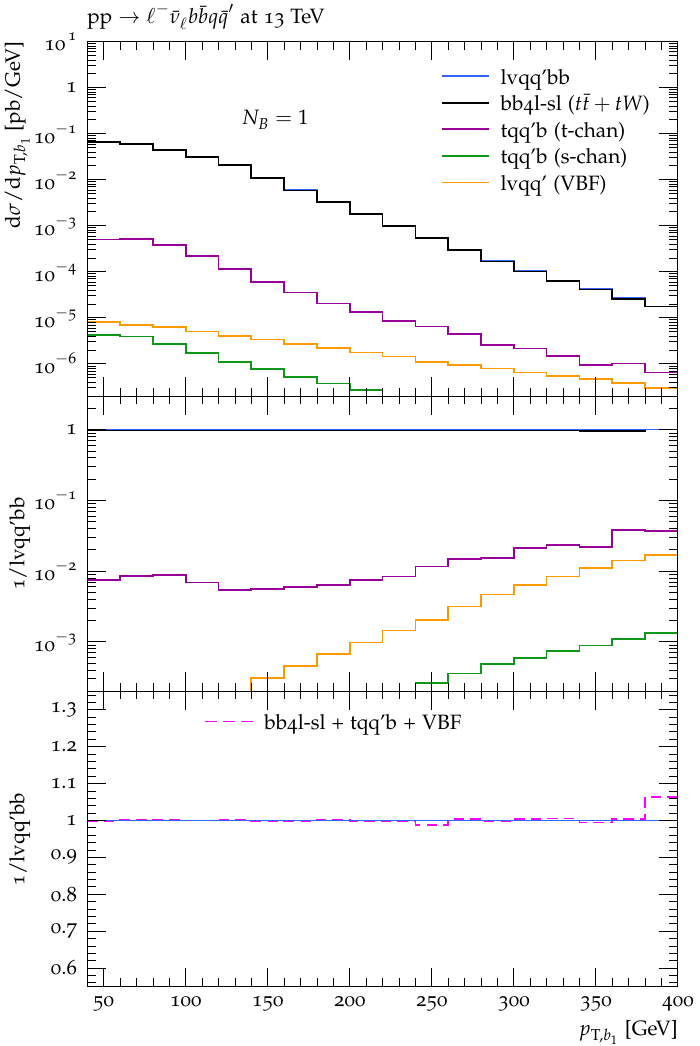}{PTB1_B}
\vspace{-5mm}
	\end{center}
	\caption{Off-shell $\ell^{-} \bar\nu_\ell q \bar q' b\bar b$ 
production at $\ord(\as^2\alpha^4)$ 
in the phase space with $N_B\ge 2$
(left) and $N_B= 1$ (right)
$B$-jets:
distributions in the missing $p_\rT$ (\subref{MET_BB}--\subref{MET_B})
and in the $p_\rT$ of the leading $B$-jet
(\subref{PTB1_BB}--\subref{PTB1_B}).
Same predictions and ratios as 
in~\reffi{fig:lo_comparison_2b_mjj}\subref{mjj_BB}.
}
\label{fig:NB21_comparison_1}
\end{figure}

\begin{figure}
	\begin{center}
\appBplot{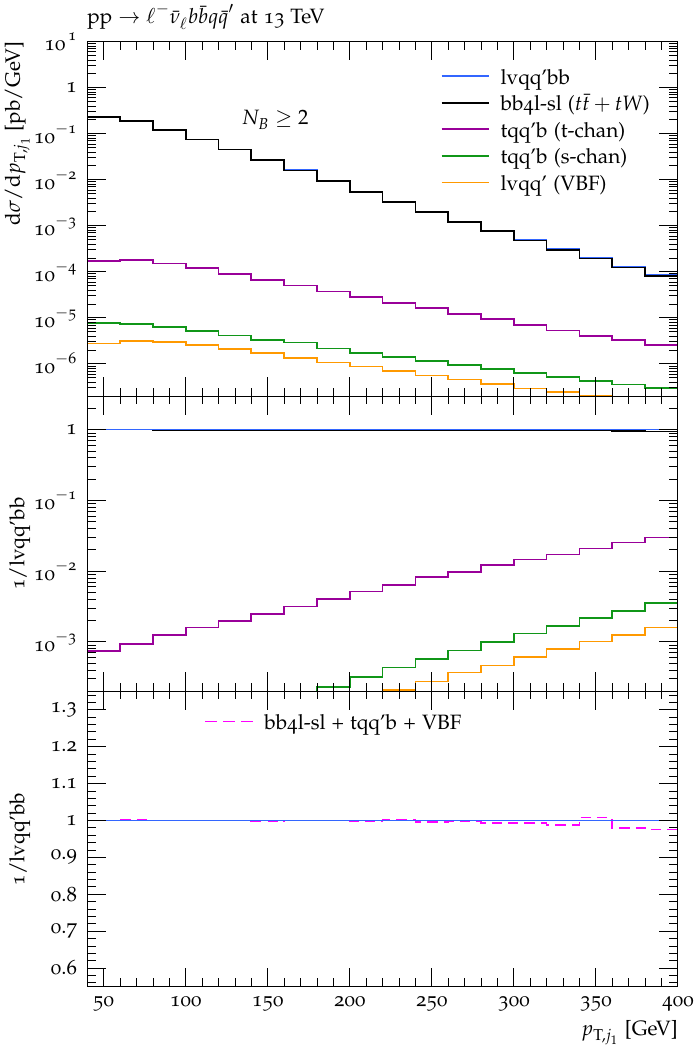}{PTJ1_BB}
\appBplot{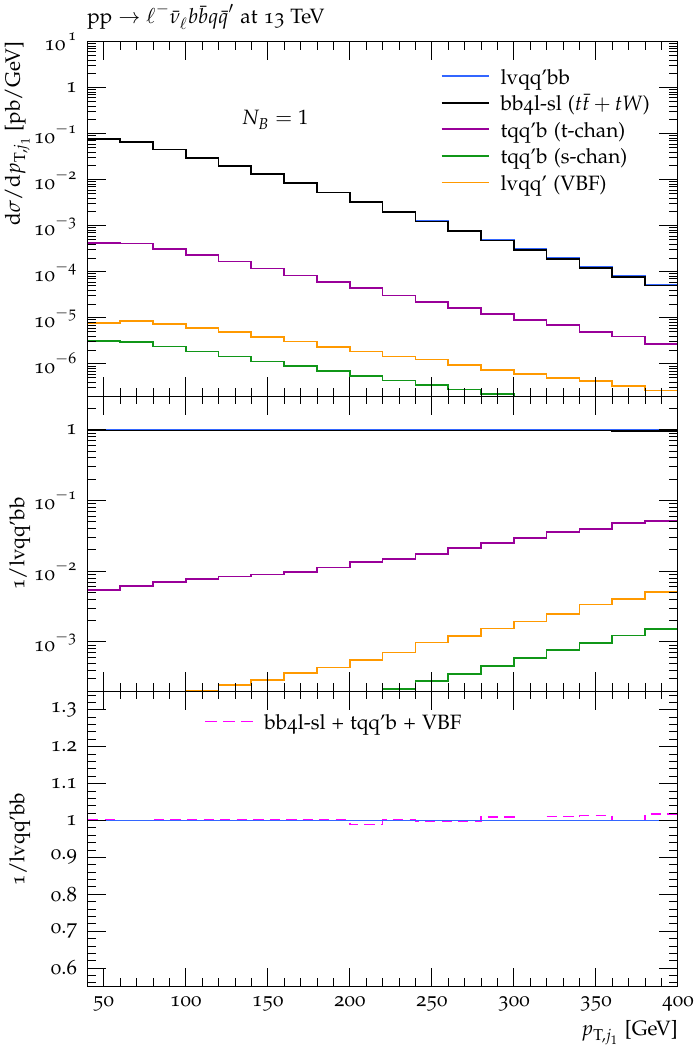}{PTJ1_B}\\
\appBplot{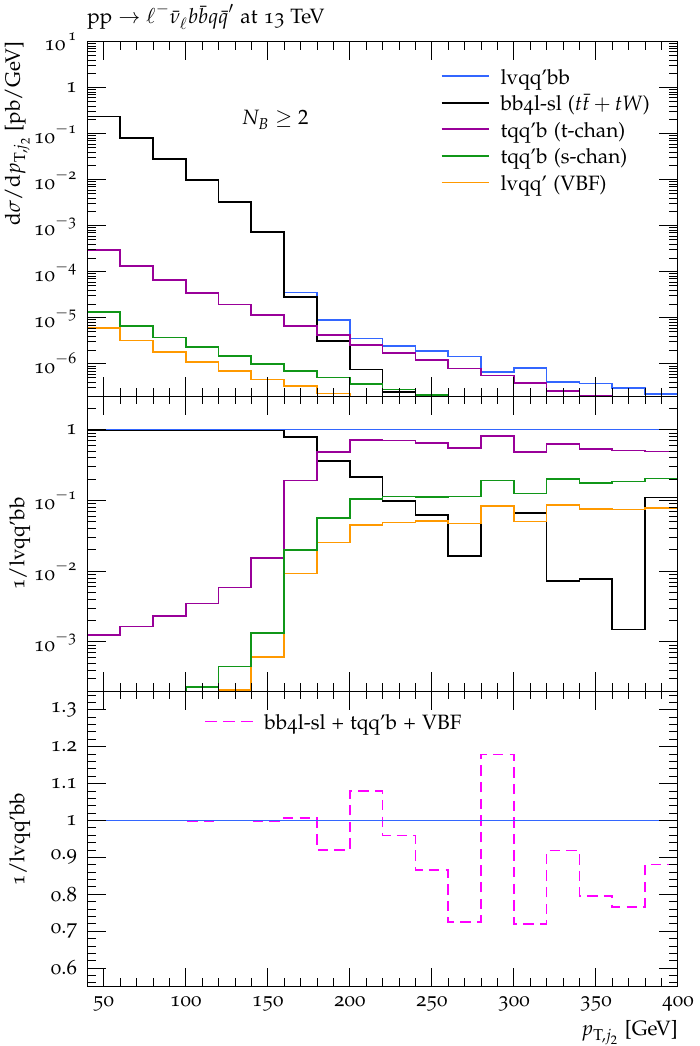}{PTJ2_BB}
\appBplot{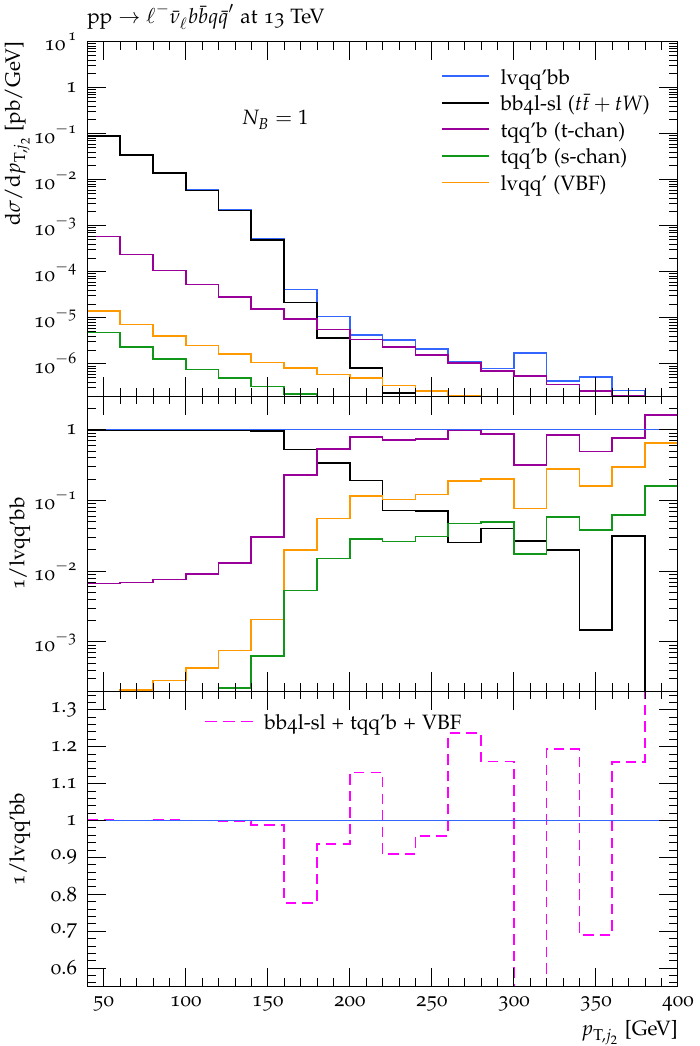}{PTJ2_B}
\vspace{-4mm}
	\end{center}
	\caption{Off-shell $\ell^{-} \bar\nu_\ell q \bar q' b\bar b$ 
production at $\ord(\as^2\alpha^4)$:
distributions in the $p_\rT$ of the first 
(\subref{PTJ1_BB}--\subref{PTJ1_B})
and of the second 
(\subref{PTJ2_BB}--\subref{PTJ2_B}) light jet.
Same predictions and ratios 
as in \reffi{fig:NB21_comparison_1}.
}
\label{fig:NB21_comparison_2}
\end{figure}

Focussing on the \bbtwoltwoq signature,
in \reffis{fig:NB21_comparison_1}{fig:NB21_comparison_2} 
the above comparisons are extended to four different
transverse-momentum distribution in the
phase-space regions with two light jets plus
$N_B\ge 2$ or $N_B=1$ $B$-jets.

For the distributions in the 
missing $p_\rT$ 
(\reffi{fig:NB21_comparison_1}\subref{MET_BB}--\subref{MET_B}), 
in the $p_\rT$ of the leading $B$-jet 
(\reffi{fig:NB21_comparison_1}\subref{PTB1_BB}--\subref{PTB1_B}),
and in the $p_\rT$ of the leading light jet
(\reffi{fig:NB21_comparison_2}\subref{PTJ1_BB}--\subref{PTJ1_B}),
the relative weights of the various subprocesses 
vary more smoothly as compared to the case of the $m_{j_1j_2}$
distribution.
At moderate transverse momenta, the hierarchy of the 
different contributions is similar as in the
$W$-resonance region of \reffi{fig:lo_comparison_2b_mjj},
while in the high-$p_\rT$ tails the relative importance of the 
single-top and VBF subprocesses can increase, depending on the observable.
For these three $p_\rT$ distributions we observe a fairly similar behaviour 
in the $N_B\ge 2$ and $N_B=1$ regions, and
we find that the \bbfourlsl contribution 
agrees with the full $\ell^{-} \bar \nu_\ell q \bar q' b\bar b$ 
prediction
at the level of one percent or better,
with the only exception of the tail of the
leading-jet $p_\rT$ distribution, where 
above 250\,GeV the weight of the
$t$-channel single-top subprocess can 
amount to several percent.

The distribution in the $p_\rT$ of the
second light jet, shown in 
\reffi{fig:NB21_comparison_2}\subref{PTJ2_BB}--\subref{PTJ2_B}, 
reveals a somewhat different picture.
In the region with $p_{\rT,j_2} < 150\,$GeV, 
which contains the bulk of the 
cross section,
the \bbfourlsl contribution still
agrees with the fully off-shell description at the 
sub-percent level.
However, at $p_{\rT, j_2} \simeq 150\,$GeV
the single-top and VBF contributions
become rapidly more important,
and at $p_{\rT, j_2} > 200\,$GeV 
they dominate the entire cross section.
We observe that this transition is due to the abrupt
suppression of $t\bar t+tW$ production, 
which goes down by one to two orders of magnitude
within two bins.
For this reason, the region above 200\,GeV suffers from 
limited Monte Carlo statistics.

We investigated a large number of
further kinematic distributions and we observed 
a consistent picture throughout: whenever single-top and VBF processes are 
suppressed with respect to $\ttbar+tW$ production, 
the \bbfourlsl{} contribution 
agrees with the complete off-shell calculation
at or below the percent level.
Moreover generally, the incoherent 
superposition of $\ttbar+tW$ production 
with the single-top and VBF subprocesses
provides an excellent approximation in the full phase space.

These observations justify the treatment of
off-shell $\ttbar+tW$ production 
with semileptonic decays, as defined in \bbfourlsl,
as a separate process, i.e.~neglecting 
interferences with the other ingredients of 
$pp\to \ell^{-} \bar\nu_\ell q \bar q' b\bar b$. 

\subsection{Effect of the S3 selection at NLO}
\label{sec:NLOapproximation}

In the order to investigate the consistency of 
the \bbfourlsl{} process definition beyond LO,
we have extended the studies of~\refse{sec:LOapproximation} 
to the real-emission process
\beq
\label{eq:sl-qqpj_signature}
pp \to \ell^{-} \bar \nu_{\ell} q \bar q' b\bar b j\,.
\eeq
In order to avoid IR divergences,
for the additional light jet
we require $p_{{\rm T},j} > p_{{\rm T},{\rm min}}$,
and to minimise the bias due to this technical cut 
we compare three different variants with 
$p_{{\rm T},{\rm min}} = 20\,\GeV, 5\,\GeV,$ and 1\,GeV.

The above process corresponds to the real-emission contribution to the 
outcome of the S1+S2 selection steps, and in the following 
we investigate the effect of the additional S3 selection,
which isolates the $\ttbar+tW$ contribution as define in
\bbfourlsl.
In \reffis{fig:nlo_comparison_A}{fig:nlo_comparison_C} we
present the same set of distributions as in~\refse{sec:LOapproximation}
in the $N_B \geq 2$ and $N_B = 1$ regions.
Exact off-shell predictions
for the process~(\ref{eq:sl-qqpj_signature}) 
are compared to the contributions of the most relevant subprocesses, which are
the same as in~\refse{sec:LOapproximation}
with the exception of the VBF subprocess, which turns out to be completely 
negligible and is thus omitted.
In the case of the $\ttbar+tW$ subprocess, i.e.~\bbfourlsl,
we also show the effect of disabling 
QCD radiation in $W\to q\bar q'$ decays.

\newcommand{\appBBplot}[2]{
\begin{subfigure}{0.42\linewidth}
\includegraphics[width=\textwidth]{./figures/loplots/#1}
\caption{\phantom{\hspace{-12mm}}}
\label{#2}
\end{subfigure}}

\begin{figure}
	\begin{center}
\appBBplot{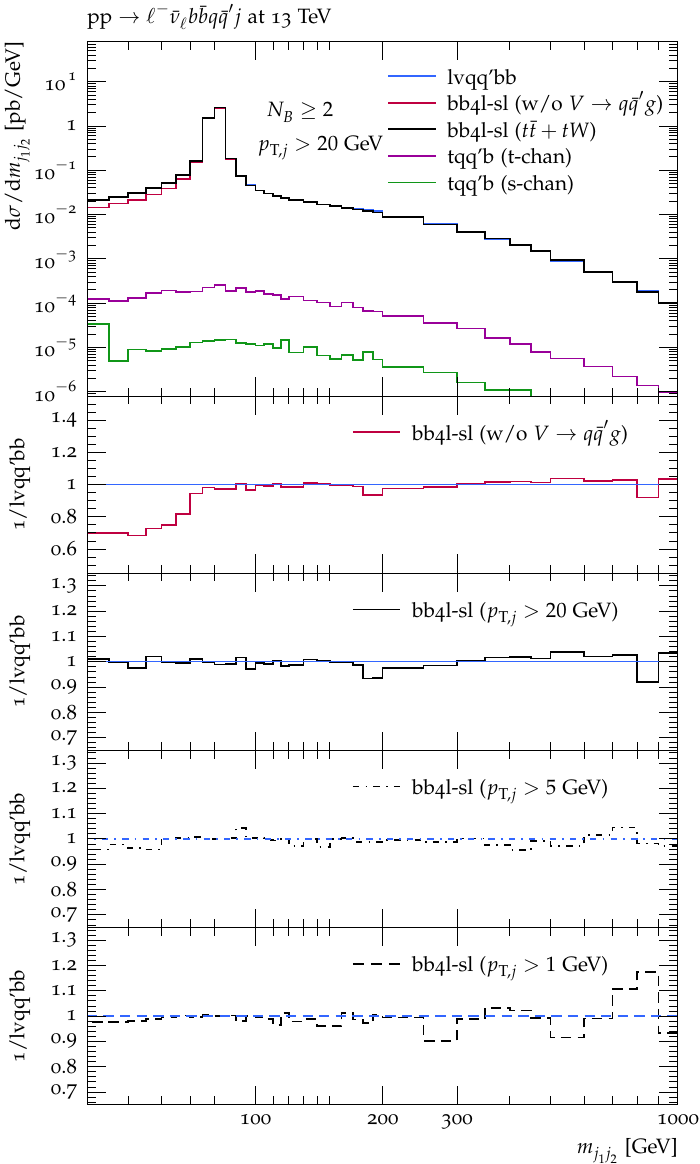}{MJJ_BBJ}
\appBBplot{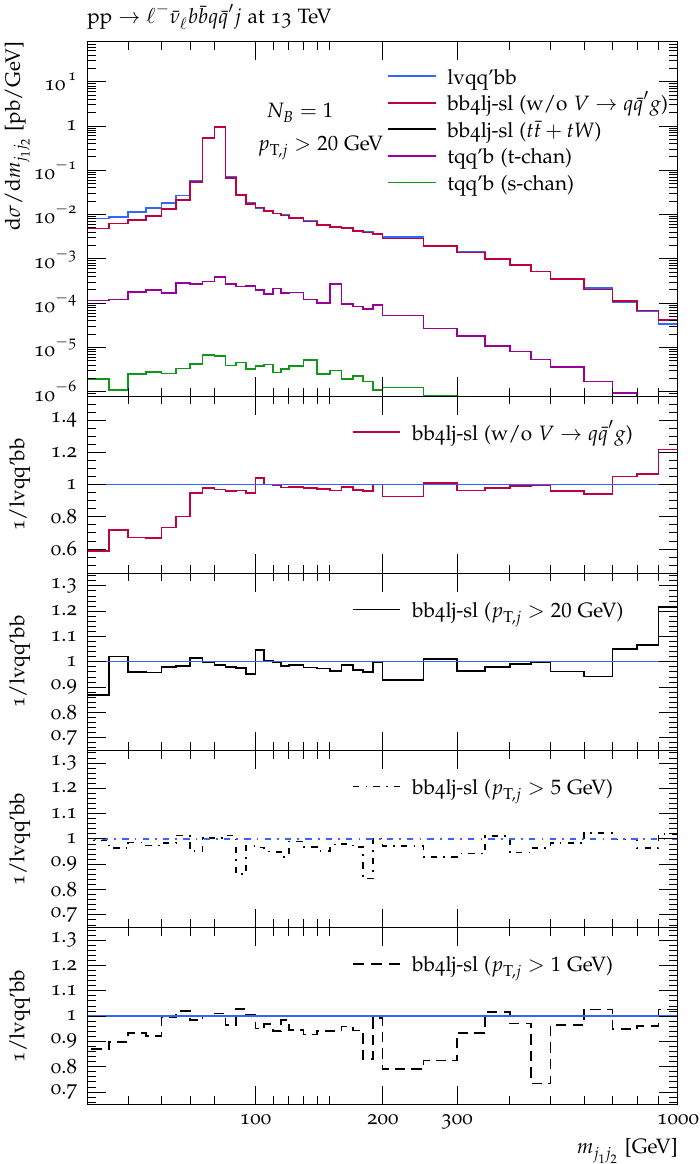}{MJJ_BJ}
\vspace{-4mm}
	\end{center}
	\caption{Distribution in the 
dijet invariant mass of the first two 
light jets for off-shell $\ell^{-} \bar \nu_\ell q \bar q' b\bar b+$jet 
production at $\ord(\as^3\alpha^4)$ 
in the phase space with 
$N_B\ge 2$ (\subref{MJJ_BBJ}) and
$N_B=1$ (\subref{MJJ_BJ}) $B$-jets.
The precise definition of the dijet mass
is given in the main text.
Complete off-shell predictions are compared to the contributions
of the most relevant subprocesses:
$\ttbar+tW$ production (i.e.~\bbfourlsl)
including (black) or excluding (red) radiation 
in the $W\to q\bar q'$ decay, and $t$-channel (purple) plus $s$-channel
(green) single-top production.
The upper panels show 
absolute predictions with $p_{\rT,{\rm min}}=20$\,GeV,
while the three lowest panels show the ratio of 
\bbfourlsl{} wrt the complete off-shell
process for three different values 
of $p_{\rT,{\rm min}}$.
The second panel shows the same ratio 
for \bbfourlsl{} without radiation 
in the $W\to q\bar q'$ decay
and with $p_{\rT,{\rm min}}=20$\,GeV.
}
\label{fig:nlo_comparison_A}
\end{figure}

\begin{figure}
  \begin{center}
\appBBplot{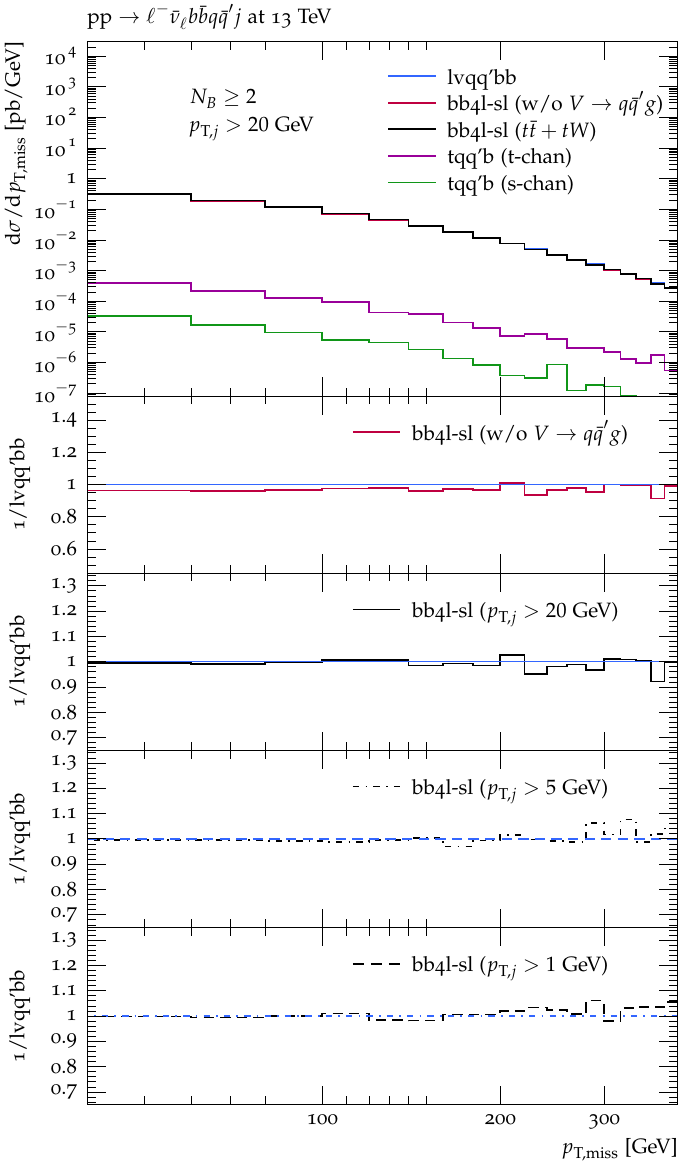}{MET_BBJ}
\appBBplot{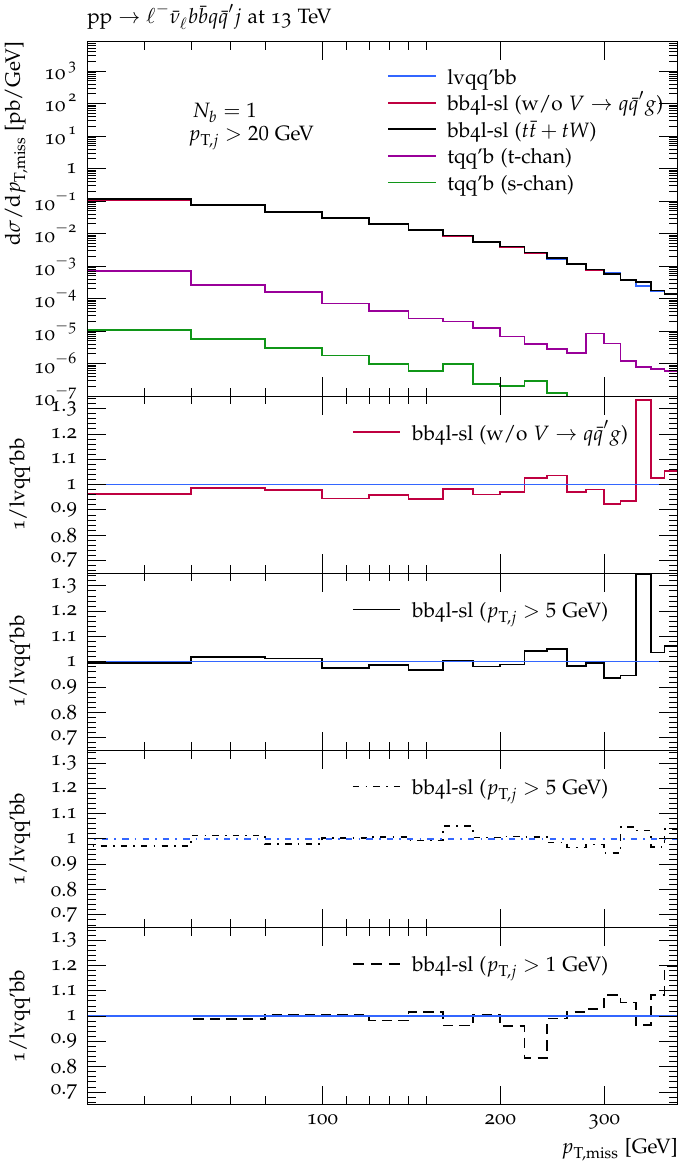}{MET_BJ}\\
\appBBplot{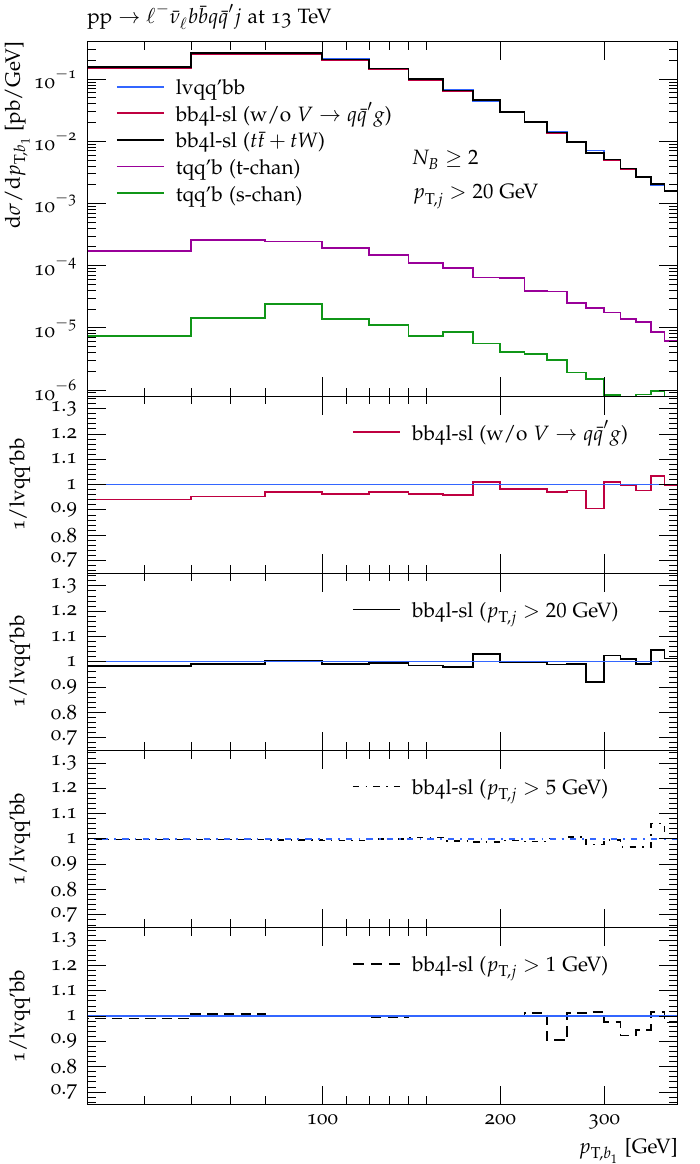}{PTB1_BBJ}
\appBBplot{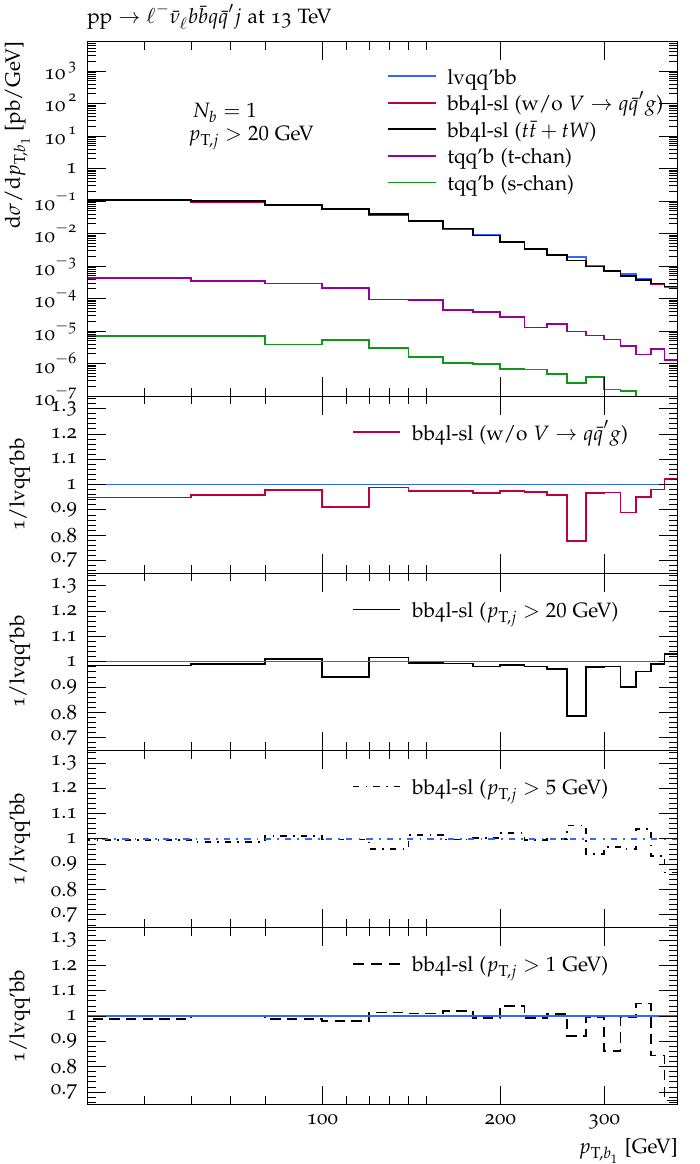}{PTB1_BJ}
\vspace{-4mm}
  \end{center}
\caption{Same predictions and ratios as in \reffi{fig:nlo_comparison_A}
for the distributions in the missing 
$p_\rT$ (\subref{MET_BBJ}--\subref{MET_BJ})
and in the $p_\rT$ of the first $B$-jet
(\subref{PTJ2_BBJ}--\subref{PTJ2_BJ}).
}
\label{fig:nlo_comparison_B}
\end{figure}

\begin{figure}
  \begin{center}
\appBBplot{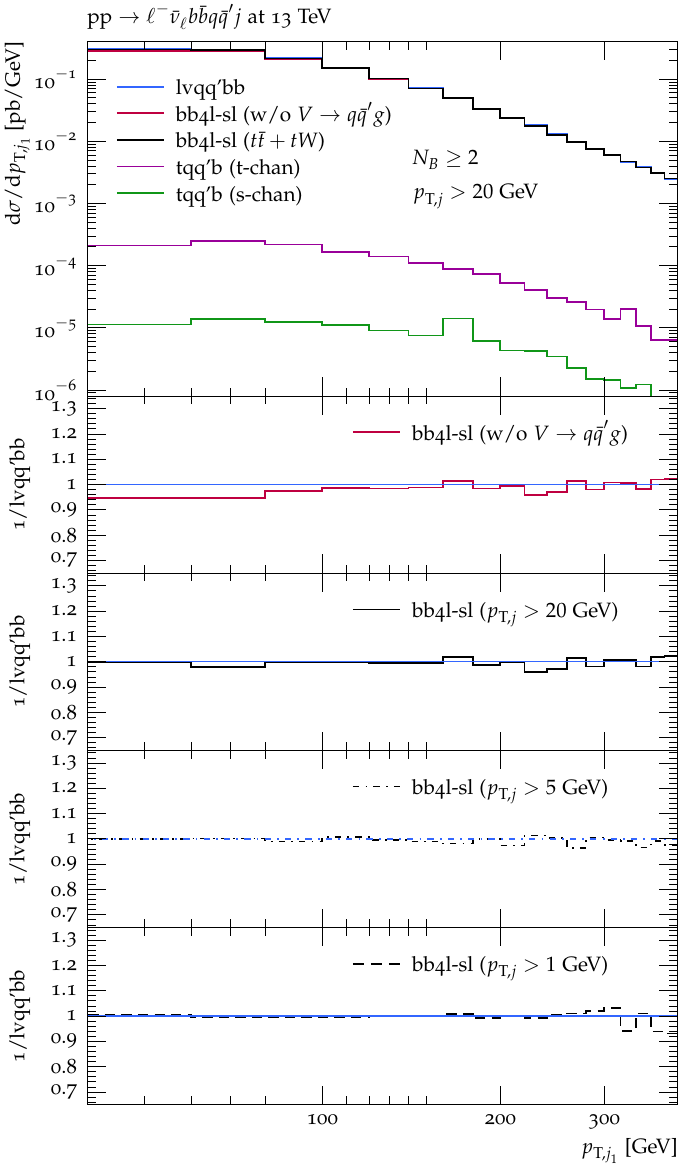}{PTJ1_BBJ}
\appBBplot{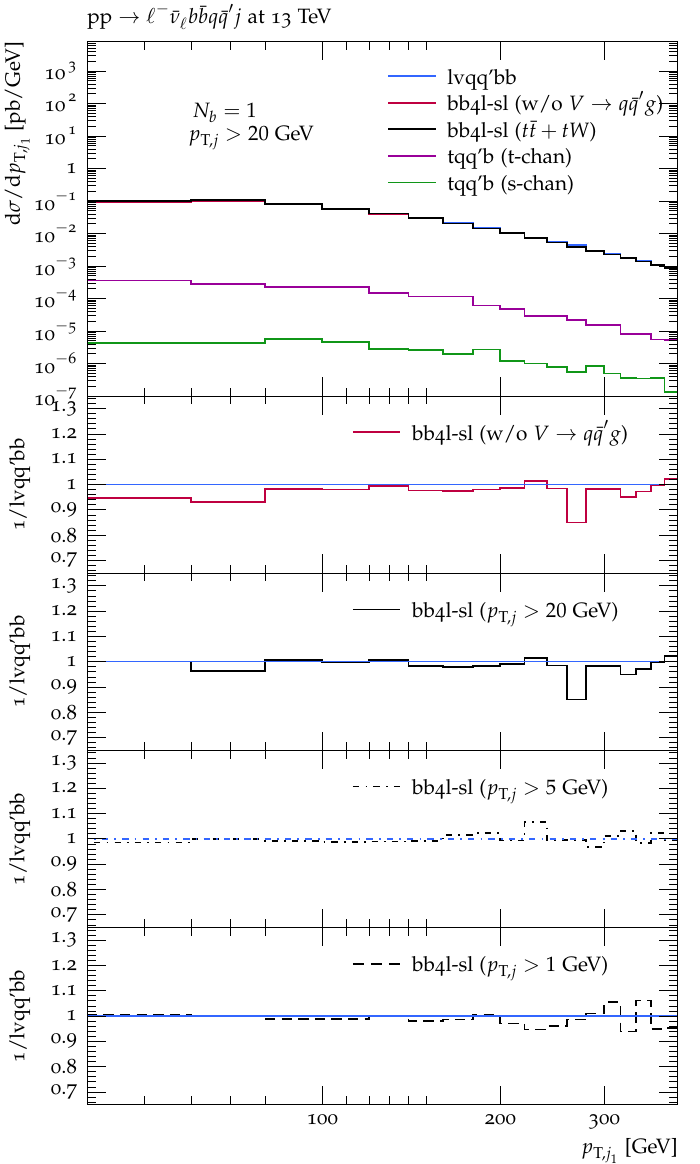}{PTJ1_BJ}\\
\appBBplot{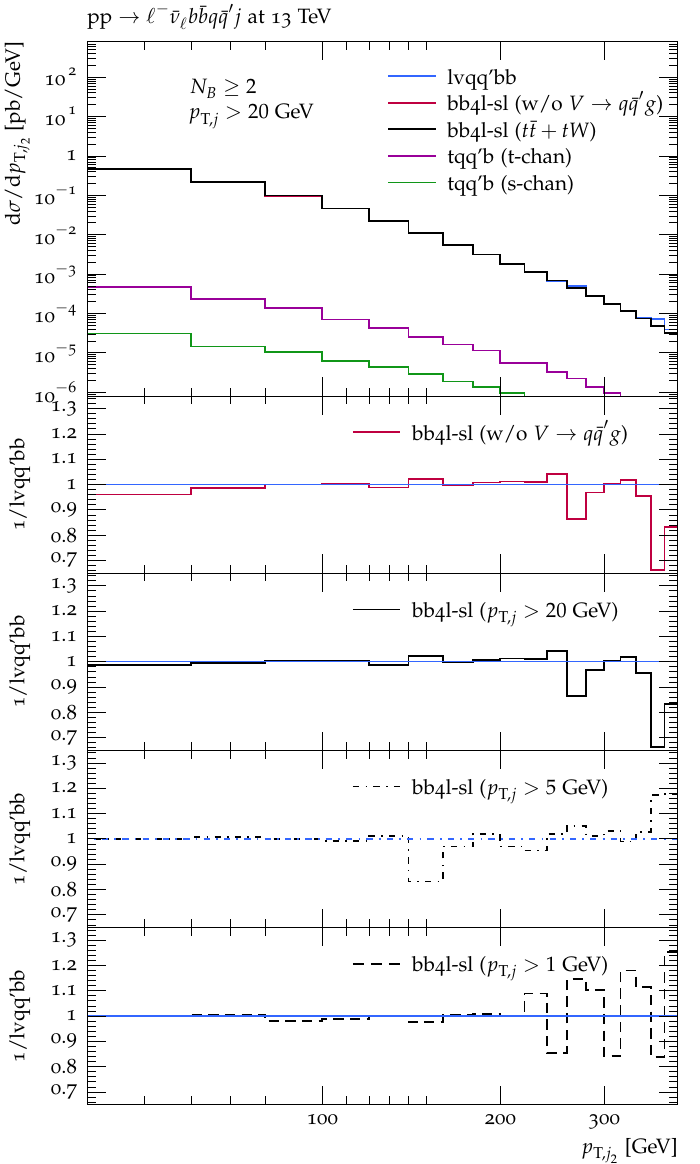}{PTJ2_BBJ}
\appBBplot{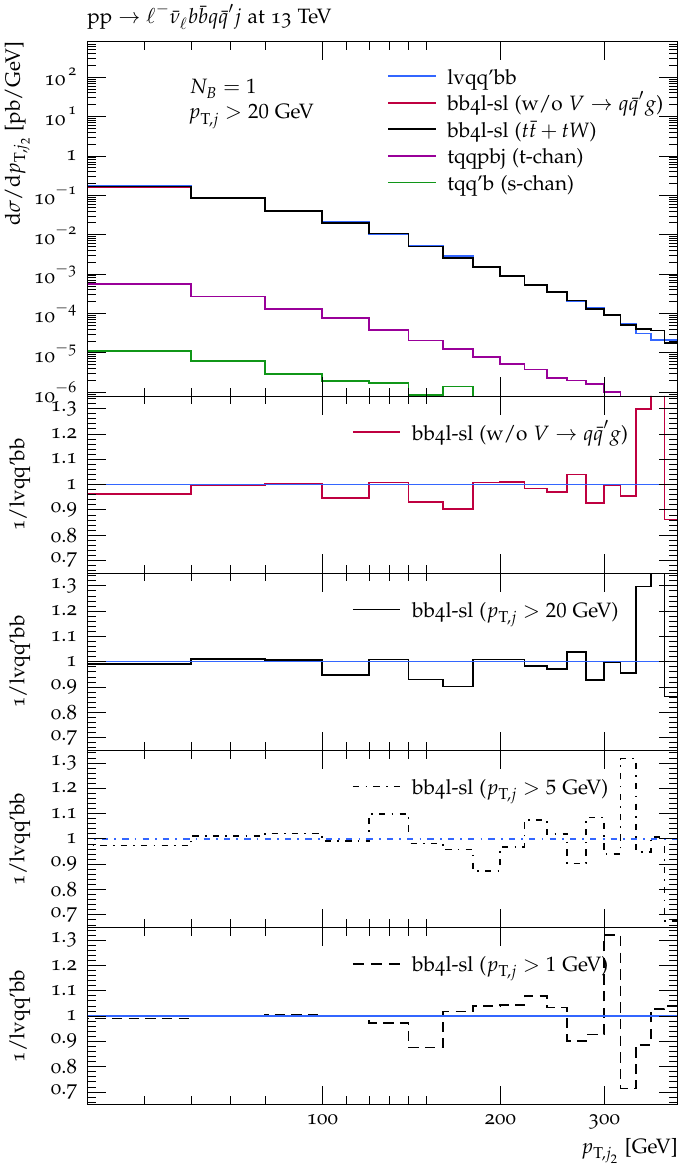}{PTJ2_BJ}
\vspace{-4mm}
  \end{center}
\caption{Same predictions and ratios as in \reffi{fig:nlo_comparison_A}
for the distributions in the $p_\rT$ of the first 
(\subref{PTJ1_BBJ}--\subref{PTJ1_BJ})
and of the second 
(\subref{PTJ2_BBJ}--\subref{PTJ2_BJ})
light jet.
}
	\label{fig:nlo_comparison_C}
\end{figure}

In~\reffi{fig:nlo_comparison_A} we present the
distribution in the dijet mass, defined 
as the invariant mass of 
the system of two or three light jets 
that is closest to $m_W$.
For this observable we note that 
the inclusion of radiation in the $W\to q\bar q'$ decay
is quite relevant, especially in the region below 
the $W$ resonance, where it can increase the 
differential cross section by up to 
$30\%$. In general, for the dijet mass distribution and for all
$p_\rT$ distributions in \reffis{fig:nlo_comparison_B}{fig:nlo_comparison_C}
we find that---for all considered 
jet-$p_\rT$ thresholds---the \bbfourlsl{} contribution 
agrees with the exact off-shell cross section
 at the $1\%$ level.
In particular, it is interesting to note that 
also the tail of the distribution in the 
$p_\rT$ of the second light jet
(\reffi{fig:nlo_comparison_C}\subref{PTJ2_BBJ}--\subref{PTJ2_BJ})
is free from any sizeable contribution beyond
the one from \bbfourlsl.
This is due to the fact that, 
thanks to the emission of NLO QCD radiation, 
the abrupt suppression 
of the $\ttbar+tW$ cross section observed 
in~\reffi{fig:NB21_comparison_2}\subref{PTJ2_BB}--\subref{PTJ2_B}
disappears, i.e.~$\ttbar+tW$ production can populate the phase space with two 
high-$p_\rT$ light jets in a similarly efficient way as the
single-top subprocesses.

These findings provide further evidence of the consistency of the 
process definition adopted for the \bbfourlsl generator.

\FloatBarrier
\bibliography{paper}

\end{document}

%% file: texmacros.tex
\def\rtb{real-to-Born\xspace}
\def\Whad{W_{\mathrm{had}}}
\def\DPA{\mathrm{DPA}}
\newcommand{\proj}{\omega}
\newcommand{\weight}{\rho}
\newcommand{\labcoll}{c}
\newcommand{\setcoll}{\mathcal{C}}
\newcommand{\labhist}{h}
\newcommand{\sethist}{\mathcal{H}}
\newcommand{\labsubp}{s}
\newcommand{\setsubp}{\mathcal{S}}
\newcommand{\weightcollal}[1]{\weight^{(\mathrm{coll})}_{#1}}
\newcommand{\weighthistal}[1]{\weight^{(\mathrm{hist})}_{#1}}
\newcommand{\rR}{\mathrm{R}}
\newcommand{\rrad}{\mathrm{rad}}
\newcommand{\Phirad}[1]{\Phi_{\rrad}}
\newcommand{\Phiradal}[1]{\Phi_{\rrad,#1}}

\def\pro{{\rm prod}}
\def\dec{{\rm dec}}
\def\sigfake{\sigma_{\rm spurious}}
\def\kappafake{\kappa_{\rm spurious}}
\def\pxd{\pro \times \dec}

\newcommand{\sigofford}[1]{\sigma_{\rm off-shell}^{#1}}
\newcommand{\sigpxd}{\sigma_{\pxd}}                
\newcommand{\sigpxdord}[1]{\sigma_{{\pro}\times{\dec}}^{#1}}
\newcommand{\sigp}{\sigma}
\newcommand{\sigpord}[1]{\sigma_{#1}}
\newcommand{\BR}{\text{BR}}
\def\NLOexp{\mathrm{NLO}_{\mathrm{exp}}}

\newcommand{\naivehist}{\mathrm{naive}}
\newcommand{\MEhist}{\mathrm{ME}}
\newcommand{\full}{\mathrm{full}}
\newcommand{\chitw}{\chi}

\newcommand{\twoLB}{2LB}
\newcommand{\LO}{\mathrm{LO}}

\newcommand{\NLO}{\mathrm{NLO}}
\newcommand{\NLOPS}{\mathrm{NLOPS}}

\newcommand{\OpenLoops}{{\rmfamily\scshape OpenLoops}\xspace}

\newcommand{\Pythia}{{\tt Pythia}\xspace}

\newcommand\POWHEGBOX{{\tt POWHEG\,BOX}\xspace}
\newcommand\POWHEG{{POWHEG}\xspace}
\newcommand\POWHEGRES{{POWHEG--RES}\xspace}
\newcommand\POWRES{\POWHEGRES}
\newcommand\MCatNLO{MC@NLO}
\newcommand\PythiaEight{{\tt Pythia8}}
\newcommand\PythiaEightInTitle{\texttt{\textbf{Pythia8}}}
\newcommand\RES{{\tt POWHEG\,BOX\,RES}\xspace}

\newcommand\bbfourl{{\tt bb4l}\xspace}
\newcommand\bbfourlInTitle{\texttt{\textbf{bb4l}}\xspace}
\newcommand\bbfourldl{{\tt bb4l-dl}\xspace}
\newcommand\bbfourlsl{{\tt bb4l-sl}\xspace}
\newcommand\bbfourlslInTitle{\texttt{\textbf{bb4l-sl}}\xspace}
\newcommand\bbfourldlInTitle{\texttt{\textbf{bb4l-dl}}\xspace}
\newcommand\bbtwoltwoq{\ensuremath{\ell \nu_l q \bar{q}' b \bar{b}}\xspace}

\newcommand\allrad{{\tt allrad}\xspace}

\newcommand\hvq{{\tt hvq}}
\newcommand\STwtch{${\tt ST}_{\tt wtch}$}
\newcommand\ST{\STwtch}
\newcommand\STDR{\ST{\tt-}$\tt{DR}$}
\newcommand\STDS{\ST{\tt-}$\tt{DS}$}

\newcommand\DS{{\tt DS}}

\newcommand{\MCFM}{{\tt MCFM}\xspace}

\newcommand{\mathd}{\mathrm{d}}

\newcommand{\tmop}[1]{\ensuremath{\operatorname{#1}}}

\newcommand{\GeV}{\text{GeV}\xspace}
\newcommand{\GF}{{G_{\sss\mu}}}
\newcommand{\ri}{\mathrm{i}}

\newcommand\sss{\mathchoice%
{\displaystyle}%
{\scriptstyle}%
{\scriptscriptstyle}%
{\scriptscriptstyle}%
}
\newcommand{\rB}{\mathrm{B}}
\newcommand{\rad}{\mathrm{rad}}
\newcommand{\rT}{\mathrm{T}}
\newcommand{\rF}{\mathrm{F}}

\newcommand{\ord}{\mathcal{O}}

\newcommand{\ttbar}{\ensuremath{t \bar t}\xspace}
\newcommand{\bbbar}{\ensuremath{b \bar b}\xspace}

\newcommand{\Wt}{\ensuremath{tW}\xspace}
\newcommand{\tW}{\Wt}

\newcommand{\muF}{\ensuremath{\mu_{\mathrm{F}}}}
\newcommand{\muR}{\ensuremath{\mu_{\mathrm{R}}}}

\def\beq{\begin{equation}}
\def\beqn{\begin{eqnarray}}
\def\eeq{\end{equation}}
\def\eeqn{\end{eqnarray}}

\def\({\left(} 
\def\){\right)} 
 
\newcommand     \MSB            {\ifmmode {\overline{\rm MS}} \else
                                 $\overline{\rm MS}$\fi}

\newcommand\as{\ensuremath{\alpha_{\rm S}}\xspace}
\newcommand\gs{\ensuremath{g_{\rm S}}\xspace}

\newcommand\aew{\ensuremath{\alpha}\xspace}

\newcommand\pT{\ensuremath{p_{\rm T}}}

\newcount\minutes 
\newcount\scratch 
\def\timestamp{%
\scratch=\time 
\divide\scratch by 60 
\edef\hours{\the\scratch} 
\multiply\scratch by 60 
\minutes=\time 
\advance\minutes by -\scratch 
---$\,$\hours:\null 
\ifnum\minutes< 10 0\fi 
\the\minutes}

\def\refeq#1{\mbox{(\ref{#1})}}
\def\refeqs#1#2{\mbox{(\ref{#1})--(\ref{#2})}}
\def\reffi#1{\mbox{Fig.~\ref{#1}}}
\def\reffis#1#2{\mbox{Figs.~\ref{#1}--\ref{#2}}}
\def\refta#1{\mbox{Tab.~\ref{#1}}}

\def\refse#1{\mbox{Sect.~\ref{#1}}}
\def\refses#1#2{\mbox{Sects.~\ref{#1}--\ref{#2}}}
\def\refapp#1{\mbox{App.~\ref{#1}}}
\def\citere#1{\mbox{Ref.~\cite{#1}}}
\def\citeres#1{\mbox{Refs.~\cite{#1}}}

\def\rd{\mathrm{d}}

\newcommand{\calR}{\mathcal{R}}
\newcommand{\calA}{\mathcal{A}}

\newcommand{\calC}{\mathcal{C}}

\newcommand{\ie}{i.e.~}

\newcommand{\bit}{\begin{itemize}}
\newcommand{\eit}{\end{itemize}}
